\documentclass[12pt]{iopart}
\usepackage[dvips]{graphicx}
\usepackage{amssymb,amsopn, verbatim,latexsym}
\usepackage[bookmarks,dvips,colorlinks, 
pdfauthor={Martin Horvat},
pdftitle={The bends in quantum waveguides and crossproducts of Bessel functions}]{hyperref}

\begin{document}

\frenchspacing

\def\picbox#1#2{\fbox{\vbox to#2{\hbox to#1{}}}}
\def\bra#1{\langle#1|}
\def\ket#1{|#1\rangle}
\def\braket#1#2{\langle#1|#2\rangle}
\def\ave#1{\left\langle #1 \right\rangle}
\def\parc#1#2{\frac{\partial #1}{\partial #2}}
\def\pa{\partial} 
\def\scalar#1#2{\langle#1|#2\rangle}
\def\eps{\varepsilon}

\def\bC{{\mathbb{C}}}
\def\bZ{{\mathbb{Z}}}
\def\bR{{\mathbb{R}}}
\def\bN{{\mathbb{N}}}

\def\dd{{\rm d}}
\def\id{{\rm id}}
\def\ii{{\rm i}}
\def\rL{{\rm L}}
\def\rR{{\rm R}}

\def\asin{\operatorname{arcsin}}
\def\acos{\operatorname{arccos}}
\def\asinh{\operatorname{arcsinh}}
\def\acosh{\operatorname{arccosh}}
\def\atan{\operatorname{arctan}}
\def\atanh{\operatorname{arctanh}}
\def\tr{\operatorname{tr}}
\def\vol{\operatorname{vol}}
\def\rot{\operatorname{rot}}
\def\grad{\operatorname{grad}}
\def\diag{\operatorname{diag}}
\def\card{\operatorname{card}}
\def\const{\operatorname{const}}

\def\beq{\begin{equation}}
\def\eeq{\end{equation}}
\def\beqa{\begin{eqnarray}}
\def\eeqa{\end{eqnarray}}
\def\beqaa{\begin{eqnarray*}}
\def\eeqaa{\end{eqnarray*}}

\def\mymat#1#2#3#4{%
{\left[\begin{array}{cc} #1 & #2 \cr #3 & #4\end{array}\right]}%
}
\def\myvec#1#2{%
{\left[\begin{array}{c} #1 \cr #2 \end{array}\right]}%
}
\def\rNo{N_{\rm o}}
\def\rNc{N_{\rm c}}
\def\roo{{\rm oo}}

\title{The bends on a quantum waveguide and cross-products of Bessel functions}

\author{Martin Horvat and Toma\v z Prosen}

\address{Physics Department, Faculty of Mathematics and Physics, 
University of Ljubljana, Slovenia}

\eads{\mailto{martin.horvat@fmf.uni-lj.si},
      \mailto{tomaz.prosen@fmf.uni-lj.si}}  
\begin{abstract}
A detailed analysis of the wave-mode structure in a bend and its incorporation into a stable algorithm for calculation of the scattering matrix of the bend is presented. The calculations are based on the modal approach. The stability and precision of the algorithm is numerically and analytically analysed. The algorithm enables precise numerical calculations of scattering across the bend. The reflection is a purely quantum phenomenon and is discussed in more detail over a larger energy interval. The behaviour of the reflection is explained partially by a one-dimensional scattering model and heuristic calculations of the scattering matrix for narrow bends. In the same spirit we explain the numerical results for the Wigner-Smith delay time in the bend.
\end{abstract}
\submitto{\JPA}
\pacs{02.30.Gp, 
02.60.-x,
03.65.Nk,
05.60.Gg,
52.25.Tx,
84.40.Az}


\section{Introduction}
The wave propagation in bent waveguides has a long and rich history of research that dates back to Lord Rayleigh \cite{reyleigh:phil_mag:1897} and continues to the present days. Initially bends have been investigated in the framework of the electromagnetic theory, but more recently also quantum mechanical aspects attracted a lot of attention. The bends are popular subject of investigation, because they are typical elements incorporated into designs of waveguides. The computation of their properties to sufficiently high precision seems to be a difficult problem in the regimes of high energies and high curvatures even today. \par
\begin{figure}[!htb]
 \centering
 \includegraphics[width=8cm]{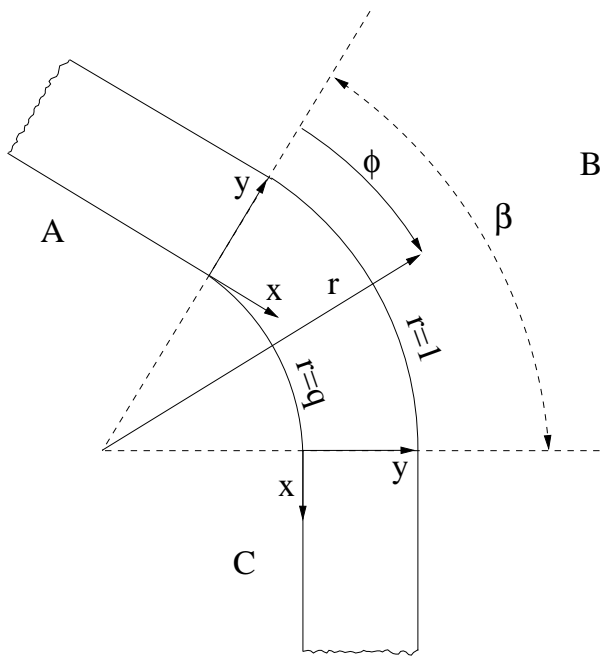}
\caption{A schematic picture of a finite bend of the inner radius $r=q$ and the outer radius $r=1$ on a straight waveguide of width $a=1-q$.}
\label{pic:schema}
\end{figure}

We are discussing a bend as a scatterer of non-relativistic quantum waves on a two-dimensional ideal straight waveguide as shown in figure \ref{pic:schema}. Such a structure is referred to as {\it an open billiard}. The past research of quantum aspects of bends can be separated into two branches. These are studies of bound states, their existence \cite{jensen:ann:71, exner:jmp:89, exner:jmp:93} and spectra \cite{lin:prb:96} and the scattering properties, which are both reviewed in reference \cite{londergan:book:99}. In order to describe quantum phenomena over our open billiard several approaches have been used in the past: Green function approach \cite{spivack:wrm:02}, finite difference mesh calculations \cite{lent:apl:90} and mode-matching techniques (MMT) using natural modes i.e. eigenfunctions of the Laplacian in the bend \cite{cochran:radsci:66, accatino:proc:90, sols:prb:90, sprung:jap:92, lin:prb:96, rashid:proc:02} and other bases \cite{amari:proc:00}. The work \cite{lin:prb:96} is particularly interesting as it raises the question on how to stabilise the calculations and gives a MMT method that is stable, but unfortunately a bit ambiguous. The MMT based on natural modes is called {\it the modal approach} and is the main topic of discussion in the present paper. The modal approach looks the most promising to deal with bends, because of its simplicity, power of interpretation and precision of results. But it also hides some problems that we examine here in detail.\par
Let us introduce the modal approach in our open billiard composed of a bend, with the inner radius $r=q$ and the outer radius $r=1$, and a straight waveguide of width $a=1-q$ as shown in figure \ref{pic:schema}. The area of the billiard, denoted by $\Omega$, can be separated into three sections: A -- the left lead, B -- the bend and C -- the right lead. We are searching for the wave function $\psi({\bf r}\in\Omega)$, which solves the stationary Schr\"{o}dinger (Helmholtz) equation on $\Omega$ with Dirichlet boundary conditions reading
\beq
  -\Delta \psi ({\bf r}) = k^2 \psi ({\bf r})\>,\qquad 
  \psi|_{{\bf r} \in \partial \Omega} = 0\>,
  \label{eq:eigen_problem}
\eeq
where $E = k^2$ is the energy and $k$ the corresponding wavenumber. The Helmholtz equation (\ref{eq:eigen_problem}) is written in Cartesian coordinates ${\bf r} =(x,y)$ in the asymptotic regions A and C as
\beq
  -\left(\frac{\pa^2}{\pa x^2} + 
  \frac{\pa^2}{\pa y^2}\right) \psi = k^2 \psi\>,
  \quad
  \psi|_{y=0,1-q} = 0\>, 
\label{eq:eigen_straight}
\eeq
and in polar coordinates ${\bf r} =(r,\phi)$ across the bend B  as
\beq
  -\left(\frac{\pa^2}{\pa r^2} +  \frac{1}{r}\frac{\pa}{\pa r}  + 
  \frac{1}{r^2} \frac{\pa^2}{\pa \phi^2}\right)\psi = k^2 \psi\>,
  \quad
  \psi|_{r=q,1} = 0\>.
  \label{eq:eigen_bend}
\eeq
The modal approach suggests that we first solve Helmholtz equation (\ref{eq:eigen_problem}) on each region of the open billiard separately. Thereby we obtain partial solutions called {\it mode functions}, which are then used to describe the solution across the whole billiard. In the infinite straight waveguide and in the bend the mode functions are given by the ans\"atze $\psi \propto u(y) \exp(\ii g x)$ and $\psi \propto  U (r) \exp(\ii \nu \phi)$, respectively. The ans\"atze in equations (\ref{eq:eigen_straight}) and (\ref{eq:eigen_bend}) give the following equations for the mode functions:
\beqa
 &\frac{\dd^2 u}{\dd y^2} + (k^2 -g^2) u = 0\>,
  \quad 
  &u|_{y=0,a}= 0\>,\quad  y\in[0,a]\>, \label{eq:straight_cross_eq0}
\\
&\frac{\dd^2 U}{\dd r^2} + 
 \frac{1}{r}\frac{\dd U}{\dd r} + \left(k^2-\frac{\nu^2}{r^2}\right)U = 0\>,
 \quad
 &U|_{r=q,1}= 0\>, \quad r\in[q,1]\>, \label{eq:bend_cross_eq0}
\eeqa
where $a=1-q$ is the channel's width and the scalars $g,\mu\in\bC$ are called {\it mode numbers}. The equations (\ref{eq:straight_cross_eq0}) and (\ref{eq:bend_cross_eq0}) have a discrete set of solutions i.e. the mode functions and the corresponding mode numbers, which we refer to as {\it modes}. 

The modes in the straight waveguide and in the bend are denoted by pairs $(g_n, u_n(y))$ and  $(\nu_p, U_p(r))$, respectively, where $p,n\in\bN$. The modes in 
the straight waveguide are explicitly written as
\beq
  k^2 = g_n^2 + \left(\frac{\pi n}{a}\right)^2\>,\quad
  u_n(x) = \sqrt{\frac{2}{a}} \sin\left( \frac{\pi}{a} n x \right)\>,
  \label{eq:straight_mode}
\eeq
whereas modes in the bend are more complicated. They are discussed in Section \ref{sec:bess}, where we also show that mode numbers are either real or imaginary. The mode functions $u_n(y)$ and $U_p(r)$ are then used in the bases of functions in which we expand waves over parts of the waveguide. The basis in the straight waveguide is given by
\beq
  e_n^\pm({\bf r})= u_n(y) \frac{\exp(\pm\ii g_n x)}{\sqrt{g_n}}\>,
  \label{eq:mode_fun_straight}
\eeq
and in the bend by
\beq
  f_p^\pm({\bf r}) = U_p (r)  \frac{\exp(\pm\ii \nu_p \phi)}{\sqrt{\nu_p}}\>,
  \label{eq:mode_fun_bend}
\eeq
where the sign $\pm$ labels the two directions of phase (probability) flux propagation.  We define the square-root of a complex number $z=|z|\exp(\ii \phi)$, $\phi\in [0,2\pi)$ as $\sqrt{z} = |z|^{1/2} \exp (\ii \phi/2)$. The basis functions are called {\it wave modes} or {\it modes of the Laplacian}.  We distinguish two types of wave modes. The wave modes corresponding to real and imaginary mode numbers are called {\it open modes or travelling waves} and {\it closed modes or decaying (evanescent) waves}, respectively. The wave-function $\psi({\bf r})$ (\ref{eq:eigen_problem}) in the entire open billiard region $\Omega$ is expressed in terms of the wave modes as 
\beqa
  \psi({\bf r}) &= \sum_n a^+_n e_n^+ ({\bf r})+ a^-_n e_n^- ({\bf r})\>,
  \quad &{\bf r} \in \Omega_{\rm A}\>,
  \label{eq:param_A}\\
 \psi({\bf r}) &= \sum_p \lambda^+_p f_p^+({\bf r})+\lambda^-_p f_p^-({\bf r})\>,
 \quad &{\bf r} \in \Omega_{\rm B}\>,
  \label{eq:param_B}\\
 \psi({\bf r}) &= \sum_n b^+_n e_n^+({\bf r})+ b^-_n e_n^-({\bf r})\>,
 \quad &{\bf r} \in \Omega_{\rm C}\>,
 \label{eq:param_C}
\eeqa
where $\Omega_{A, B, C}$ are regions corresponding to sections A, B and C, respectively. The expansion coefficients $a^\pm_n$,  $\lambda^\pm_p$ and  $b^\pm_n$ are determined by the condition that the wave function $\psi(\vec r)$ is smooth everywhere in $\Omega$, in particular on the boundaries
between different regions $\Omega_{A,B,C}$. The solution of the presented problem will be discussed in Section \ref{sec:scatt}. \par
The paper is organised as follows.
In Section \ref{sec:bess} we present a detailed study of the mode structure in the bend, which is closely related to the work of Cochran \cite{cochran:jciam:64, cochran:pcps:66, cochran:qjmam:66}. In comparison to the work of others, ours is directed more towards the application of the mode structure to scattering calculations. In addition, we write explicit formulae for the mode functions in the bend, where we give special attention to the closed modes. In Section \ref{sec:scatt} we outline a numerically stable MMT for calculation of the scattering matrix \cite{newton:book:02} of a single bend. The section \ref{sec:scatt} is concluded with the presentation of numerical results obtained by our method and compared to analytic estimates of the quantum transport properties of the bend. By considering the analogy between the quantum theory and EM theory we can connect our work to the EM wave propagation of longitudinal magnetic waves \cite{cochran:radsci:66}.

\section{The cross-product of Bessel functions}\label{sec:bess}
In this section we analyse the properties of the mode numbers and the corresponding mode functions for a given wavenumber $k$ and inner radius $q$. The mode functions in the bend $U_p(r)$ are proportional to well known {\it cross-products of Bessel functions} \cite{cochran:jciam:64} of the first kind, t $J_\nu$, and Bessel functions of the second kind, $Y_\nu$, \cite{olver:72} written as
\beq
  Z_\nu (k,r) = 
  J_\nu(kr)Y_\nu(k) - Y_\nu(kr) J_\nu(k)\>,
  \label{eq:bcp1}
\eeq
or
\beq
  Z_{\nu,k} (r) = 
  \frac{J_{-\nu}(kr) J_\nu(k) - J_\nu(kr)J_{-\nu}(k)}{\sin(\nu\pi)}\>,\quad
  \nu\notin\bZ \>,
 \label{eq:bcp2} 
\eeq
where the allowed values of mode numbers $\nu$ are determined by the Dirichlet boundary conditions $Z_{\nu,k}(q) = 0$. In equation (\ref{eq:bcp2}) we have used the relation $Y_\nu (z) = (J_\nu(z)\cos(\nu\pi)- J_{-\nu}(z))/\sin(\nu\pi)$ valid for orders $\nu\notin\bZ$. The understanding of the mode structure in the bend is essential for calculations of the scattering over our open billiard in the modal approach.
\subsection{The properties of mode numbers}
The set of mode numbers at a given wave-number $k\in \bR,\, k>0$ and inner radius $q\in (0,1)$ is denoted by ${\cal M}_{k,q}=\{\nu\in\bC: Z_{\nu,k}(q) = 0\}$. The functions $Z_{\nu,k}(r)$ are even $Z_{-\nu,k}(r) = Z_{\nu,k}(r)$ and analytic in the order $\nu$ \cite{cochran:pcps:66}. These properties yield the following symmetry of the set of mode numbers:
\beq
 {\cal M}_{k,q} =  - {\cal M}_{k,q}\>,\qquad  
 {\cal M}_{k,q} ^* = {\cal M}_{k,q}\>.
 \label{eq:bcp_symm0}
\eeq
In addition we conclude that mode numbers are either purely real or purely imaginary 
\beq
{\cal M}_{k,q} \subset \bR\cup\ii\bR\>.
\label{eq:bcp_symm1}
\eeq 
The number of real modes is finite, whereas the number of imaginary modes is infinite at a finite wavenumber $k$. The proof of the later is given in \ref{sec:mode_symm}. The properties (\ref{eq:bcp_symm0}) and (\ref{eq:bcp_symm1}) enable a decomposition of ${\cal M}_{k,q}$ into two disjoint subsets of mode numbers laying on the positive ${\cal M}_{k,q,+}$ and the negative ${\cal M}_{k,q,-}$ real and imaginary axes:
\beqa
 {\cal M}_{k,q,+} &=& 
 \{\nu\in {\cal M}_{k,q}\,:\, \Re{\nu} \ge 0 \textrm{ or } \Im{\nu} \ge 0 \}\>,\\
 {\cal M}_{k,q,-} &=&
 \{\nu\in {\cal M}_{k,q}\,:\, \Re{\nu} < 0 \textrm{ or } \Im{\nu} < 0 \}  \subseteq -{\cal M}_{k,q,+}\>. \label{eq:mode_set_minus}
\eeqa
It easy to see that ${\cal M}_{k,q} = {\cal M}_{k,q,+} \cup  {\cal M}_{k,q,-}$. We call $\Re\{ {\cal M}_{k,q,+}\}$ and $\Im \{{\cal M}_{k,q,+}\}$ the set of {\it real modes} and {\it imaginary  modes}, respectively. The number of real modes in the bend $N_{\rm b}=\card \Re\{{\cal M}_{k,q,+}\}$ is equal or one more than the number of real modes in the straight waveguide $N_{\rm s}= \lfloor ka/\pi\rfloor$:
\beq
  0 \le  N_{\rm b}  - N_{\rm s} \leq 1 \>.
  \label{eq:mode_ineq} 
\eeq
where $\lfloor x\rfloor$ denotes the largest integer smaller than $x$.
Taking into account the analyticity of $Z_{\nu,k}(r)$ in the order $\nu$ and in the wavenumber $k$ \cite{cochran:pcps:66} we find that $N_{\rm b}$
can be computed in the semi-classical limit, for $q\neq 0$, as
\beq
  N_{\rm b} = 
  \left\lfloor\frac{ka}{\pi} + \frac{a}{8\pi q k} + O(k^{-2})\right\rfloor\>,
  \label{eq:number_of_modes}    
\eeq
and $N_{\rm s}$ is asymptotically, as k$\to\infty$, close to $N_{\rm b}$.
The expressions (\ref{eq:mode_ineq}) and (\ref{eq:number_of_modes}) are explained in \ref{sec:mode_diff}. The asymptotic form of $Z_{\nu,k}(r)$ in the order parameter \cite{olver:62} reads as
\beq
  Z_{\nu,k}(r) = 
  \frac{1}{\pi\nu}[r^\nu O(1) - r^{-\nu} O(1)]\>.
  \qquad 
  |\nu| \gg 1\>.
  \label{eq:tay_asim}
\eeq
From equation (\ref{eq:tay_asim}) we learn that $Z_{\nu,k}(q)$ diverges exponentially with increasing order parameter on the real axis as $O(q^{-|\nu|})$ and oscillates along the imaginary axis. This bounds real mode numbers $\Re\{{\cal M}_{k,r,+}\}$ from above and indicates that there is an infinite number of almost periodic imaginary mode numbers.\par
\begin{figure}[!htb]
\centering
\includegraphics[width=7.5cm]{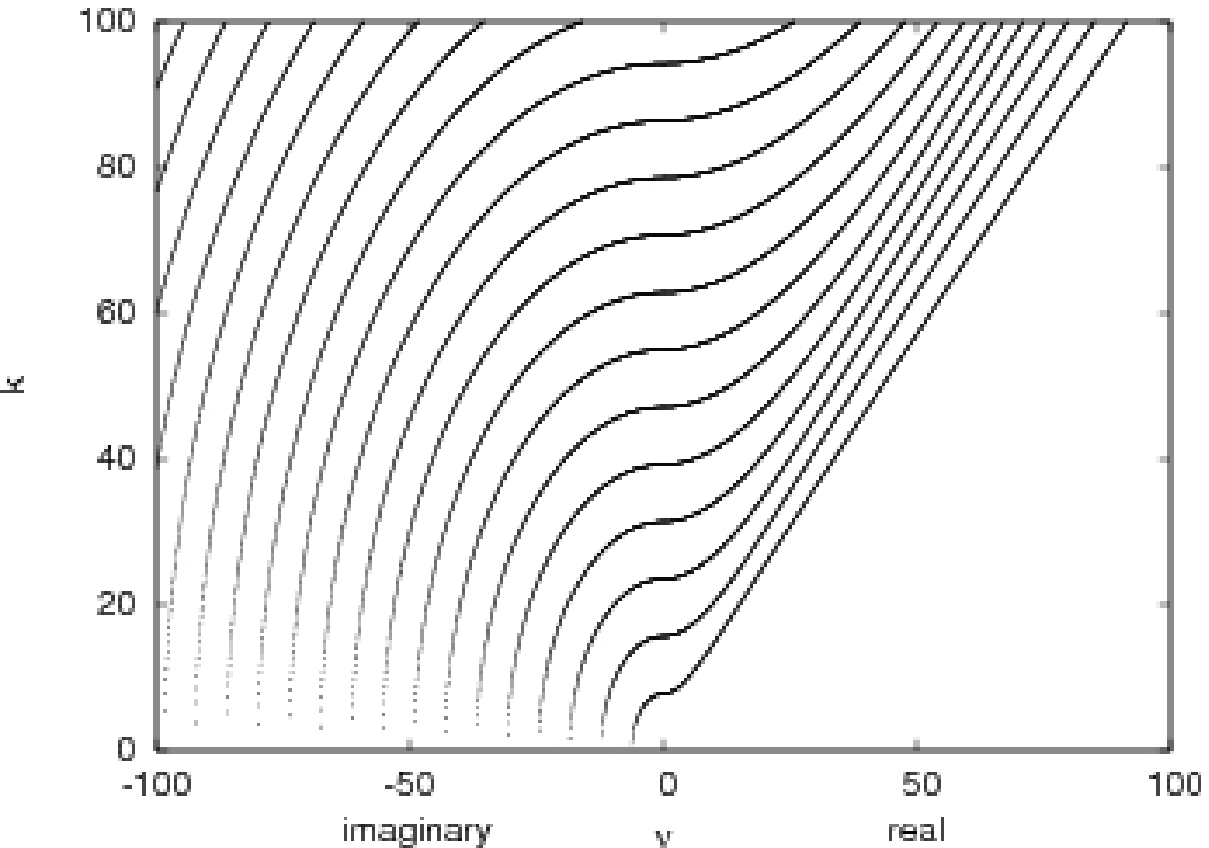}%
\includegraphics[width=7.5cm]{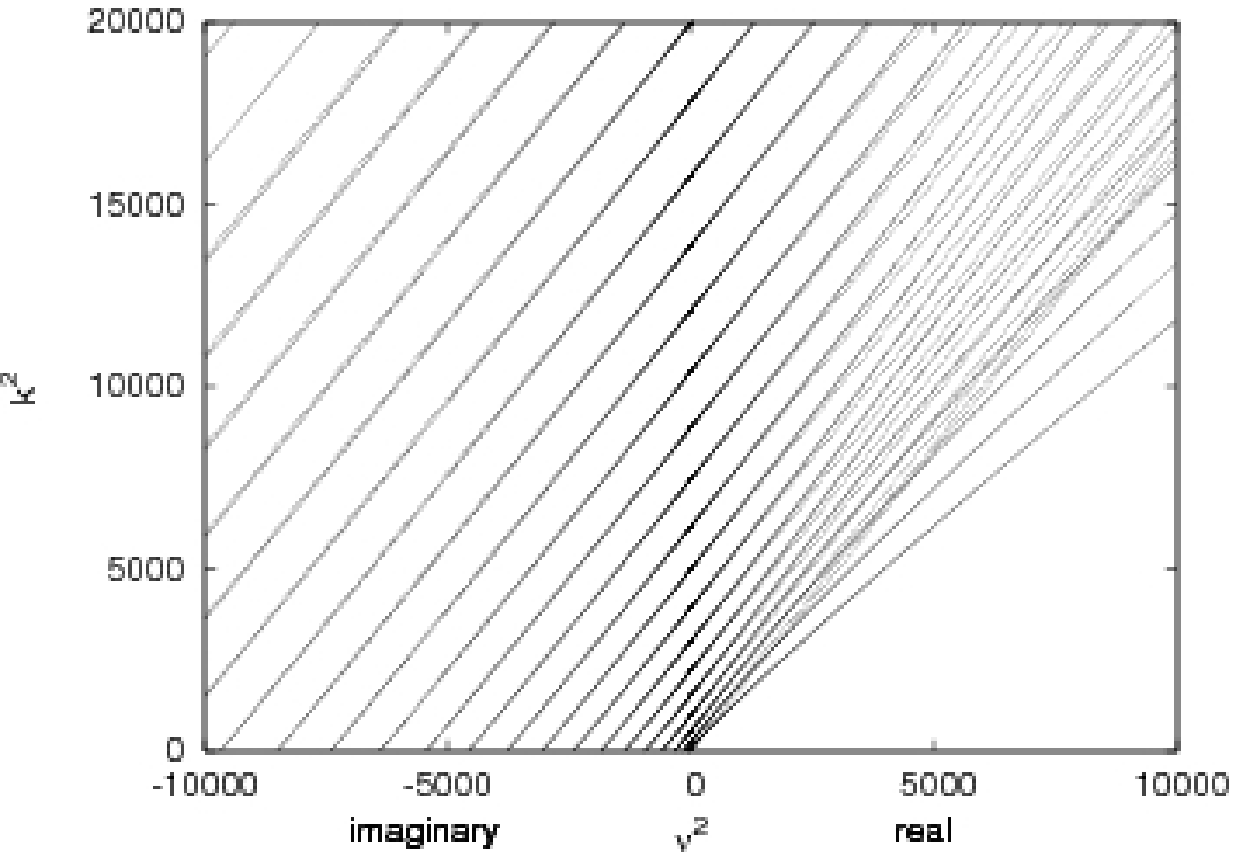}
\hbox to15cm{\hfil(a)\hfil\hfil(b)\hfil}
\caption{Two different representations of mode numbers $\nu$ at the corresponding wavenumbers $k$ in the bend with $q=0.6$. In (a) we plot $k$ vs. $\nu$, where we separately discuss real and imaginary mode numbers depicted on the right and left side of the abscissa, respectively, and in (b) we plot $k^2$ vs. $\nu^2$. The dashed lines represent solutions of equation (\ref{eq:useful_bess_apr}).}
\label{pic:bess_value}
\end{figure}

The mode numbers $\nu\in{\cal M}_{k,r}$ and consequently the mode functions can be in general obtained only numerically. We use different approximations of mode numbers to improve their numerical computation. 
By using the Debye approximation of Bessel functions for imaginary orders $\nu=\ii y$ ($y\in\bR$) valid for $y^2 + (qk)^2 \gg 1$ and the Dirichlet condition $Z_{\nu,k}(q)=0$ we obtain the following relation
\beq
\fl\hspace{1cm} 
  \sqrt{y^2 + k^2} - \sqrt{y^2 + (kq)^2}  + 
  y \log\left[ q\frac{y + \sqrt{y^2 + k^2}}{y + \sqrt{y^2 + (kq)^2}}\right] =
  \pi n\>,
  \quad 
  n\in\bN \>.
  \label{eq:WKB_imag}
\eeq
The solution of the equation (\ref{eq:WKB_imag}) in variable $y\gg k$ represents an asymptotic approximation of imaginary mode numbers $\nu_n=\ii y_n$ and is written as
\beq
  y_n = \frac{\pi n}{|\log q|} -  \frac{(ka)^2 }{4 \pi n} + O(n^{-3})\>,
  \quad 
  n\gg 1\>.
  \label{eq:asym_imag}
\eeq
The first term in equation (\ref{eq:asym_imag}) is already well known and can be also obtained from equation (\ref{eq:tay_asim}), see \cite{cochran:jciam:64}. The divergence of $Z_{\nu,k}(q)$ for $\nu\to\infty$ makes the finding of high real mode numbers, especially at large $k$, extremely difficult. We stabilize the search by using an analytic approximation of the highest real mode number $\nu_{\rm max} (k, q) = \max\{\nu: Z_{\nu,k}(q) =0\}$ for $q$ sufficiently far away from 0. This is achieved by using the asymptotic expansion of Bessel functions \cite{olver:72} in the transitional regime yielding
\beq 
  \nu_{\rm max}(k,q)  = k - 
   \root 3 \of {\frac{k}{2}}
    \left [ a_0  + 
            a_1 \exp \left(-\frac{2^{7\over 3} a^{3\over 2}}{3} k \right)
    \right ] + O(k^{-{1\over 3}})\>,  
  \label{eq:nu_bess_asym}
\eeq
where  
\beq
  a_1 = -\frac{{\rm Bi}(-a_0)}{2 {\rm Ai}'(-a_0)} \doteq 0.323685\>, 
  \qquad 
  a_0 \doteq 2.3381074\>.
\eeq
The constant $a_0$ is the negative first zero of the Airy function, ${\rm Ai}(-a_0)=0$. The exact implicit formula $Z_{\nu,k}(q) = 0$ for mode numbers at given $k$ and $q\neq 0$ has an interesting simple first order approximation reading
\beq
  \left(\frac{k}{k_0(n,q)}\right)^2 - 
  \left(\frac{\nu \log q}{\pi n}\right)^2  \approx 1\>,
  \label{eq:useful_bess}
\eeq
which is asymptotically exact in two independent limits: $\nu={\rm fixed}$, $k\to\infty$, and $k={\rm fixed}$, $|\nu|\to\infty$. The relation (\ref{eq:useful_bess}) represents a useful approximation of mode numbers  and is to our knowledge a new uniform approximation of modes in a bend. The expression $k_0(n,q)$ is the $n$-th zero of $Z_{0,k}(q)$, which can be easily found numerically. In the limit of large $n$, where we can use $k_0 (n,q) \approx \pi n/a$, the relation (\ref{eq:useful_bess}) in simplified to
\beq
  (k a)^2 - (\nu\log q)^2 \approx (\pi n)^2\>.
  \label{eq:useful_bess_apr}  
\eeq
The validity of this formula is illustrated in figure \ref{pic:bess_value}, where we compare mode numbers obtained from the approximate relation (\ref{eq:useful_bess_apr}) with the exacts ones.
The highest real mode $\nu_{\rm max}$ for small wave-numbers $k < \nu$ can be approximated using  equation (\ref{eq:useful_bess}) as
\beq
  \nu_{\rm max}(k,q) \approx
  \frac{\pi}{|\log q|} \sqrt{\left(\frac{k}{k_0(1,q)}\right)^2 -1}\>.
  \label{eq:nu_bess_min}
\eeq
In practical applications it is important that below the wave-number $k_{\rm low}(q)= k_0(1,q)$ there are no real modes. In wide bends with $q \approx 0$ one can expand the cross-product of Bessel functions around $q=0$ and obtain the formula
\beq
  k_{\rm low}(q)  = b_0 +  \frac{b_1}{|\log q|} + O(q^2)\>,
  \label{eq:k_low_qmin}
\eeq  
where $b_0$ is the smallest zero of the Bessel function $J_0(x)$, $J_0(b_0)=0$ and
\beq
  b_0\doteq 2.404825558\>, \qquad 
  b_1 = -\frac{\pi Y_0(b_0)}{2 J'_0(b_0)} \doteq 1.54288974\>.
\eeq
In narrow bends, where $a (=1-q) \to 0$, we can use standard stationary perturbation theory \cite{morse:book:53}, see equation (\ref{eq:schrod_1d}) in \ref{sec:mode_diff}, to approximate $k_{\rm low}(q)$. By introducing the matrix elements 
\beq
  V_{nm} = -\frac{1}{2}\int_0^1 \dd x\,
  \frac{\sin(\pi n x) \sin(\pi m x)}{(x + \gamma)^2}\>, \quad
  \gamma = \frac{q}{1-q}\>,
\eeq
we can express the lowest wave-number as
\beq
  k_{\rm low}(q) = \frac{\pi}{a} \left [
      1 + \frac{1}{\pi^2}V_{11} 
      + \frac{1}{\pi^4}\sum_{l>1} \frac{|V_{1l}|^2}{1-l^2} 
      + O\left(|V|^3\right)  \right ]^{1/2}\>.
  \label{eq:k_low_qmax}    
\eeq
The formula (\ref{eq:k_low_qmax}) has a simple first order expansion in $a=1-q$ 
\beq
  k_{\rm low}(q) = \frac{\pi}{a}  - \frac{a}{8\pi} + O\left (a^2\right)\>.
  \label{eq:k_low_qmax_exp}    
\eeq
From the expression (\ref{eq:k_low_qmax_exp}) we learn that the lowest wave-number at which real modes exists increases with increasing $q$ and converges to $\frac{\pi}{a}$.

\subsection{Numerical evaluation of mode functions in a bend}
\begin{figure}[!htb]
\centering
\includegraphics*[bb=55 66 390 300, height=4cm]{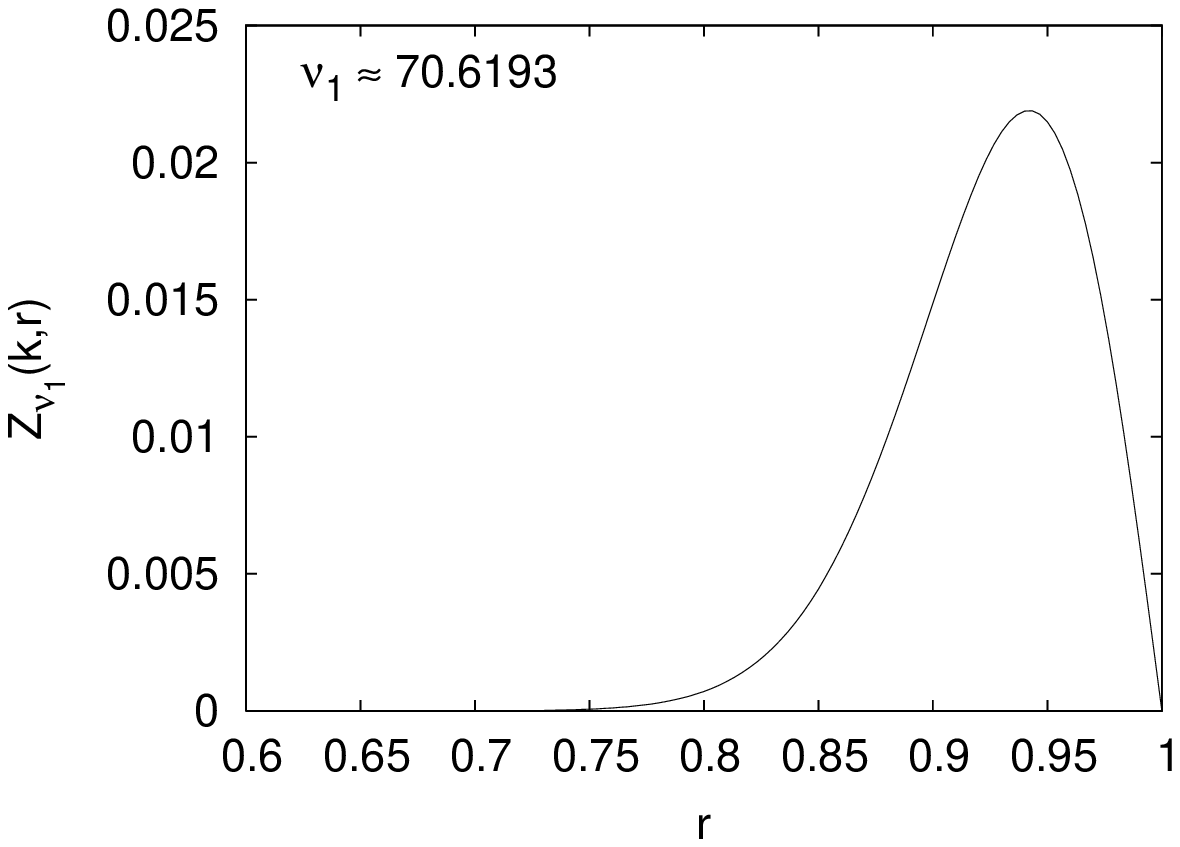}%
\includegraphics*[bb=75 66 390 300, height=4cm]{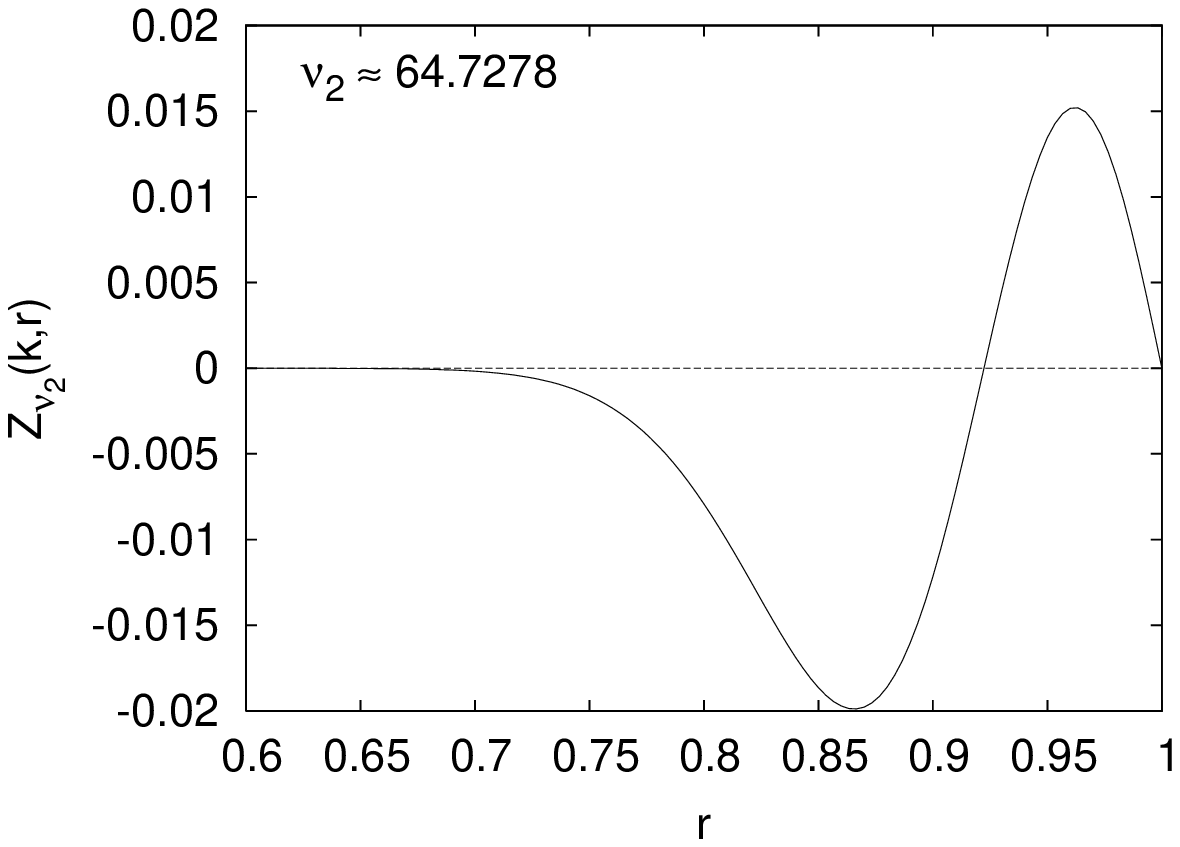}%
\includegraphics*[bb=75 66 390 300, height=4cm]{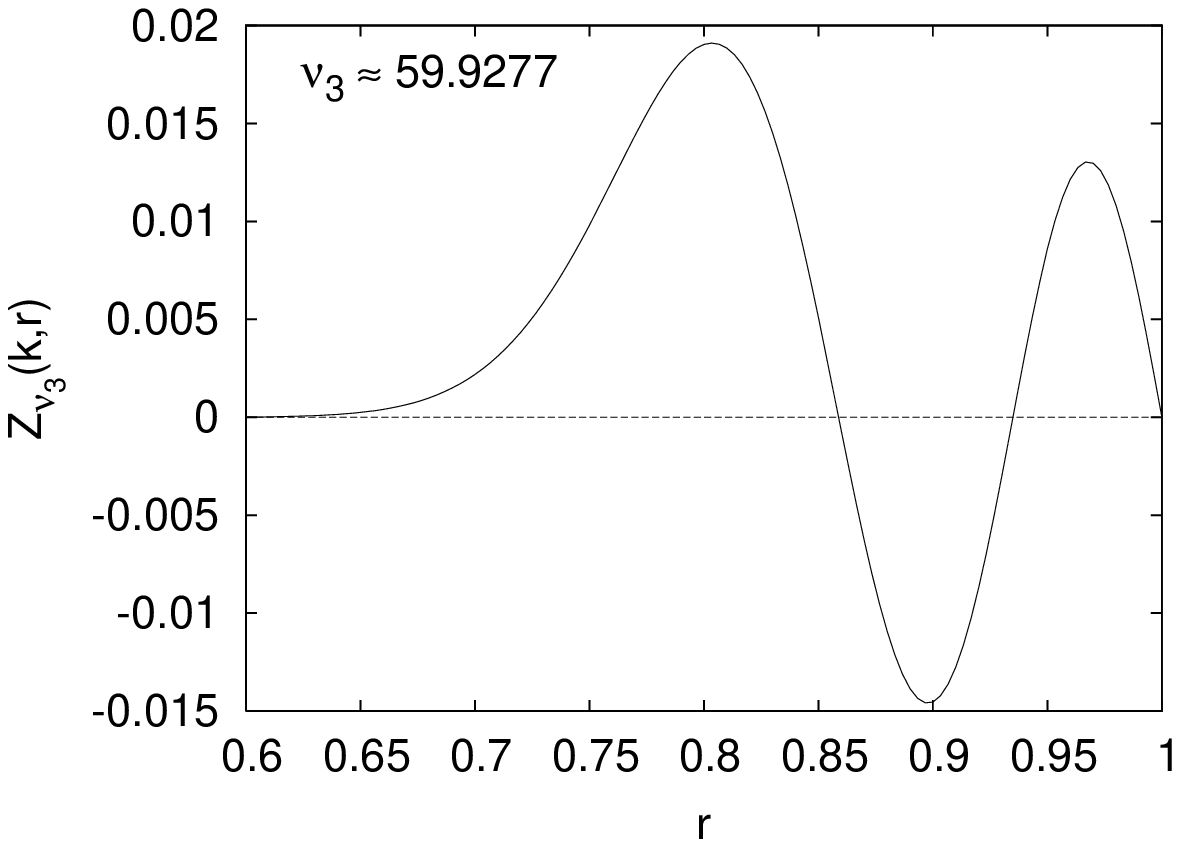}\\
\includegraphics*[bb=55 66 390 300, height=4cm]{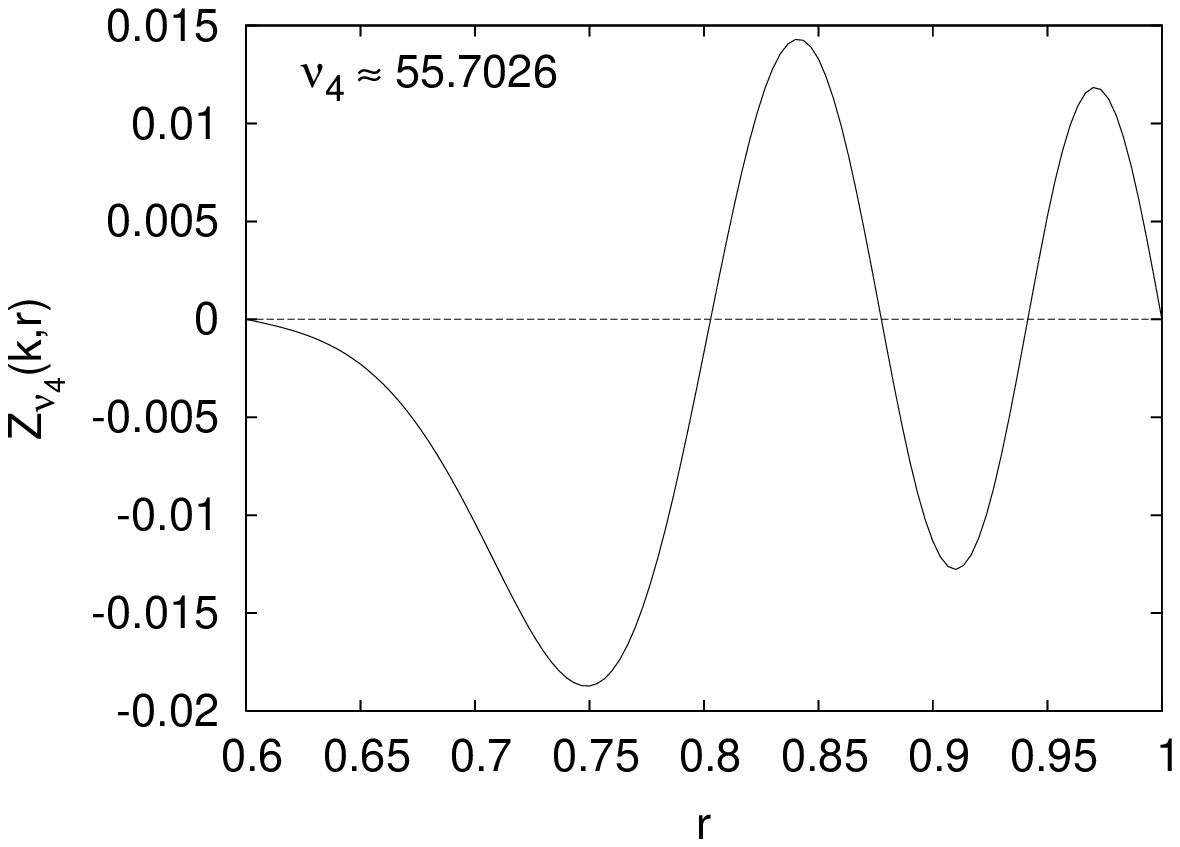}%
\includegraphics*[bb=75 66 390 300, height=4cm]{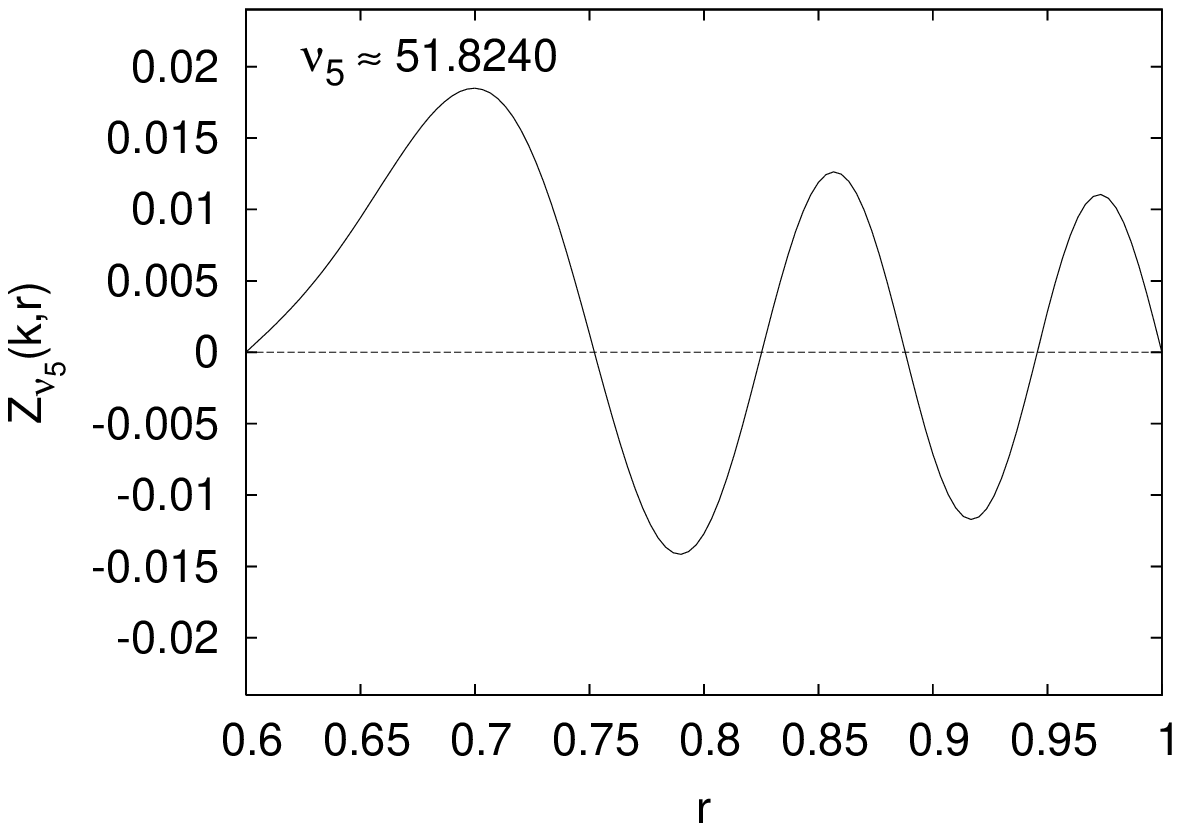}%
\includegraphics*[bb=75 66 390 300,height=4cm]{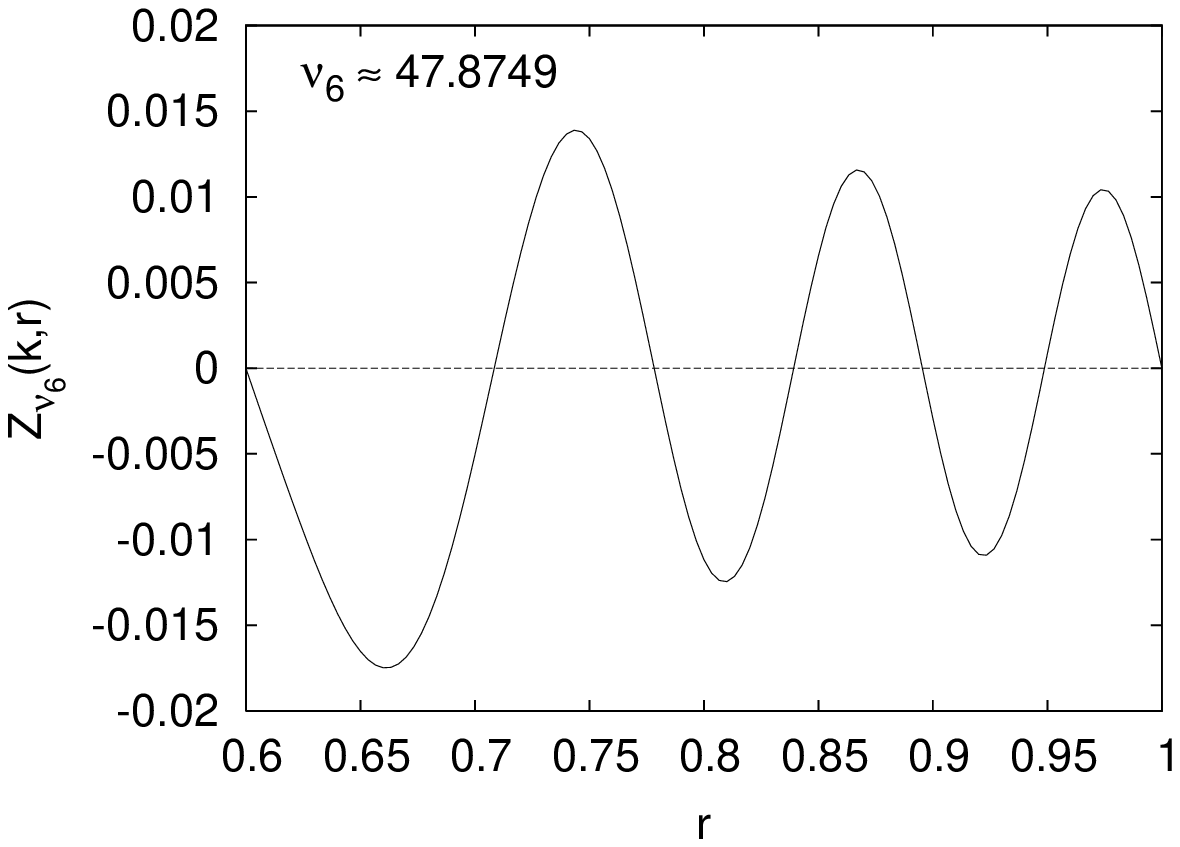}\\
\includegraphics*[bb=55 66 390 300, height=4cm]{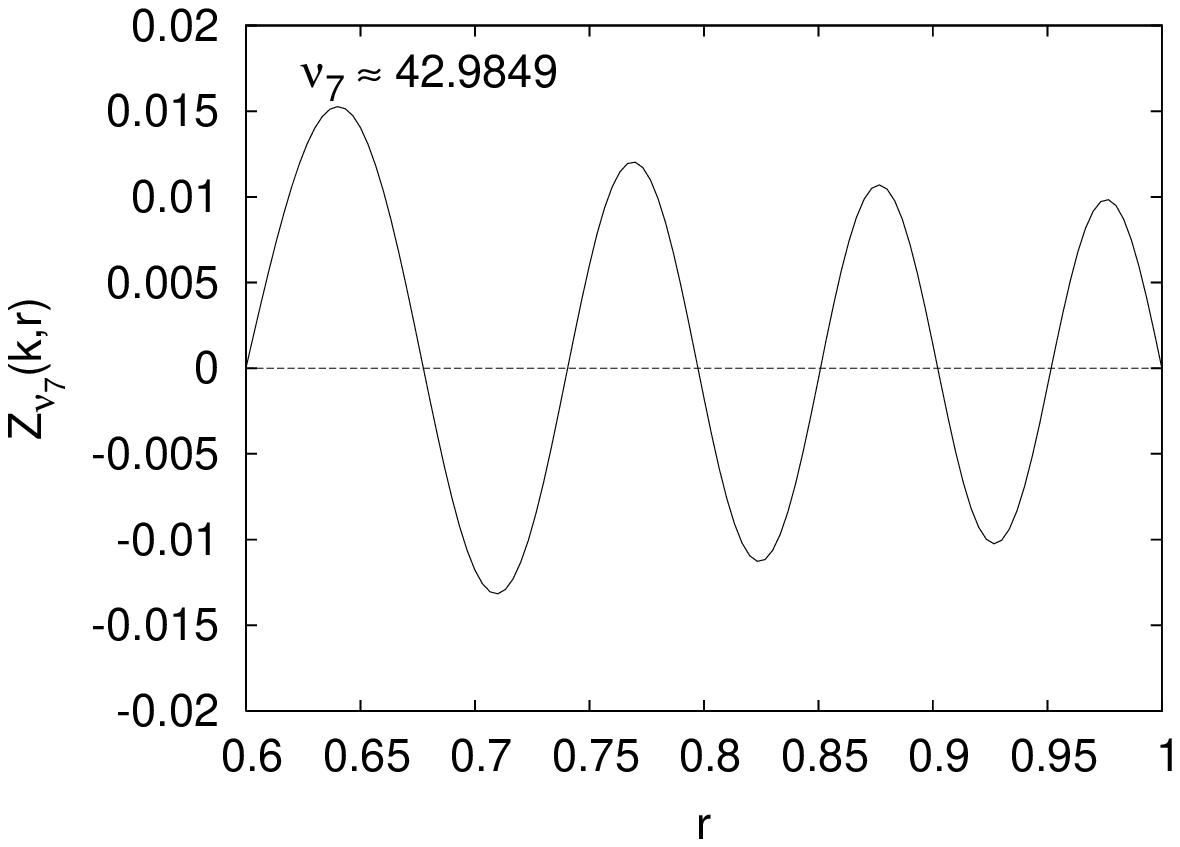}%
\includegraphics*[bb=75 66 390 300,height=4cm]{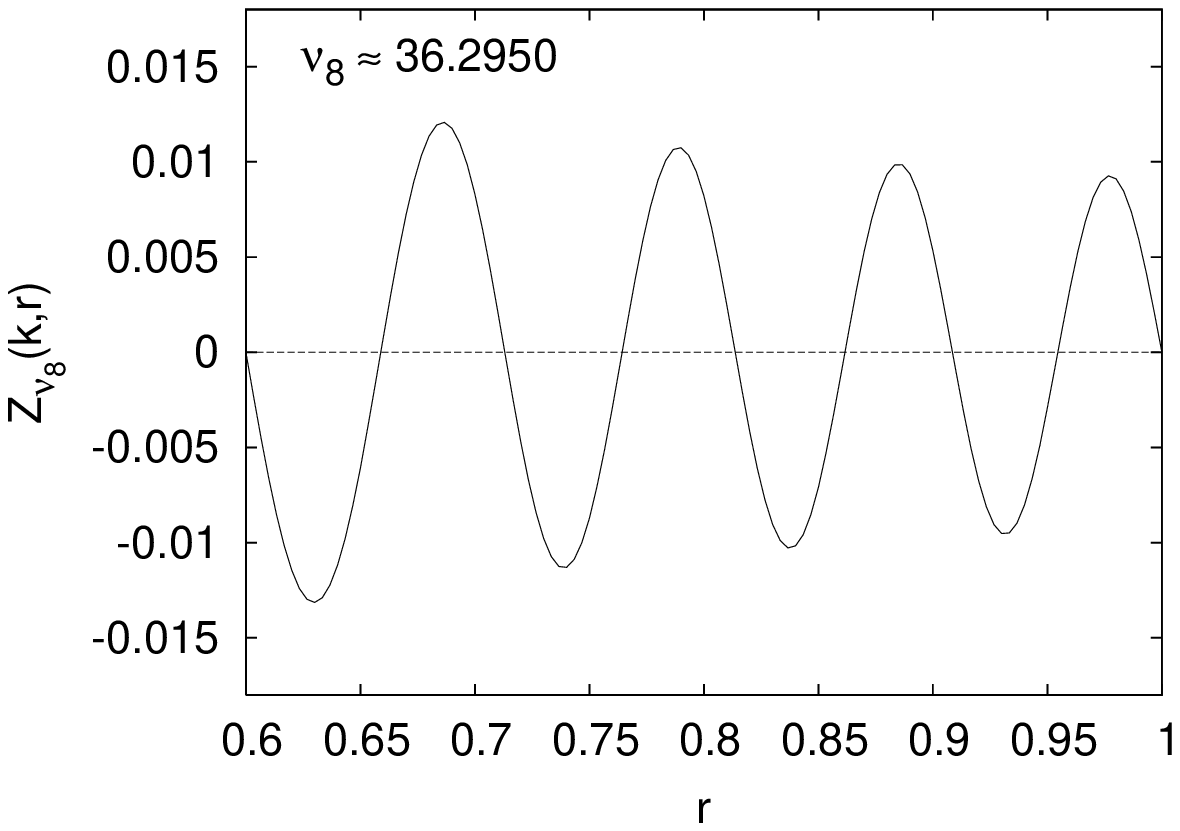}%
\includegraphics*[bb=75 66 390 300,height=4cm]{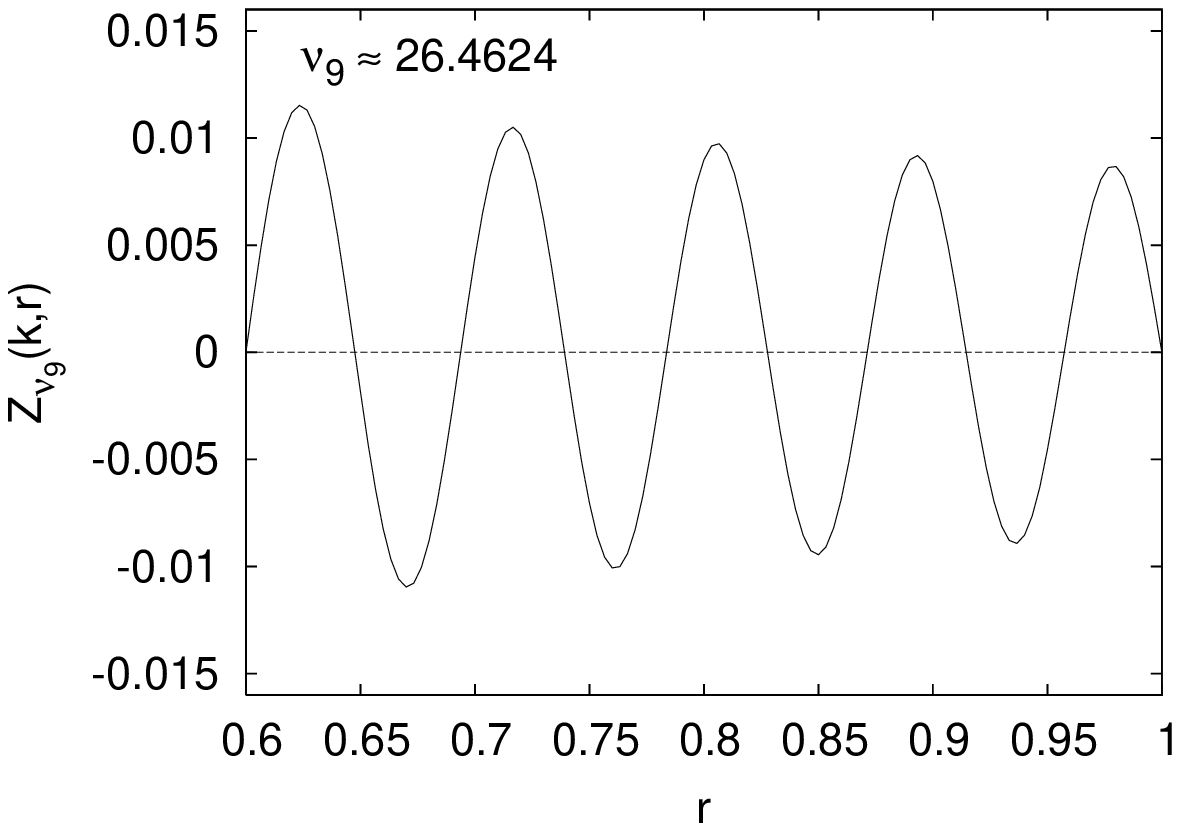}\\
\includegraphics*[bb=50 50 390 300, height=4.25cm]{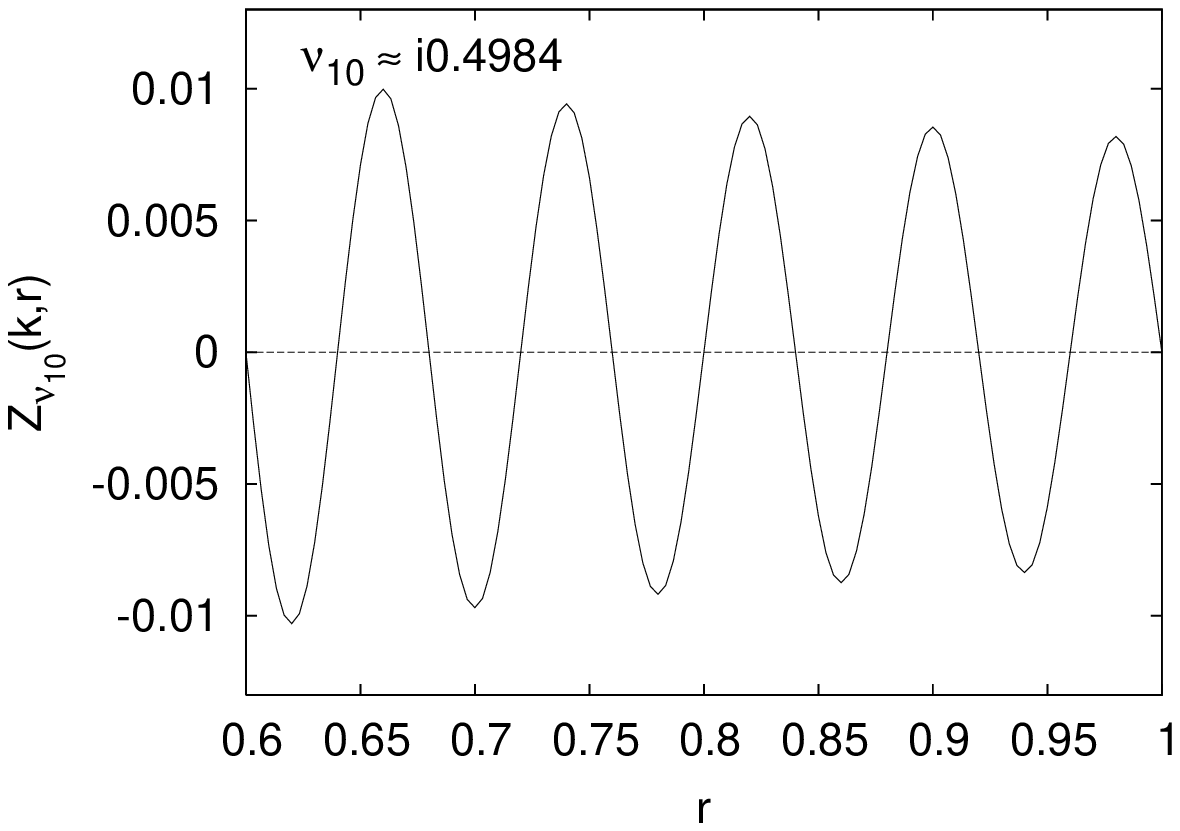}%
\includegraphics*[bb=75 50 390 300, height=4.25cm]{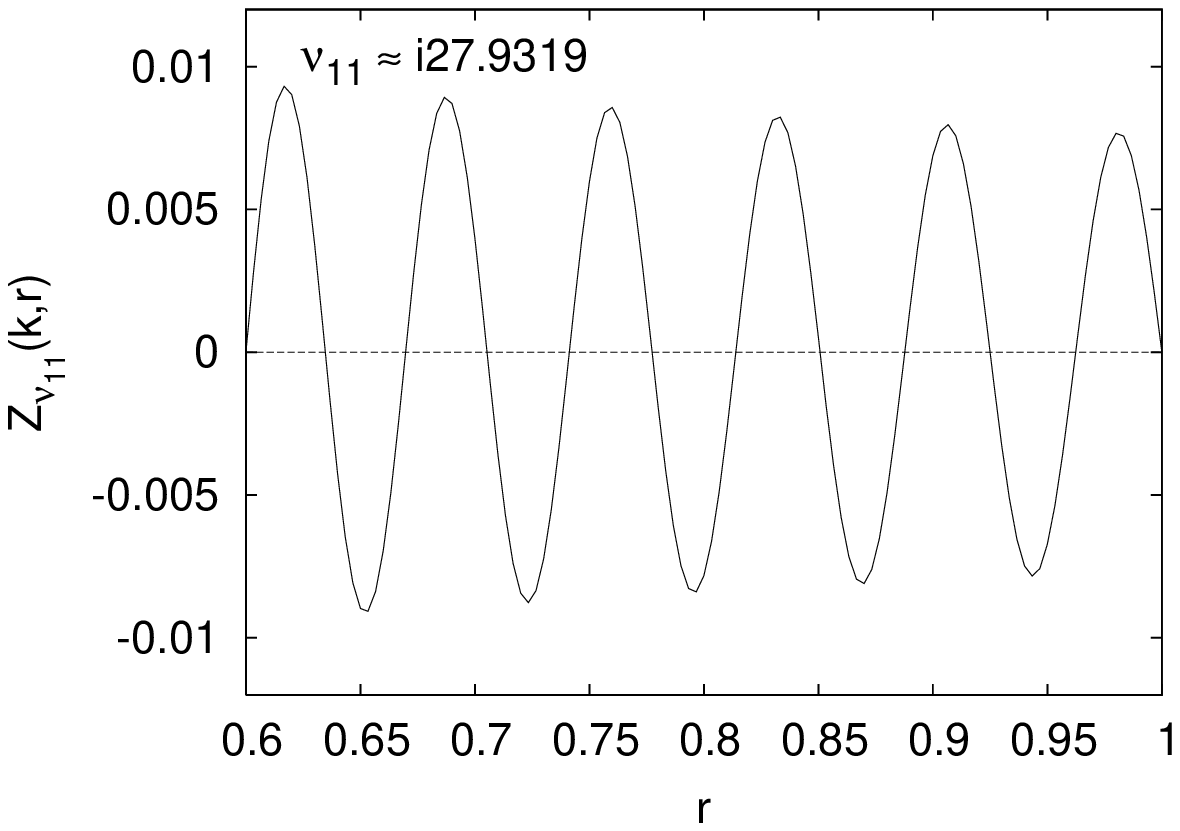}%
\includegraphics*[bb=75 50 390 300, height=4.25cm]{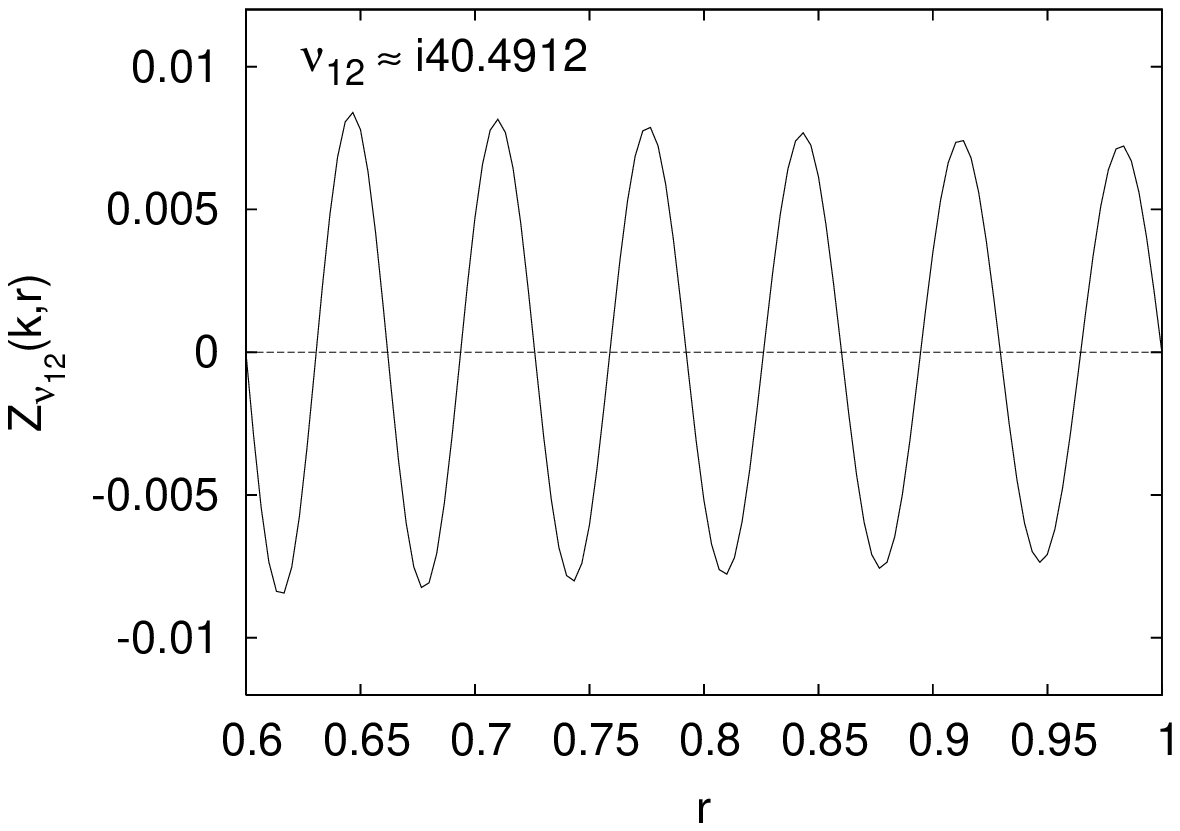}%
\caption{Plots of $Z_{\nu_n,k}(r)$ for first 12 mode-numbers $\nu_n \in {\cal M}_{k,q,+}$ $(\nu_n^2 > \nu_{n+1}^2)$ calculated at the wave-number $k = 78.5398$ $(N_{\rm b}=10)$ and inner radius $q=0.6$.}
\label{pic:bess_fun}
\end{figure}

The mode functions in the bend at a given wave-number $k$ and inner radius $q$ are proportional to cross products of Bessel functions $Z_{\nu,k}(r)$ (\ref{eq:bcp1}), where the order parameter $\nu$ takes values from the set ${\cal M}_{k,q}$. Because of the symmetry $Z_{-\nu,k}(r) = Z_{\nu,k}(r)$, we only consider mode numbers from the set ${\cal M}_{k,q,+}=\{\nu_n : \nu_n^2 > \nu_{n+1}^2\>,\; n \in\bN\}$, which are ordered by decreasing square. To illustrate the basic properties of mode functions, we plot in figure \ref{pic:bess_fun} the functions $Z_{\nu,k}(r)$ for real and first few imaginary mode numbers for $q=0.6$ and some low wave-number $k$. We see that the first mode function  $Z_{\nu_1,k}(r)$ has no zeroes on the interval $r\in(q,1)$ and each consecutive mode function has one additional zero. In the following, we present formulae and numerical recipes for stable calculation of mode functions, where we assume that the mode numbers are given.\par
The Bessel functions for real orders are well implemented in the currently available numerical libraries, i.e. \mbox{SLATEC} \cite{slatec:lib:93}. From the definitions of Bessel functions it is not unexpected that we encounter problems at evaluating $Z_{\nu,k}$ for large wavenumbers $k\gg 1$ and high real orders $\nu\gg 1$. In order to overcome these problems we apply the stable forward recursions in the order parameter \cite{olver:72} written as
\beqa
\fl\hspace{1cm}  p_{\mu+ 1} &=& 
  p_{\mu - 1} -  2\mu (x q_\mu + y r_\mu)\>,\\
\fl\hspace{1cm}  q_{\mu + 1} &=&  
  (2\mu(\mu+1)y^2  -1) r_\mu + 2\mu(\mu +1) x y q_\mu - (\mu+1)y p_{\mu-1} + \mu x p_\mu\>, \\
\fl\hspace{1cm}  r_{\mu + 1} &=&  
  (2\mu(\mu+1)x^2 -1) q_\mu + 2\mu(\mu +1) x y r_\mu - (\mu+1)x p_{\mu-1} + \mu y p_\mu\>, 
\eeqa
where we write $x=(kr)^{-1}$ and $y=k^{-1}$ and define the following symbols
\beqa
  p_\mu &=& J_\mu(kr) Y_\mu(k) - J_\mu(k) Y_\mu(kr)\>,\\
  q_\mu &=& J_\mu(kr) Y'_\mu(k) - J'_\mu(k) Y_\mu(kr)\>,\\
  r_\mu &=& J'_\mu(kr) Y_\mu(k) - J_\mu(k) Y'_\mu(kr)\>.
\eeqa
The initial conditions for the recursion, at low orders, are calculated using standard routines and
the expressions for the derivatives of Bessel functions, e.g. a relation valid for any Cylindrical function: $ C'_\nu = ( C_{\nu-1} - C_{\nu+1} ) / \nu$. However, we encountered a problem at high wavenumbers and low orders due to the lack of precision in \mbox{SLATEC} routines. Therefore in that regime we use the Hankel approximation \cite{olver:72} to evaluate $Z_{\nu,k}$ 
\beq
\fl\hspace{1cm}  Z_{\nu,k}(r) = \frac{2}{\pi \sqrt{r} k} 
                \left [ (AC+BD)\sin (k(1-r)) + (AD-BC)\cos (k(1-r))\right]\>,
  \label{eq:bcp_hankel_real}		
\eeq
where we write $A = P_\nu(kr)$, $B = Q_\nu(kr)$, $C = P_\nu(k)$ and $D = Q_\nu(k)$, which are expressed in terms of the asymptotic series:
\beq
\fl\hspace{1cm}
  P_\nu(z) = 
           1 - \frac{(\mu -1) (\mu -9)}{2!(8z)^2} + 
	   \frac{(\mu -1) (\mu -9) (\mu - 25) (\mu - 49)}{4!(8z)^4} +
	   \ldots
\label{eq:bess_hank_P}	   
\eeq
and 
\beq%
\fl\hspace{1cm}
  Q_\nu(z) = 
           \frac{\mu -1}{8z} -
	   \frac{(\mu -1)(\mu -9)(\mu-25)}{3!(8z)^3} +
           \ldots \>,\qquad \mu = 4\nu^2\>,
\label{eq:bess_hank_Q}	   
\eeq
which we sum up to $l_{\rm max} =\lfloor\nu/2\rfloor+1$ terms. There is another difficulty occurring at high wave-numbers, which can not be corrected. The first few real modes $\nu_p\in {\cal M}_{k,q,+}$ scale linearly with the wavenumber $k$ and functions $Z_{\nu,k}(r)$ diverge with increasing order $\nu$. Consequently, the values $Z_{\nu,k}(q)$ are exponentially sensitive on the precision of the first few mode numbers $\nu_p$:
\beq
  |Z_{\nu_p + \delta \nu,k}|(q) 
  \sim q^{-|\nu_p|} \sinh (|\log q|\, \delta\nu)\>.
  \label{eq:bcp_nu_asym}
\eeq
This problem can not be solved completely, but only partially corrected by manually setting the values of $Z_{\nu_p,k}(r)$ to zero around the inner radius. This can be done without any real loss of precision, because the mode functions are localised near outer radius. In practice we calculate the left side of equation (\ref{eq:bcp_nu_asym}) in a finite range arithmetic $[-m,m]$ with the maximal number $m$ (e.g. $m\approx 10^{308}$ in double precision). By considering that together with the known property $\nu_{\rm max}\sim k$ (\ref{eq:nu_bess_asym}), we find that in practice our modal approach breaks down above some wave-number $k_{\rm break} \sim \log m /|\log q|$ and consequently bounding the number of open modes in our numerical analysis below 
\beq
  N_\mathrm{o, {\rm break}} \sim  \frac{a}{\pi} \frac{\log m}{|\log q|}\>,
  \label{eq:max_open_modes}
\eeq
where we used the relation $\rNo\sim ka/\pi$ valid for narrow bends and high wave-numbers.\par
The numerical evaluation of $Z_{\nu_p,k}(r)$ at imaginary orders $\nu\in\ii \bR$ is almost unsupported in currently available numerical libraries. Therefore we have developed procedures for their evaluation ourselves and give here a summary of our work. 
For each regime of order parameters $\nu=\ii y$ and wavenumbers $k$ we use a different strategy to evaluate $Z_{\ii y,k}(r)$ in order to achieve an optimal precision control and a CPU time consumption. 
The formula for the cross-products of Bessel functions (\ref{eq:bcp1}) takes for imaginary orders a simple form
\beq
  Z_{\ii y, k}(r) = \frac{2}{\sinh(\pi y)}
  \Im\left\{J_{\ii y}(k)J_{\ii y}^*(k r)\right\}\>.
  \label{eq:bcp_imag}
\eeq
At small wave-numbers $k$ or more generally for $k\ll y$, we use the Taylor expansion of the Bessel function \cite{olver:72} and rewrite equation (\ref{eq:bcp_imag}) into 
\beq
  Z_{\ii y,k}(r) = \frac{2}{\pi y}
  \Im\left\{r^{-\ii y} u_{\ii y}(k) u_{\ii y}^*(kr)\right\}\>,
\label{eq:bcp_taylor_imag}
\eeq
where we use the series  
\beq
 u_\nu(z) = 
 \sum_{l=0}^\infty  
 \frac{(-1)^l \left(\frac{z}{2}\right)^{2l}}{l!\,(\nu+1, \nu+l+1)}\>, 
 \qquad 
 (x,y) = \frac{\Gamma(y)}{\Gamma(x)}\>.
\label{eq:bcp_taylor_imag_series}
\eeq
The series (\ref{eq:bcp_taylor_imag_series}) is summed up to the index $l_{\rm max}= \lfloor\sqrt{z^2/\epsilon + |\nu|^2/4}\rfloor$, where $\epsilon$ is the desired accuracy of the expression. 

At higher wave-numbers $k$ and orders $y \lesssim k$ we combine the backwards recursion valid for Cylindrical functions 
\beq
  a_{l-1}(z) = \frac{2(l + \ii y)}{z} a_{l}(z) - a_{l + 1}(z)\>,
  \label{eq:bess_it}
\eeq
with an appropriate normalisation formula for $a_l(z)$ \cite{olver:72} and thereby obtain an expression for $u_{\ii y}(k)$ (\ref{eq:bcp_taylor_imag}) given by
\beq
  u_{\ii y}(z) = a_0(z) \left[
    \sum_{l=0}^\infty 
    \frac{(2l + z)(1+z, l+z)}{l!} a_{2l}(z)\right]^{-1}\>.
  \label{eq:bcp_rec_imag}
\eeq
The terms $a_{2l}$ in the series (\ref{eq:bcp_rec_imag}) are given with the recursion (\ref{eq:bess_it}) started at the index $l_{\rm max}=2\lfloor(x + 1)/2\rfloor$ with initial conditions $a_{l_{\rm max}} = \eps$ and $a_{l_{\rm max}+1} = 0$, where constant $\eps$ is the smallest number supported by the CPU architecture and $x$ is determined by the equation
\beqa 
  |\log\eps| -1 + \frac{1}{2}\log(1+y^2) - y  \atan\, y = \nonumber \\
  \frac{1}{2} x \log (x^2 + y^2) - 
  x \left(\log\frac{z}{2} +1 \right) 
  - x \atan \frac{y}{x} \>.
  \label{eq:bcp_rec_eq}
\eeqa
The later equation (\ref{eq:bcp_rec_eq}) is meaningful only if the right side is positive, yielding that presented approach with the iteration formula is valid only for orders $y$ below some value scaling as $O(|\log\eps|)$.
By increasing the wavenumber $k$ further up and keeping orders small $y \ll k$ we can use the Hankel approximation (\ref{eq:bcp_hankel_real}) with the order parameter $\nu=\ii y$. The asymptotic series (\ref{eq:bess_hank_P}) and (\ref{eq:bess_hank_Q}) in the expression (\ref{eq:bcp_hankel_real}) are summed at least up to $l_{\rm max}=\lfloor z/2 + \sqrt{2 z^2 - y^2/2}\rfloor$ terms. 
At large enough wavenumbers $k$ and imaginary orders $y>k$ we can make use of the Debye approximation of Bessel functions \cite{erdelyi:book:55} and write 
\beq
Z_{\ii y}(k,r) =
\frac{2}{\pi} 
  \frac{
  \Im 
    \left\{
      G(\alpha,\xi) G^*(\beta,\zeta)
        e^{\ii 
           \left [
             \xi -\zeta - y\left(\asinh \frac{y}{k} - \asinh \frac{y}{kr}\right)
           \right]
          }
    \right\}
  }
  {(1- e^{-2\pi y}) \sqrt{\xi\zeta}}\>,
\label{eq:bcp_debye_imag}    
\eeq
with substitutions
\beq
  \fl\hspace{1cm}\xi = \sqrt{y^2 + k^2}\>,\quad
  \zeta= \sqrt{y^2 + (kr)^2}\>,\quad
  \alpha^{-1} = 1+ \frac{k^2}{y^2}\>, \quad 
  \beta^{-1} = 1+ \frac{(kr)^2}{y^2}\>.	 
\eeq
The expression $G(x,y)$ in formula (\ref{eq:bcp_debye_imag}) is given in the form of an asymptotic series 
\beq
  G(x,y) = \sum_{m=0}^\infty \frac{(-\ii)^m v_m(x)}{y^m}\>,
  \label{eq:bcp_debye_G}
\eeq
where polynomials $v_m(t)$ are generated by the following recursion 
\beq
\fl\hspace{1cm}
 v_{k+1}(t) = \frac{1}{2} (1-t) \left(k\, v_k(t) + 2t \, {v'}_k(t) \right) +
               \frac{1}{16} t^{-\frac{k+1}{2}} 
	       \int_0^t (1-5\tau) \tau^{\frac{k-1}{2}} v_k (\tau)\, \dd\tau\>.
\eeq
The first few  $v_m(t)$ read as
\beq
\fl\hspace{1cm}
  v_0(t) = 1,\quad 
  v_1(t) = \frac{1}{8} - \frac{5}{24} t,\quad
  v_2(t) = \frac{3}{128} - \frac{77}{576} t + \frac{385}{3456} t^2\>,\ldots\>. 
\eeq
The formulae (\ref{eq:bcp_hankel_real}, \ref{eq:bcp_taylor_imag}, \ref{eq:bcp_rec_imag}, \ref{eq:bcp_debye_imag}) enable a stable high precision calculation of the mode functions in the bend at imaginary mode numbers. 
\subsection{The overlap of mode-functions in different geometries}
The main ingredient in the modal description of the scattering are the overlap integrals of mode functions in the straight waveguide and in the bend. These overlap integrals ``tell'' about the compatibility of both scattering regions and are discussed in the following.\par
\begin{figure}[!htb]
\centering
\begin{minipage}[c]{10mm}
\rotatebox{90}{\small$\log_{10} |A_{np}|$}
\end{minipage}
\begin{minipage}[c]{10cm}
\centering
\vbox{\small$q=0.2$\hspace{3.5cm}$q=0.9$}
\vbox{%
\includegraphics*[height=4.7cm]{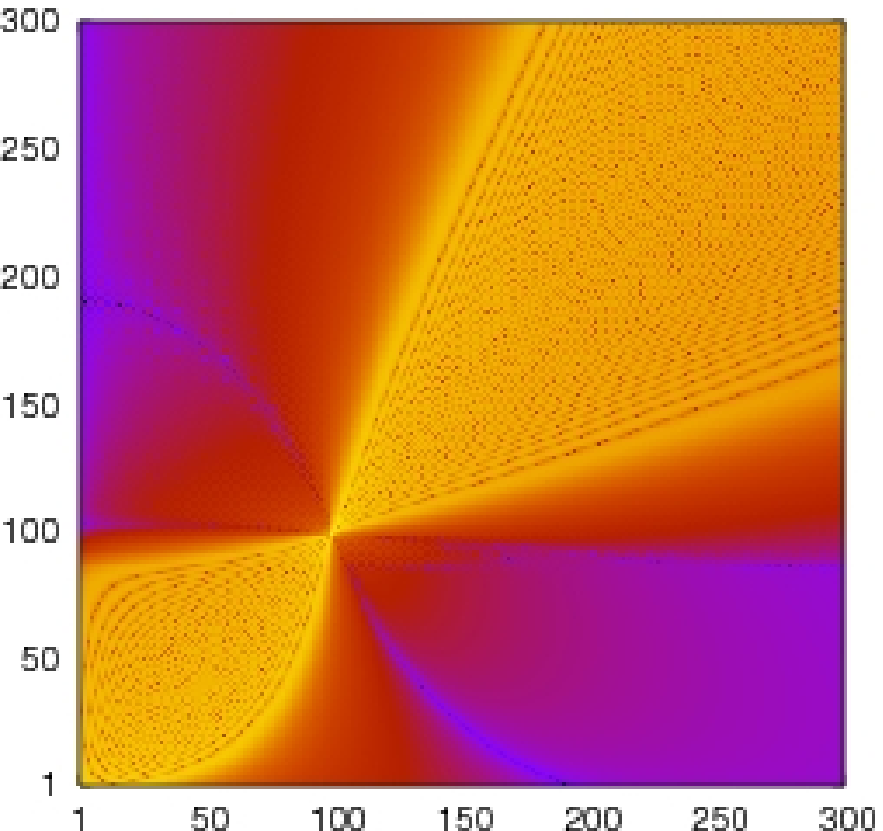}%
\includegraphics*[height=4.7cm]{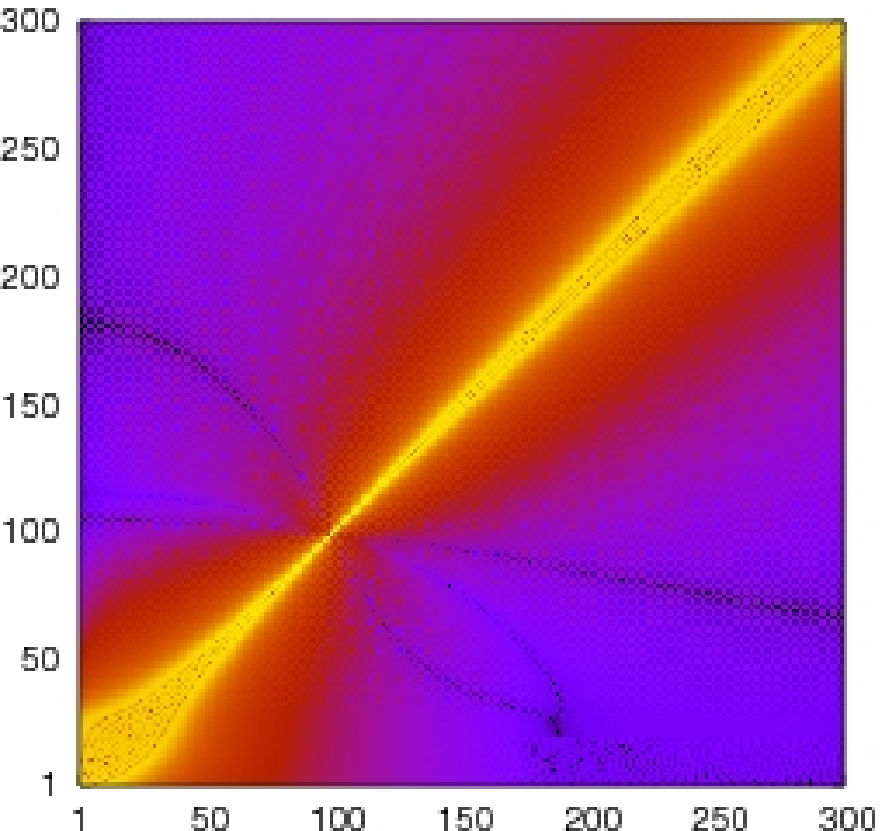}}
\end{minipage}\\*
\begin{minipage}[c]{10mm}
\rotatebox{90}{\small $\log_{10} |B_{np}|$}
\end{minipage}
\begin{minipage}[c]{10cm}
\centering
\vbox{%
\includegraphics*[height=4.7cm]{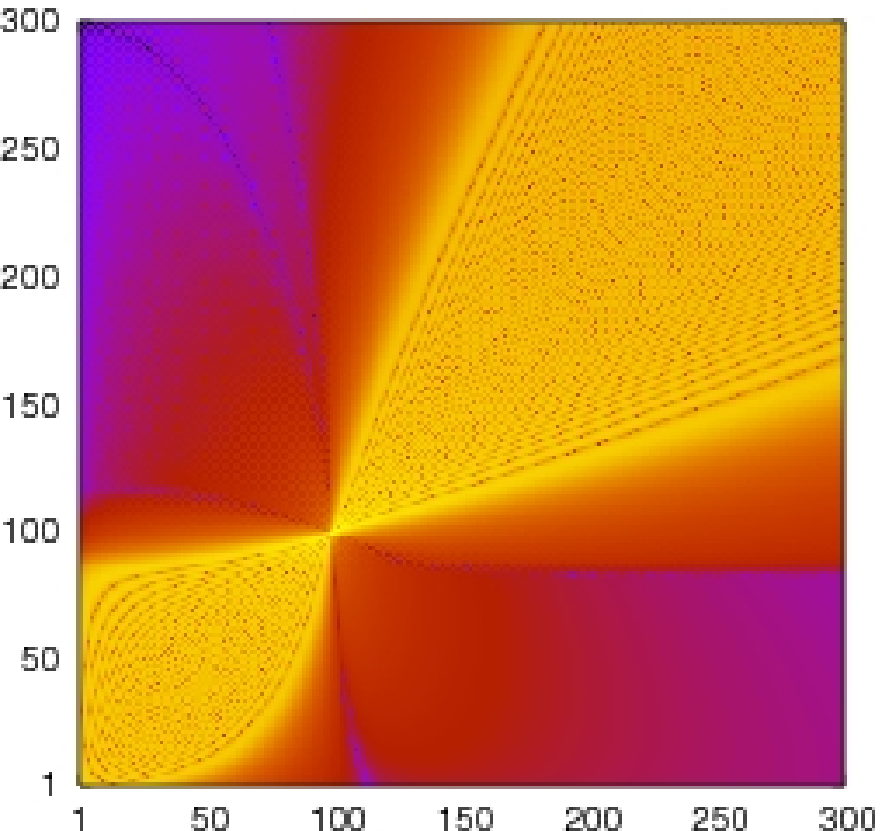}%
\includegraphics*[height=4.7cm]{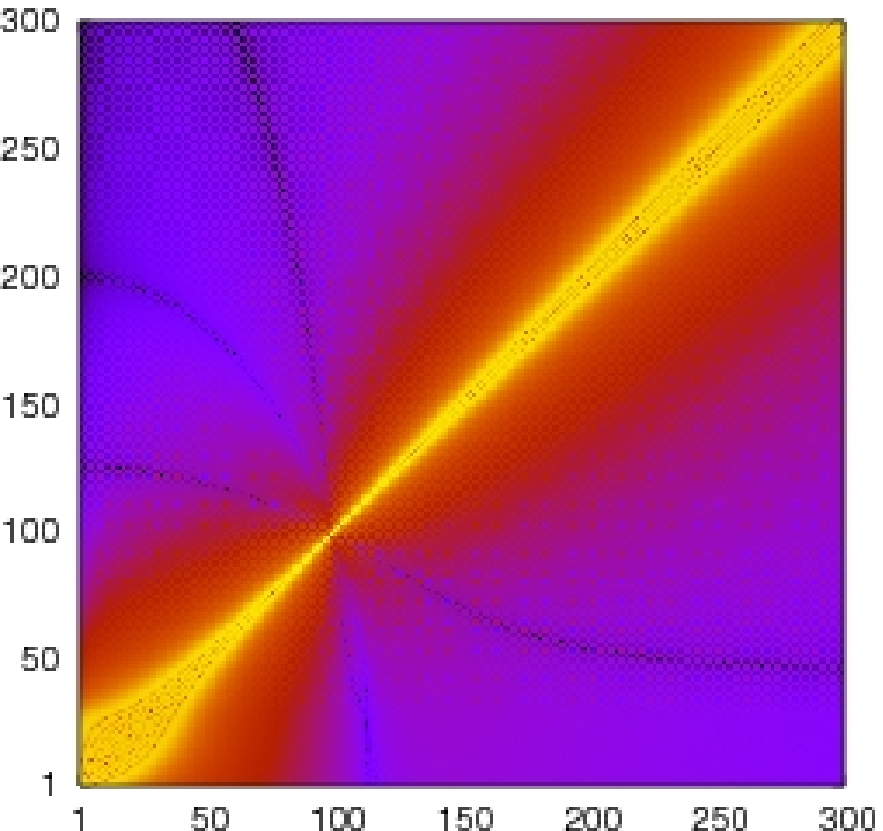}}
\end{minipage}
\hbox to15cm{\hskip34mm%
\includegraphics*[bb =80 60 388 90, width=10cm]{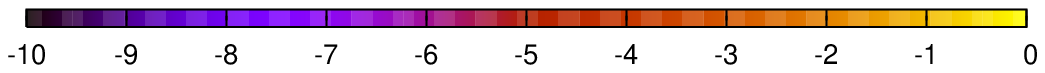}
}

\caption{Density plots of the matrix elements $\log_{10} |A_{np}|$ (top-row) and $\log_{10} |B_{np}|$ (bottom-row) for $q=0.2$ (left) and $q=0.9$ (right) as indicated in the figure. The number of open modes is $\rNo=100$ and the number of all considered modes is $N=300$.}
\label{pic:AB}
\end{figure}
\begin{figure}[!htb]
\centering
\vbox{\small
\hfil$\rNo=100$%
\hfil\hfil$\rNo=200$%
\hfil\hfil$\rNo=400$%
\hfil\hfil}
\vbox{%
\includegraphics*[height=5cm]{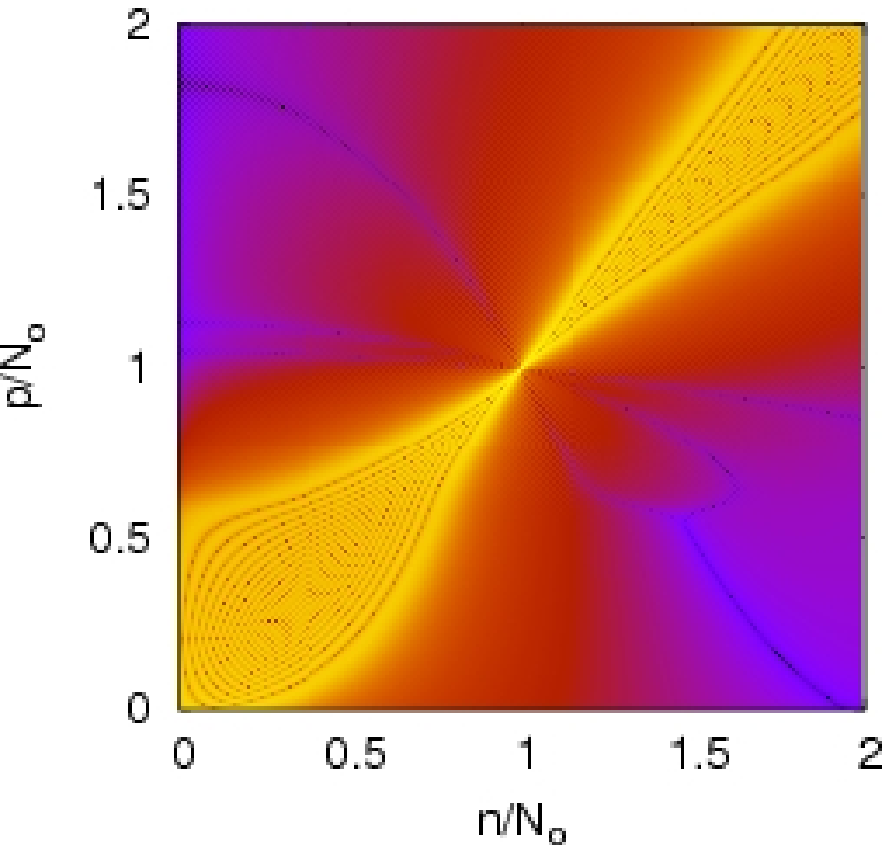}%
\includegraphics*[bb=15 0 254 241, height=5cm]{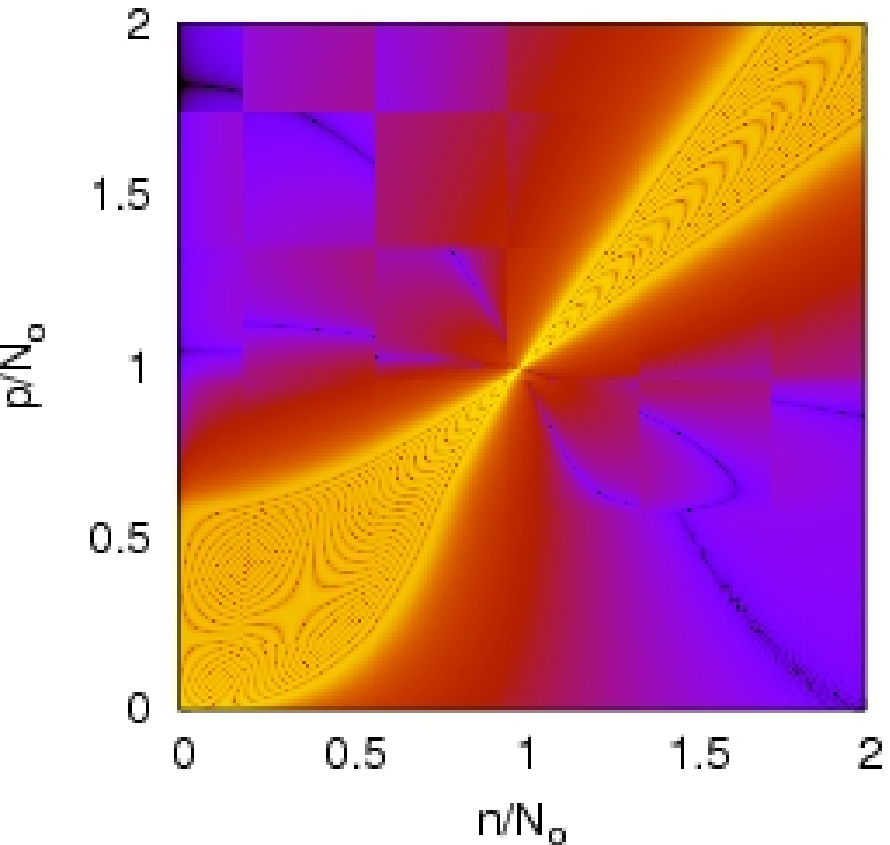}%
\includegraphics*[bb=15 0 254 241, height=5cm]{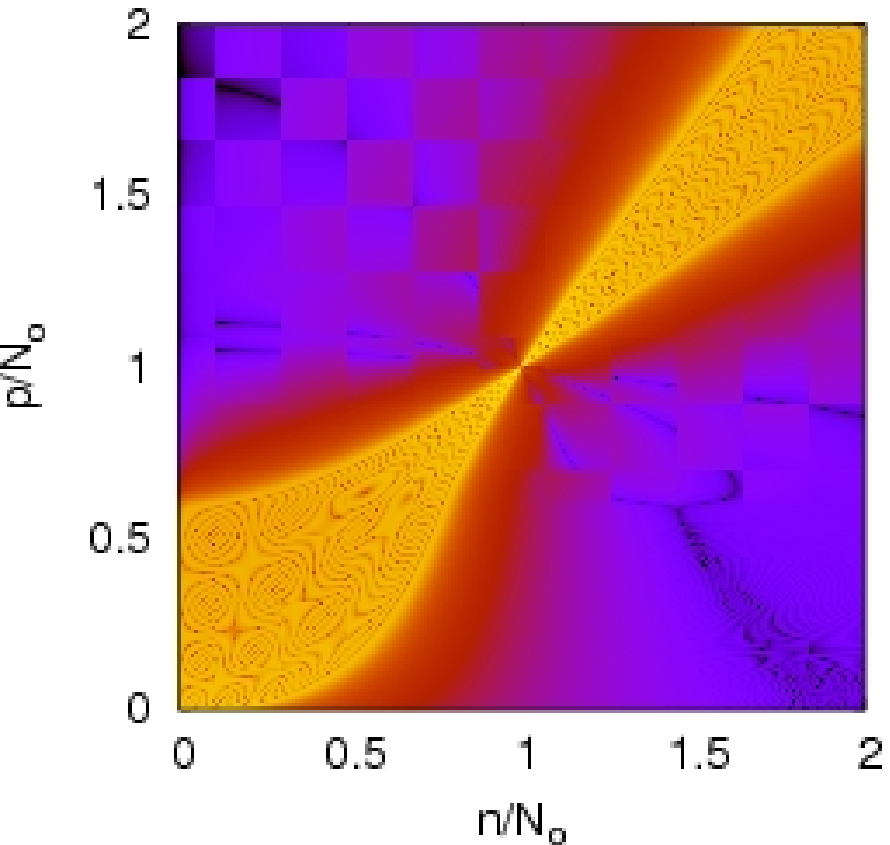}%
}
\hbox to 15cm{\hskip35mm%
\includegraphics*[bb =80 60 388 90, width=10cm]{id_pal.eps}}
\caption{Density plot of matrix elements $\log_{10}|A_{np}|$ at inner radius $q=0.6$ and different number of open modes $\rNo=$ 100, 200, and 400 as indicated in the figure.}
\label{pic:AB_amp_k}
\end{figure}
\begin{figure}[!htb]
\centering
\vbox{%
\includegraphics[width=7cm]{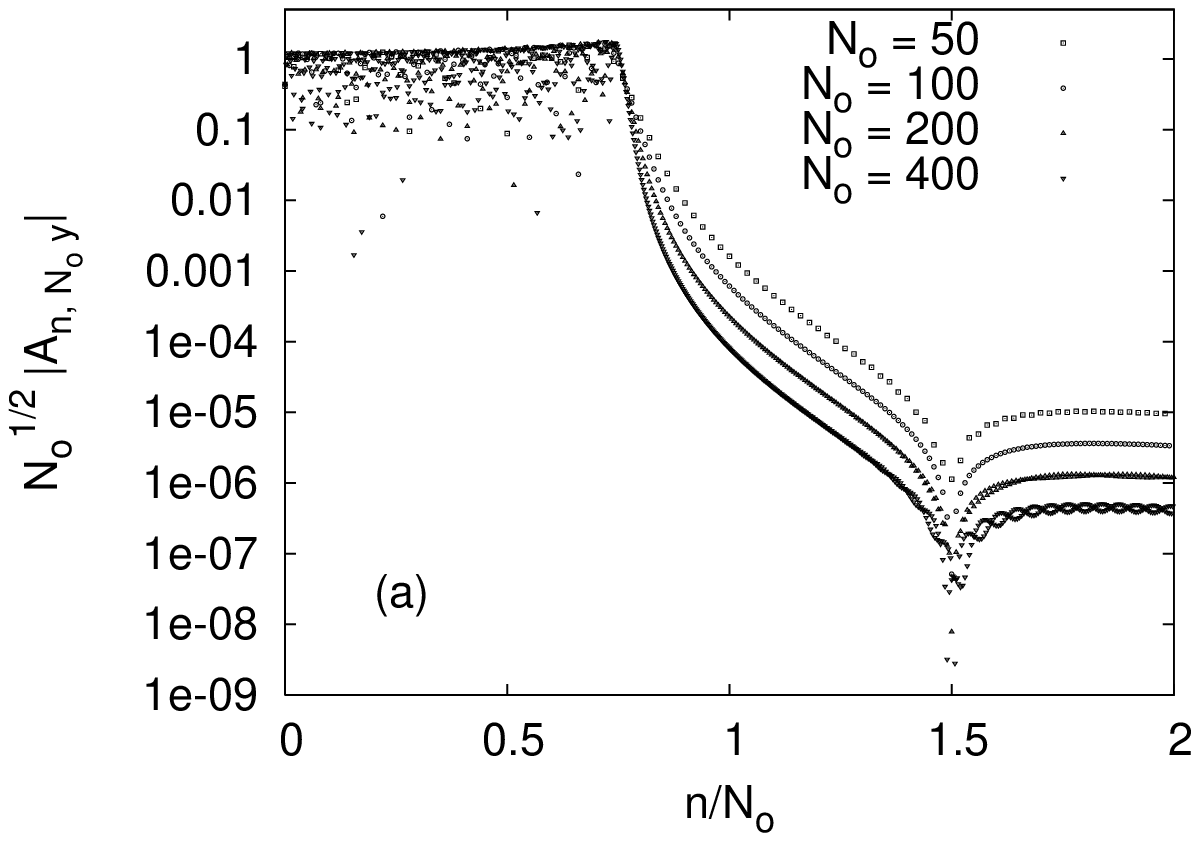}%
\includegraphics[width=7cm]{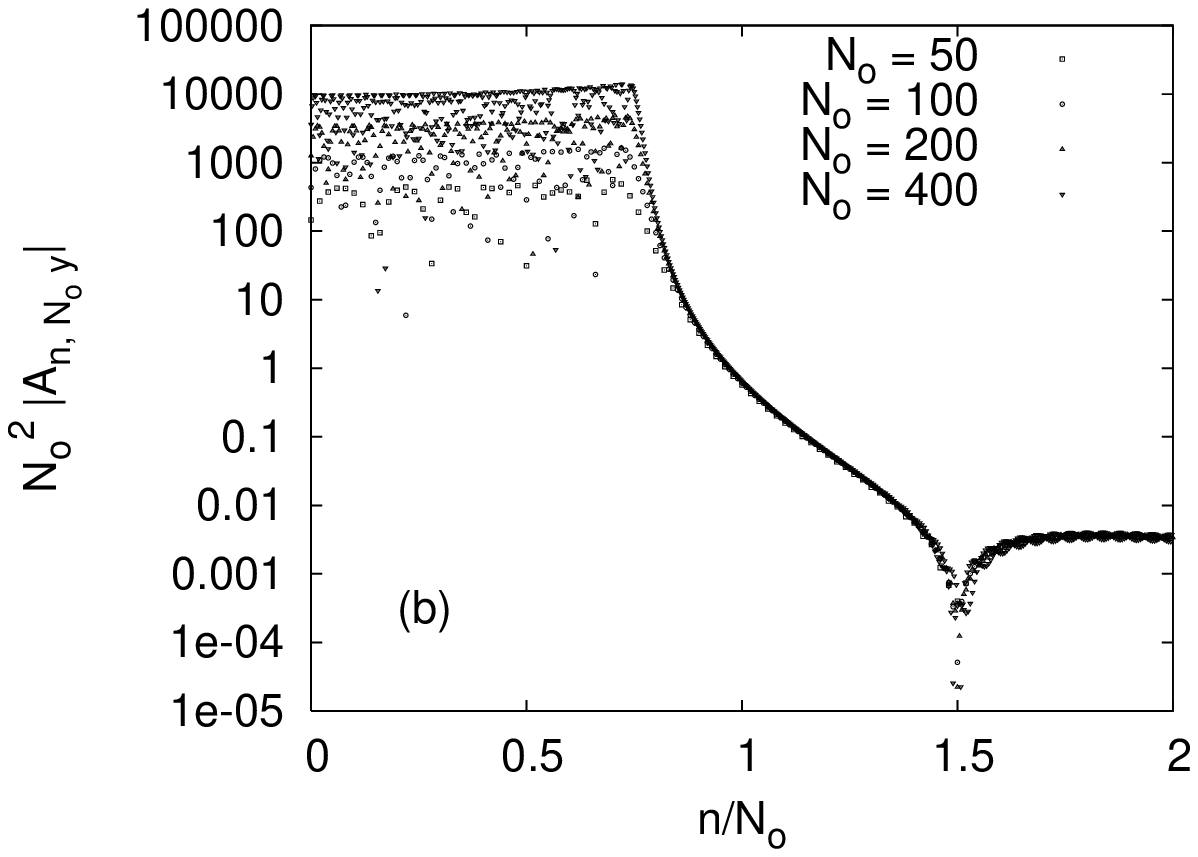}%
}
\caption{The cuts in the index space of matrix $A_{np}$ at fixed $p/\rNo = 0.5$ and different number of open modes $\rNo=$ 50, 100, 200, and 400 at inner radius $q=0.6$.}
\label{pic:AB_dif_k}
\end{figure}

The cross products of Bessel functions with order parameter $\nu\in {\cal M}_{k,q,+}$ at given wavenumber $k$ and inner radius $q$ form a set of functions
\beq 
  {\cal Z}_{k,q} = \{Z_{\nu, k} (r):\nu \in {\cal M}_{k,q,+}, r\in [q,1]\}\>,
\eeq
which is complete in $L_2[q,1]$ and orthogonal w.r.t. the weight function $w(r) = r^{-1}$. The later is derived in \ref{sec:mode_symm}. The orthogonality relation for $Z_{\nu,k}\in{\cal Z}_{k,q}$ reads 
\cite{luke:book:62}
\beq
\fl \hspace{0.5cm} 
\int_q^1 \dd r\, w(r) Z_{\nu,k} (r) Z_{\mu,k} (r) = \delta_{\mu,\nu}
  \frac{k}{2\nu}
  \left [
    Z_{\nu+1,k}(1) \frac{\pa Z_{\nu,k}}{\pa \nu} (1) -
    q Z_{\nu+1,k}(q) \frac{\pa Z_{\nu,k}}{\pa \nu} (q) 
  \right ]\>.
  \label{eq:ortho_bess}
\eeq
The line separating the bend and the straight wave-guide will be called {\it the cross-section} of our open-billiard. On the cross-section we define two different scalar products denoted by $(\cdot,\cdot)$ and $\ave{\cdot,\cdot}$, and written as
\beq
 (a,b) = \int_q^1 \frac{\dd r}{r} a(r)\, b(r)\>,\qquad 
 \ave{a,b} = \int_0^a \dd y\, a(y)\, b(y)\>.
 \label{eq:def_scalar}
\eeq
Let us now introduce modes at some fixed wave-number $k$ and inner radius $q$ for different regions of the open billiard. In the bend, mode numbers $\nu_p$ and normalised mode functions $U_p(r)$ read as
\beq
    \nu_p \in {\cal M}_{k,q,+}\>,\quad
    U_p(r) = \frac{Z_{\nu_p,k}(r)}{\sqrt{(Z_{\nu_p,k},Z_{\nu_p,k})}}\>, \qquad
    p\in \bN\>,
  \label{eq:def_modes_bend}
\eeq
where we order the mode numbers so that $\nu^2_p > \nu_{p+1}^2$, and in the straight leads connected to the bend we have mode numbers $g_n$ and mode functions $u_n(x)$ defined by
\beq
  g_n=\sqrt{k^2 - \left(\frac{\pi n}{a}\right)^2}\>,\quad  
  u_n(y) = \sqrt{\frac{2}{a}}\sin \left(\frac{\pi}{a} n y\right)\>,\qquad
   n\in\bN\>.
  \label{eq:def_modes_straight}
\eeq
The modes with real and imaginary mode numbers are called open and closed modes, respectively. The number of open modes in some geometry is denoted by $\rNo$. The overlap integrals of mode functions are given by
\beq
  A_{np} = \ave{u_n,U_p}\>,\qquad  B_{np} = (u_n, U_p)\>,
  \label{eq:mat_elem}
\eeq
where we use the relation $r = q + y$ between the coordinates. In figure \ref{pic:AB} we show a density plot of the matrix elements, in log scale, namely $\log|A_{np}|$ and $\log|B_{np}|$, at two values of inner radii with the same number of open modes $\rNo$ in both geometries. 
We see that the matrices $A_{np}$ and $B_{np}$ have a similar form for all $q$ and $k$. This is starting at small indices with a wide area of high values of matrix elements that squeezes to almost a single intensified point at $n,p \approx \rNo$ again spreading in a triangular shape with increasing indices. The parameter $q$ has a strong influence on the shape of the area with high intensities in $A$ and $B$. In the case of small values of $q$ in contrast to larger $q$, the area of high values in matrices $A$ and $B$ covers almost the whole open-open block of indices and with crossing of the narrowing at $n,p \approx \rNo$ spreads faster with increasing indices. The shape of matrices $A$ and $B$ is similar therefore in the following we only show results for the matrix $A_{np}$. We found numerically that the area of high intensities in matrices $A$ and $B$ scales with the number of open modes $\rNo$ as $A_{np}, B_{np} \sim F (n/\rNo, p/\rNo)$, where $F$ is some a well behaved function. We demonstrate this by plotting the matrix elements $|A_{np}|$ in relative indices $n/\rNo$ and $p/\rNo$ for different numbers of open modes $\rNo$ shown in figure \ref{pic:AB_amp_k}. 
In addition, we find numerical evidence that intensities of matrix elements $A_{np}$ and $B_{np}$ in the region of open modes and in the region of closed modes scale differently with $\rNo$
\beqa
  \textrm{open modes}&:&   
  |A_{np}|,\; |B_{np}| \lesssim 
  \frac{1}{\sqrt{\rNo}} F_{\rm max} \left(\frac{n}{\rNo}, \frac{p}{\rNo}\right)\>, 
  \label{eq:AB_scaling_max}\\
  \textrm{closed modes}&:&
  |A_{np}|,\; |B_{np}| \lesssim
  \frac{1}{\rNo^2} F_{\rm min} \left( \frac{n}{\rNo}, \frac{p}{\rNo}\right)\>. 
  \label{eq:AB_scaling_tails}  
\eeqa
where $F_{\rm max}$ and $F_{\rm tail}$ are some well behaved functions. Taking into account these 
phenomenological findings enables a better precision control of scattering calculations. In figure \ref{pic:AB_dif_k} we see that $F_{\rm max}$ is an envelope function for maximal values of $\rNo^{1\over 2} |A_{np}|$ and that $F_{\rm min}$ can be chosen to fit the tails of $\rNo^{2} |A_{np}|$. 
We prove the scaling relation (\ref{eq:AB_scaling_tails}) by using an asymptotic approximation of the mode function in the bend
\beq
  U_p(r) = 
  \sqrt{\frac{2}{|\log q|}} \sin\left(\frac{\pi p}{\log q} \log r \right)\>, 
  \quad
  p\gg 1\>,
\label{eq:U_p_asym}
\eeq
which yields the following asymptotic behaviour of matrix elements
\beqa
  \fl A_{np} \approx
  \frac{6}{\pi^2} \frac{|\log q|^{5 \over 2}}{(1-q)^{3\over 2}} \left(q^2(-1)^p + (-1)^{n+1}\right)  n p^{-3} + O (p^{-5}) \quad  n=\textrm{fixed},\; p\to\infty\>,
\label{eq:AB_tails_asym_p}  \\
\fl  A_{np} \approx 
  \frac{2}{\pi^2} \frac{(1-q)^{5\over 2}}{|\log q|^{3\over 2}} \left(q^{-2}(-1)^{p+1} + (-1)^n\right) p n^{-3} + O (n^{-5})\quad  p=\textrm{fixed} \gg 1,\; n\to\infty\>.
  \label{eq:AB_tails_asym_n}  
\eeqa
The off-diagonal diagonal elements $A_{np}$ and $B_{np}$ decay algebraically with increasing index. The pre-factor of the decay is decreasing with increasing $q$ and is singular at $q=0$. This means that for large enough $q$ it is possible to approximately express the open modes of the bend solely in terms of open modes of the straight waveguide and vice versa, as indicated by the relation (\ref{eq:AB_scaling_max}).\par
\begin{figure}[!htb]
\centering
\hbox to13.5cm{\small%
\hspace{2.5cm}$\rNc=1$%
\hspace{3.4cm}$\rNc=100$%
\hspace{3.3cm}$\rNc=500$\hfil}
\vbox{\centering
\includegraphics*[height=4.5cm]{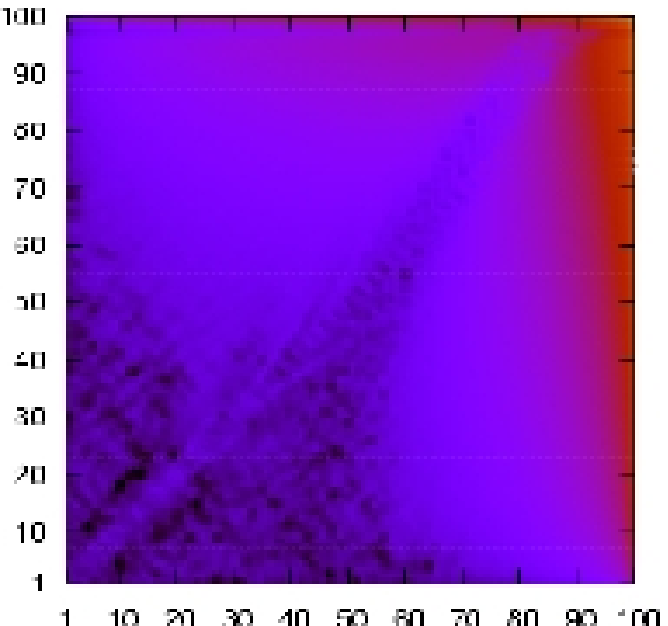}%
\includegraphics*[height=4.5cm]{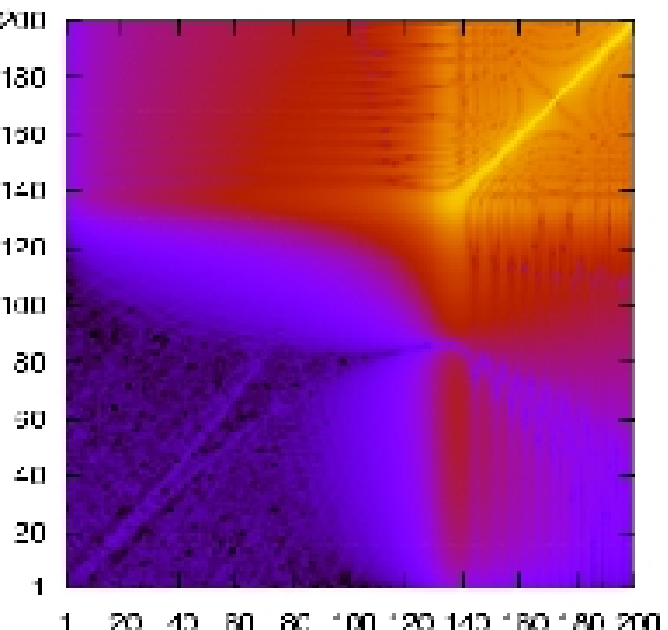}%
\includegraphics*[height=4.5cm]{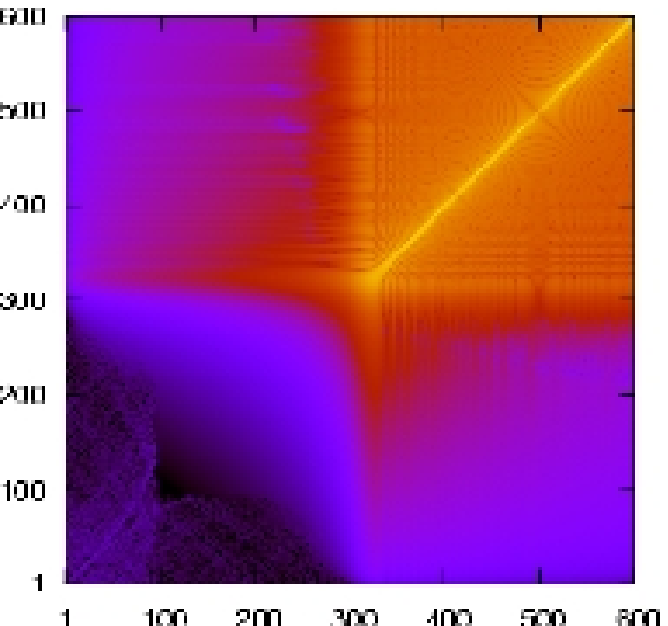}}
\hskip15mm\includegraphics*[bb = 80 60 388 90, width=10cm]{id_pal.eps}
\caption{Density plot of matrix elements $\log_{10}|(AB^T)_{nm} - \delta_{nm}|$ for different numbers of closed modes $\rNc$ = 1, 100 and 500, as indicated in the figure, at inner radius $q=0.2$ and $\rNo=100$ open modes. The labels on the abscissa and the ordinate are matrix indices $n$ and $m$, respectively.} 
\label{pic:AB_id_testing}
\end{figure}

From the definition of matrix elements $A_{np}$ and $B_{np}$ (\ref{eq:mat_elem}), and completeness of the mode functions at given $k$ and $q$, it follows that $A$ and $B$ are transition matrices between the sets of mode functions in different regions,
\beq
  u_n(x) = \sum_{p\in\bN} B_{np}\, U_p(r)\>,\qquad 
  U_p(r) = \sum_{n\in\bN} A_{np}\, u_n(x)\>,
  \label{eq:mat_trans}
\eeq
yielding the relation
\beq
  A B^T = A^T B  = \id \>.
  \label{eq:mat_AB_id}
\eeq
In practice we work with finite sets of modes, where the identity (\ref{eq:mat_AB_id}) cannot hold exactly. 
\begin{figure}[!htb]
\centering
\hbox to13.5cm{\small%
\hspace{2.5cm}$q=0.2$\hspace{3.3cm}$q=0.6$\hspace{3.3cm}$q=0.9$\hfil}
\vbox{%
\includegraphics*[height=4.5cm]{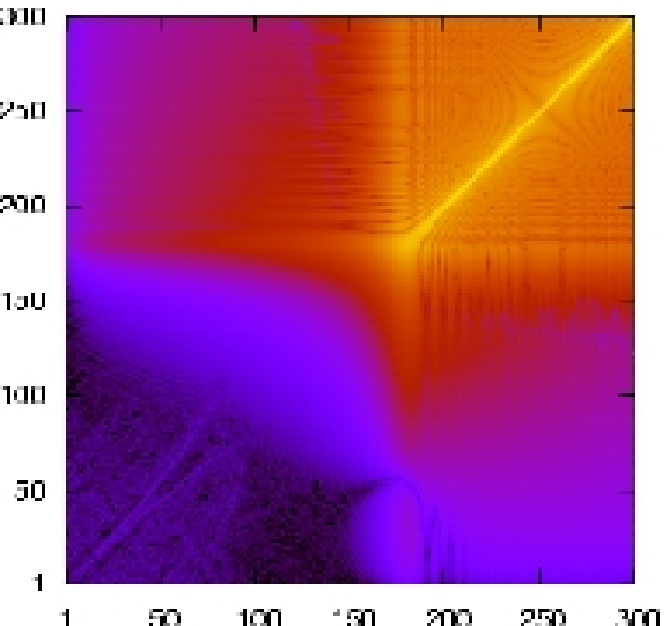}%
\includegraphics*[height=4.5cm]{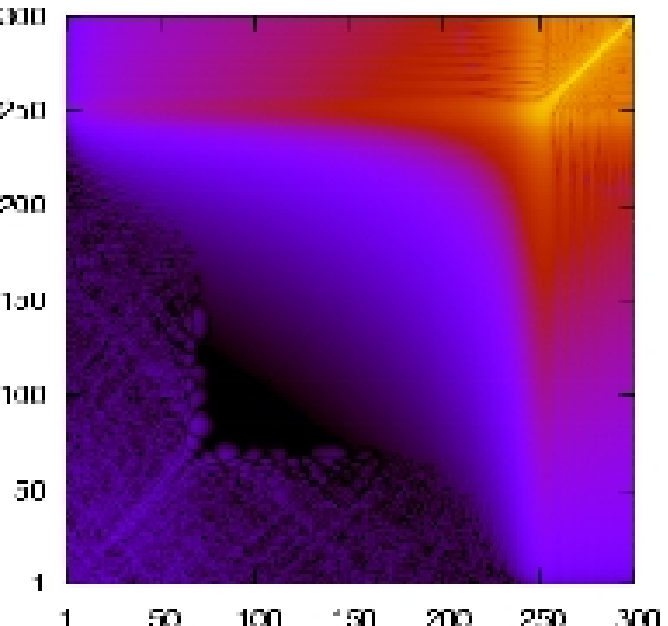}%
\includegraphics*[height=4.5cm]{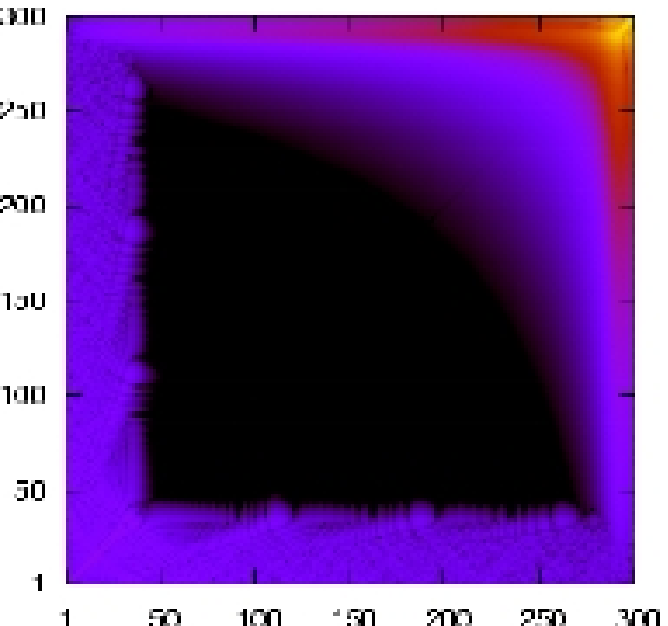}}
\includegraphics*[bb = 80 60 388 90, width=10cm]{id_pal.eps}
\caption{Density plot of matrix elements $\log_{10}|(AB^T)_{nm}-\delta_{nm}|$ for $q$ = 0.2, 0.6 and 0.9 (from left to right). On the abscissa and ordinate we plot indices $n$ and $m$, respectively. The number of open modes is $\rNo = 100$ and the total number of modes is 300.}
\label{pic:AB_id}
\end{figure}
In figure \ref{pic:AB_id} we plot $AB^T$ for different inner radii $q$ and fixed $\rNo=100$. The mismatch from the identity (\ref{eq:mat_AB_id}) on some sub-set of indices starting at the origin $n=1$, or $p=1$, increases with decreasing inner radius $q$. This means that the numerical calculation of scattering for smaller $q$ should be less accurate at finite dimensions.
The discrepancy between $AB^T$ and the identity on (truncated) finite dimensional spaces is strongly non-uniform in indices. Before going into practical aspects of this problem, we examine the convergence of matrix elements $(AB^T)_{nm}$ to $\delta_{nm}$ with increasing number of all considered modes $N=\rNo + \rNc$ at fixed $\rNo$, where $\rNc$ is the number of closed modes. An example of such convergence is shown in figure \ref{pic:AB_id_testing}.
The convergence proceeds by the standard scenario, where the agreement between $[A B^T]_{np}$ and $\delta_{np}$ propagates block-wise from low to higher indices with increasing $N$. The propagation is slow due to the triangular shaped area of high intensities in the closed-closed modes block of $A$ and $B$. The speed of propagation of accuracy to higher indices increases with increasing inner radius $q$.\par
\begin{figure}[!htb]
\centering
\includegraphics[width=7cm]{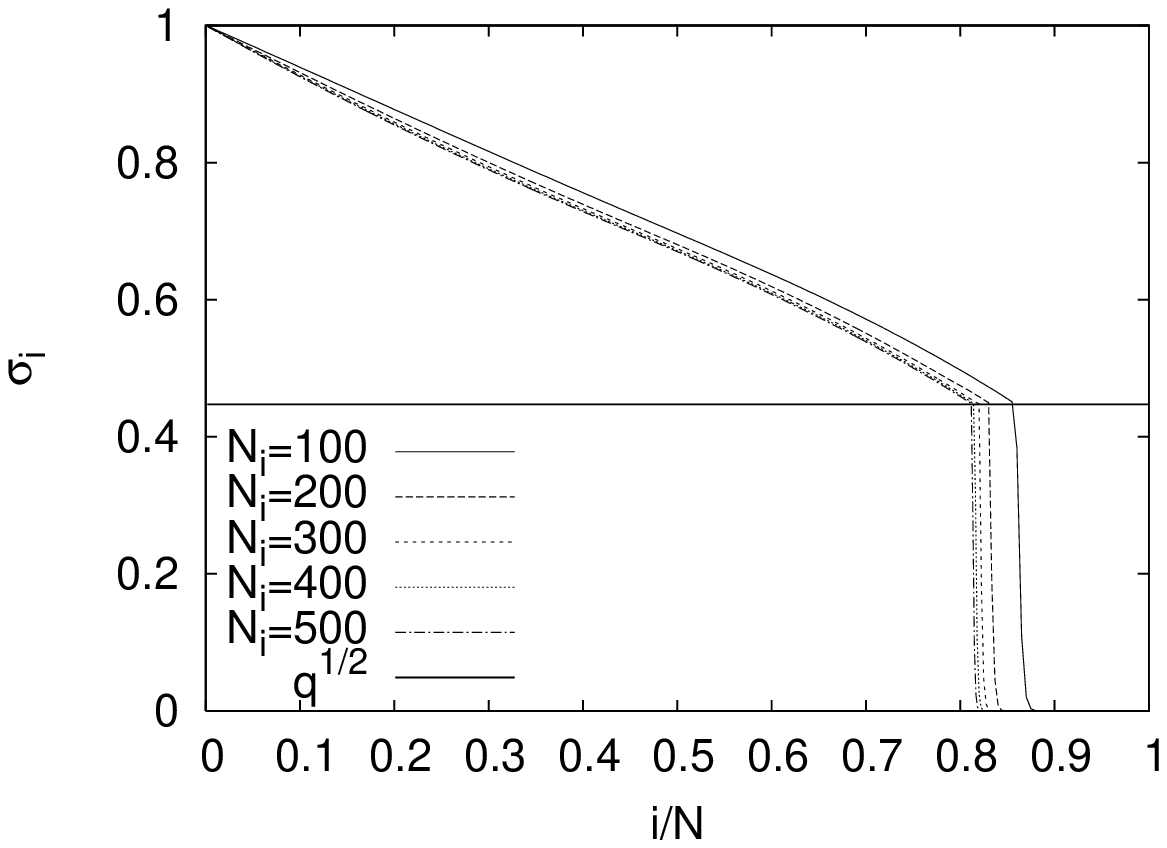}
\includegraphics[width=7cm]{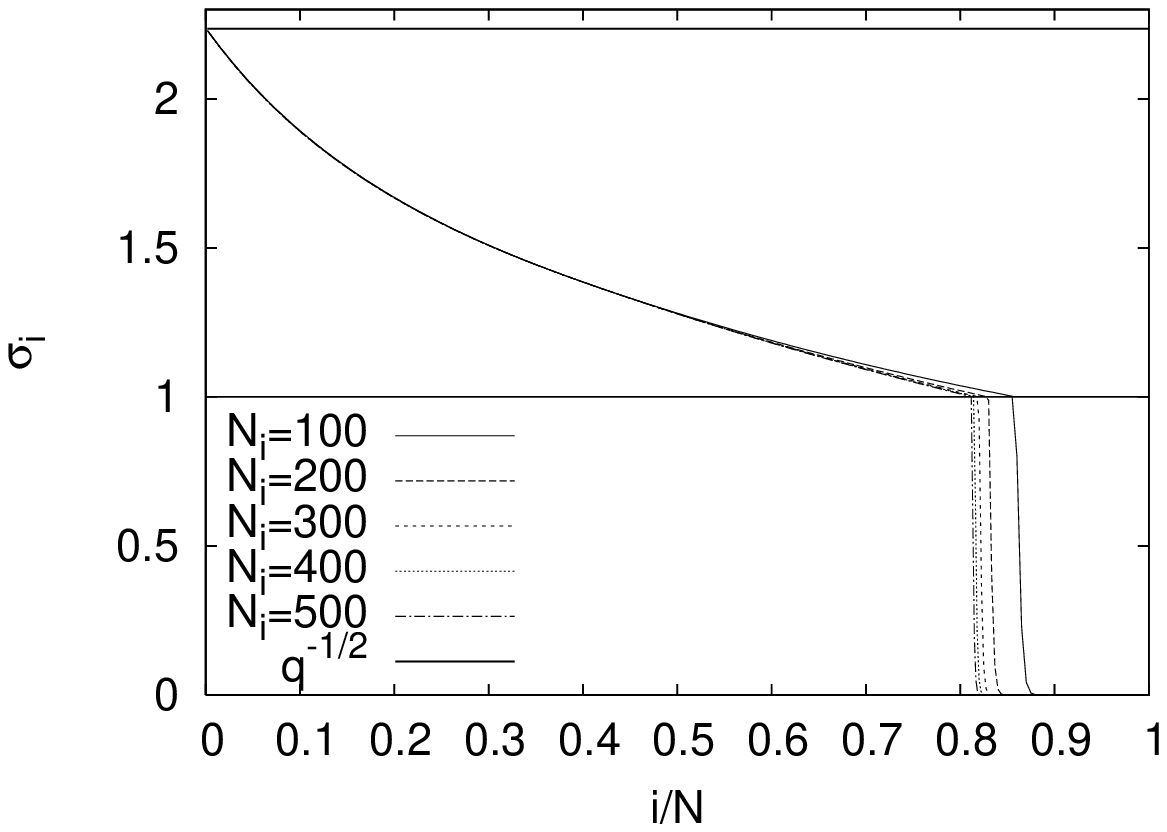}
\caption{Singular values of matrices $A$ (left) and $B$ (right) for different number of closed modes $\rNc$ = 100, 200, 300, 400 and 500 at $q$ =0.2 and $\rNo$ = 100.}
\label{pic:AB_svd_spectra}
\end{figure}
\begin{figure}[!htb]
\centering
\vbox{%
\includegraphics*[bb=10 10 264 244,height=4.7cm]{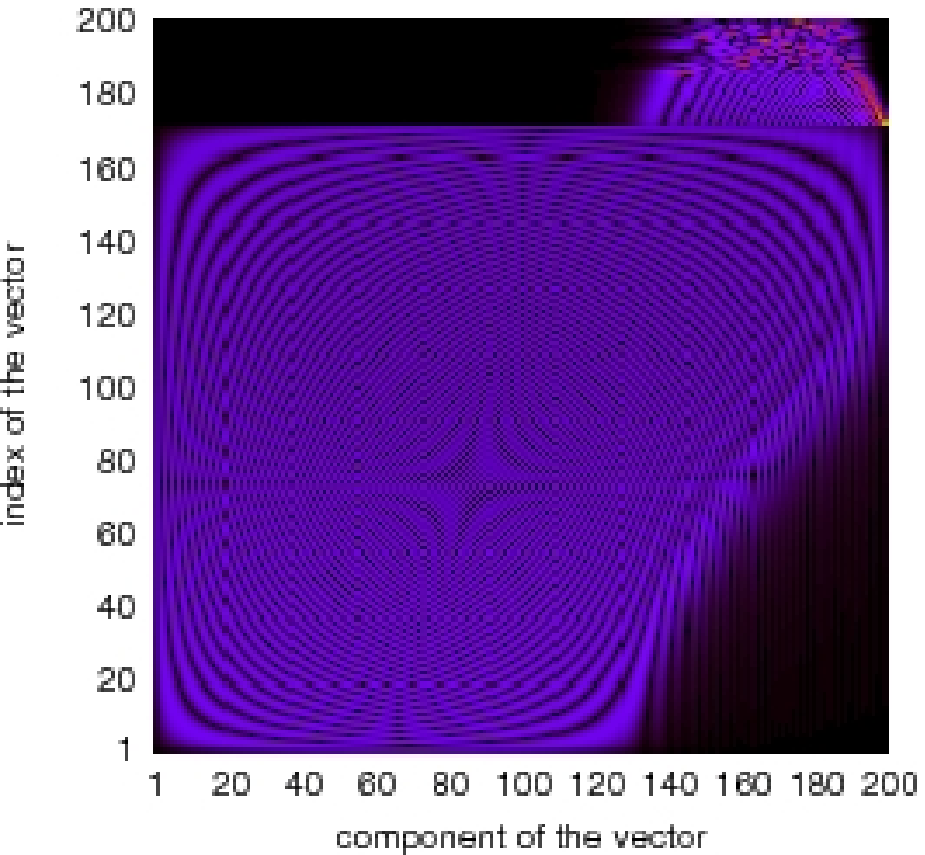}%
\includegraphics*[bb=10 10 264 244,height=4.7cm]{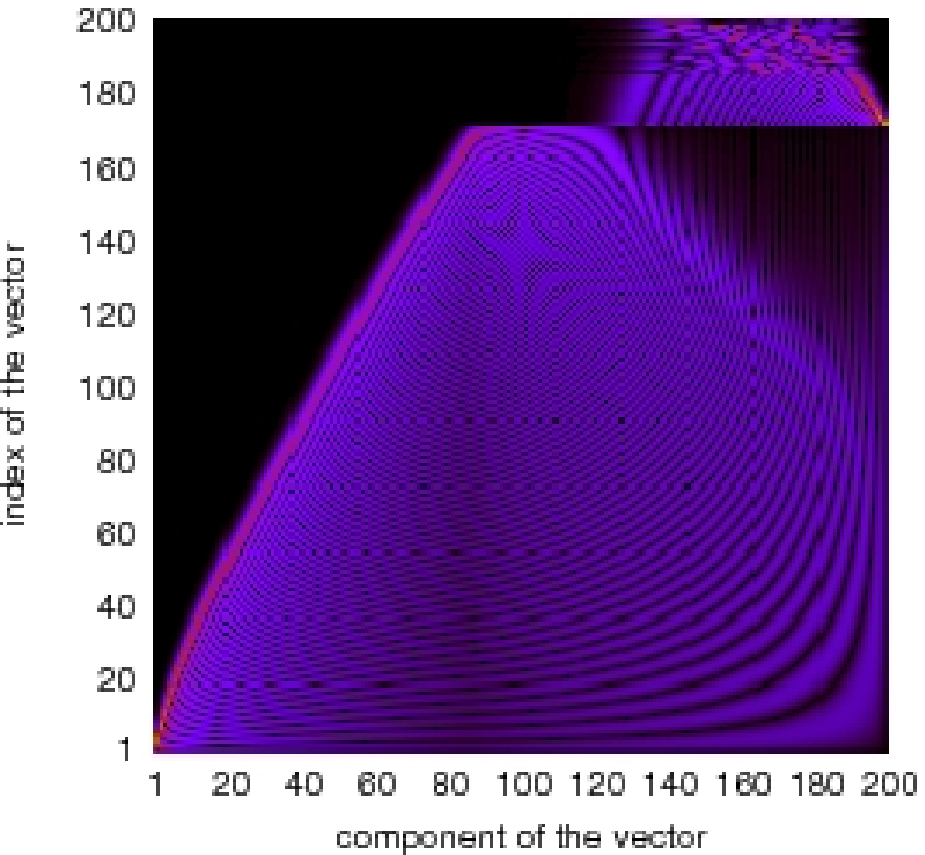}}
\vbox{%
\includegraphics*[bb=10 10 264 244,height=4.7cm]{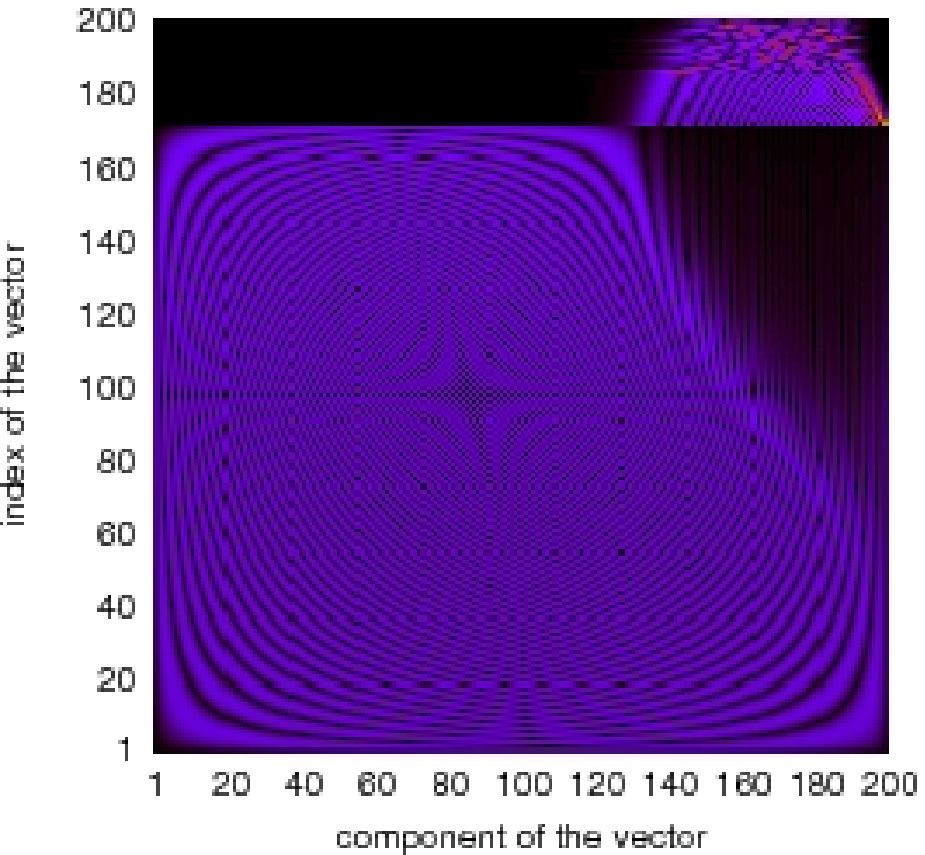}%
\includegraphics*[bb=10 10 264 244,height=4.7cm]{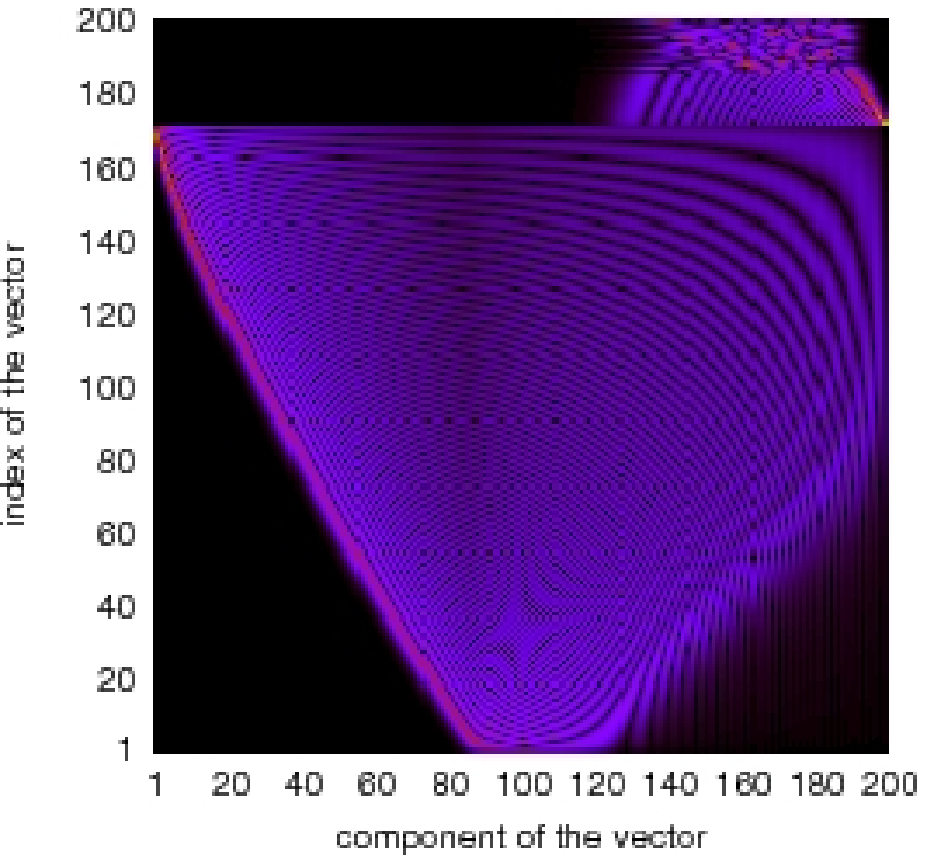}}%

\hskip5mm\includegraphics[bb=80 60 388 90, clip=, width=10cm]{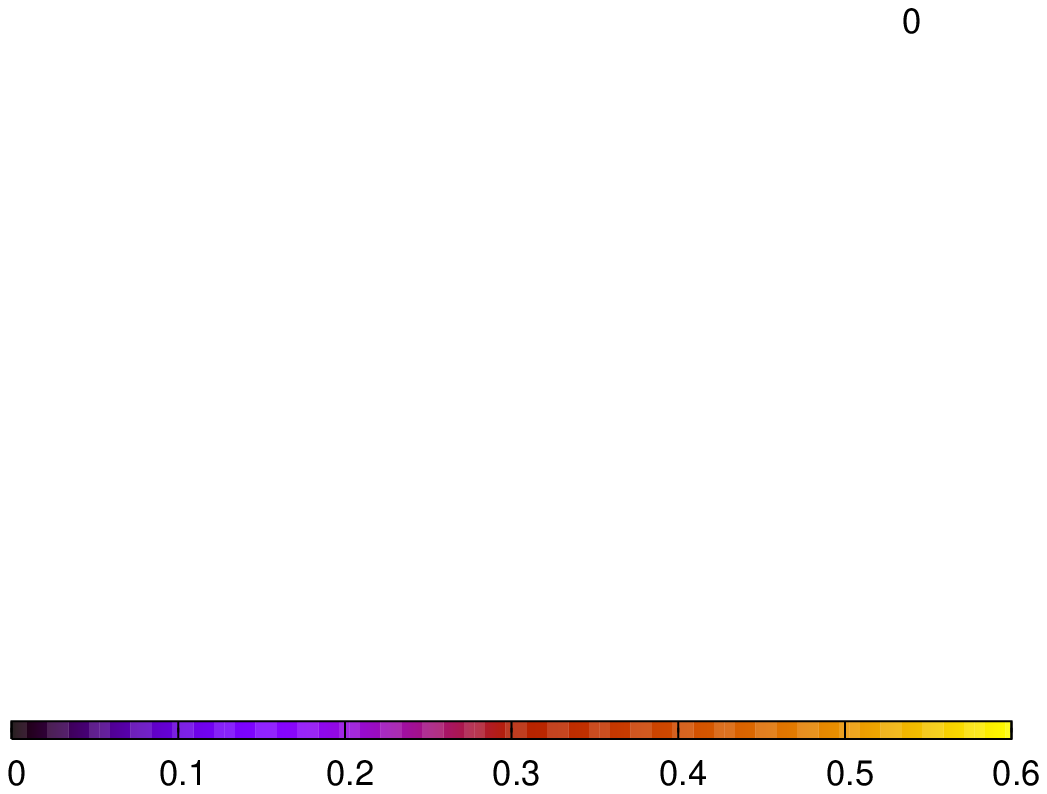}
\caption{Density plots of matrix elements $|U_{nm}|$ (left) and $|V_{nm}|$ (right) of the SVD decomposition of matrices $A$ (top row) and $B$ (bottom row) at inner radius $q=0.2$ with $\rNo =100$ open and $\rNc = 100$ closed modes. Labels on the abscissa and the ordinate are indices $n$ and $m$, respectively.}
\label{pic:AB_svd_vecs}
\end{figure}

The SVD decompositions \cite{demmel:book:97} of matrices $A$ and $B$ is useful for improving and stabilising the scattering calculations and will be used in the next section. From definitions of transition matrices (\ref{eq:mat_elem}) and completeness of mode functions we obtain
\beqa
  ( A A^T )_{nn'} &= \ave{u_n, r\, u_{n'}}\>,\qquad
  ( A^T A )_{pp'} &= (U_p, r\,U_{p'})\>, \label{eq:rel_AA}\\
  ( B B^T )_{nn'} &= \ave{u_n, r^{-1}\, u_{n'}}\>,\qquad 
  ( B^T B )_{pp'} &= (U_p, r^{-1}\,U_{p'})\>.
  \label{eq:rel_BB}
\eeqa
which we use to bound the image
\beqa
  \frac{\| A a \|_2}{ \| a \|_2 } 
  \in [q^{1\over 2},1]\>,\quad
  \frac{\| B a \|_2}{ \| a \|_2 } 
  \in \left[1,q^{-{1 \over 2}}\right]
  \quad\textrm{for}\quad
  \|a\|_2 \neq 0\>.
  \label{eq:AB_norm}
\eeqa
These results together with $AB^T=\id$ can be used to determine the form of the SVD decomposition 
\beq
   A = U \Sigma  V^T\>, \quad B = U \Sigma^{-1} V^T\>,\quad 
   \Sigma = \diag \left\{\sigma_i\in [\sqrt{q},1]\right\}_{i\in\bN}\>,
   \label{eq:AB_svd}
\eeq
where $U$ and $V$ are orthogonal matrices. We show here an example of the SVD decomposition of finite dimensional matrices $A$ and $B$ at $q=0.2$ and $\rNo=100$. In figure \ref{pic:AB_svd_spectra} we show singular values and in figure \ref{pic:AB_svd_vecs} we show density plots of the corresponding matrices $U$ and $V$, where the inner indices are ordered by decreasing magnitude of singular values. 
We see that the relative dimension of the space, which violates the bounds of the singular spectra
(\ref{eq:AB_norm}), converges with increasing space dimension $N=\rNo + \rNc$, where $\rNc$ is the number of imaginary modes. In the presented case $q=0.2$ the relative dimension is around $20\%$. It is important to see that vectors in $V$ and $U$ corresponding to singular values, which violate the bounds, have non-zero components only at closed modes. We conclude that due to a finite dimensional representation of matrices $A$ and $B$ we have deviations from the infinitely dimensional case only at high laying decaying modes that span a space of almost fixed relative dimension for some value of $q$.
%
\section{The scattering across a bend}\label{sec:scatt}
In this section we solve the on-shell scattering of a non-relativistic particle across our open billiard using the modal approach, initiated in the introduction. The scattering is discussed at fixed wave-number $k$ and inner radius $q$. 
The control of the precision of scattering calculations is studied in detail. 
In the second part of this section we investigate some interesting physical scattering properties of the bend.
 
\subsection{The scattering matrix of a bend}
The scattering matrix $S$ \cite{newton:book:02} is a linear mapping between the incoming and outgoing ``waves'' with respect to our scatterer. We reorganise the expansion coefficients of the wave function over different regions (\ref{eq:param_A}), (\ref{eq:param_B}), (\ref{eq:param_C}) into the incoming contributions denoted as $ v_{\rm in} = (a^+_1,a^+_2,\ldots,b^-_1,b^-_2,\ldots)$ and outgoing contributions $v_{\rm out} = (a^-_1,a^-_2,\ldots, b^+_1, b^+_2,\ldots)$. Then the scattering matrix $S$ can be defined as
\beq
  S v_{\rm in } = v_{\rm out}\>.
  \label{eq:S_symb}
\eeq
The S-matrix in our case  has a simple symmetric block form
\beq
  S = \mymat{R}{T}{T}{R}\>,
  \label{eq:bend_scatt_def}
\eeq
with $R$ and $T$ being the reflection and the transmission matrix, respectively. By reordering of rows and columns in $S$ so that the matrix elements concerning open (subscript $\rm o$) and closed (subscript $\rm c$) modes are separated and grouped together we obtain the matrix ${\cal S}$, reading
\beq
  S \to {\cal S} = 
  \mymat{{\cal S}_\roo } {{\cal S}_{\rm oc}}
        {{\cal S}_{\rm co}}{{\cal S}_{\rm cc}}\>.
\label{eq:S_reorder}  
\eeq
We find that blocks of matrix ${\cal S}$ (\ref{eq:S_reorder}) obey the {\it generalised unitarity} \cite{prosen:jpa:95} defined by the following relations
\beqa
{\cal S}_\roo{\cal S}_\roo^\dag &=&  {\cal S}_\roo^\dag {\cal S}_\roo = \id \>,
\label{eq:S_sim_1}\\
\ii\,{\cal S}_\roo {\cal S}_{\rm co}^\dag &=& {\cal S}_{\rm oc}\>,
\label{eq:S_sim_2}\\
\ii\, {\cal S}_{\rm oc}^\dag {\cal S}_\roo &=& {\cal S}_{\rm co}\>,
\label{eq:S_sim_3}\\
\ii\, {\cal S}_{\rm co} {\cal S}_{\rm co}^\dag &=& \ii\,  {\cal S}_{\rm oc}^\dag {\cal S}_{\rm oc} = {\cal S}_{\rm cc} - {\cal S}^\dag_{\rm cc}\>.
\label{eq:S_sim_4}
\eeqa
The relations (\ref{eq:S_sim_1} - \ref{eq:S_sim_4}) result from the probability current conservation, which is also equivalent to the condition $AB^T=\id$. Due to the time-reversal symmetry of the physical problem the scattering matrices $S$ and ${\cal S}$ are symmetric
\beq
S^T = S\>,\qquad  {\cal S}^T = {\cal S}\>.
\label{eq:S_sim_T}
\eeq
The block symmetry of the scattering matrix (\ref{eq:bend_scatt_def}) simplifies its calculation. We may consider individual incoming waves $e_n^+$ represented by the following wave function ansatz
\beqa
  \psi({\bf  r}) &= 
  e^+_n({\bf  r}) + \sum_m  e^-_m({\bf  r})\; R_{mn}\>, 
  \qquad 
  &{\bf  r} \in \Omega_{\rm A}\>, \label{eq:scatt_A_rt} \\
  \psi({\bf  r}) &=
  \sum_p f^+_p({\bf  r})\;\Lambda^+_{pn} +f^-_p({\bf  r})\;\Lambda^-_{pn}\>,
  \qquad
  &{\bf  r} \in \Omega_{\rm B}\>, \label{eq:scatt_B_rt} \\
  \psi({\bf  r}) &= 
  \sum_m e^+_m({\bf  r})\; T_{mn}\>.
  \qquad 
  &{\bf  r} \in \Omega_{\rm C}\>, \label{eq:scatt_C_rt} 
\eeqa
The continuity of $\psi({\bf r})$ and its normal derivative on the connecting cross-sections between regions $\Omega_{\rm A,B,C}$ determines the matrix elements $R_{mn}$, $T_{mn}$ and $\Lambda^{\pm}_{pn}$ and yields the following system of matrix equations
\beqa
  \id + R = M( \Lambda^+ + \Lambda^-)\>,\qquad
  T = M ( {\cal F} \Lambda^+ +  {\cal F}^{-1} \Lambda^- )\>,
  \label{eq:scatt_sys_1}\\
  \id - R = N ( \Lambda^+ - \Lambda^- )\>,\qquad
  T =  N ( {\cal F} \Lambda^+ -  {\cal F}^{-1} \Lambda^- )\>,
  \label{eq:scatt_sys_2}\\
  M = G^{\frac{1}{2}} A V^{-\frac{1}{2}}\>,\qquad 
  N = G^{-\frac{1}{2}} B V^{\frac{1}{2}}\>,
\eeqa
which we write using diagonal matrices $V= \diag\{\nu_n\}_{n\in\bN}$, $G = \diag\{g_n\}_{n\in\bN}$ and ${\cal F} = \exp(\ii V)$, and transition matrices $A$ and $B$ (\ref{eq:mat_elem}). The elimination of matrices $\Lambda^\pm$ from equations (\ref{eq:scatt_sys_1}) and (\ref{eq:scatt_sys_2}) yields the blocks of the scattering matrix $S$, reading
\beqa
  T 
  = 
  (C - {\cal F} D C^{-1} {\cal F} D)^{-1} {\cal F} (C - DC^{-1}D)\>,
  \label{eq:scatt_T}\\
  R 
  = 
  (C - {\cal F} D C^{-1} {\cal F} D)^{-1} ({\cal F} DC^{-1} {\cal F}C - D)\>,
  \label{eq:scatt_R}
\eeqa
that we express by using the following auxiliary matrices
\beqa
  C =  M^{-1} + N^{-1} 
  = 
  V^{1\over 2} B^T G^{-{1\over 2}} +  
  V^{-{1\over 2}} A^T G^{1\over 2}\>,
  \label{eq:mat_C}\\
  D = M^{-1} - N^{-1} 
  = 
  V^{1\over 2} B^T G^{-{1\over 2}} -  
  V^{-{1\over 2}} A^T G^{1\over 2}\>,
\label{eq:mat_D}
\eeqa
where we take into account the relation $AB^T=\id$. The presented form of the matrix $T$ (\ref{eq:scatt_T}) and $R$ (\ref{eq:scatt_R}) is chosen in order to increase its numerical stability i.e. minimising the use of inverses and avoiding direct computation of ${\cal F}^{-1}$.
\subsection{Numerically stable scheme for scattering matrix calculation}
A bend on a straight waveguide is a paradigmatic example for testing numerical schemes and ideas on how to accurately calculate the scattering matrix. In particular, the high curvature case $q\to 0$ turns to be highly non-trivial. Here we give a simple and stable procedure to obtain the scattering matrix with a clear precision control for practically all curvatures.\par
The scattering across a bend of angle $\beta$ and inner radius $q$ back to asymptotic region at some wavenumber $k$ is described by the scattering matrix $S(\beta)$ (\ref{eq:bend_scatt_def}), which is composed of the reflection matrix $R(\beta)$ (\ref{eq:scatt_R}) and the transmission matrix $T(\beta)$ (\ref{eq:scatt_T}). In practice, we work with finite dimensional matrix approximations, denoted by
\beq
  R_N(\beta),\; T_N(\beta) \in \bC^{N\times N}\>, 
  \quad 
  S_N(\beta) \in \bC^{2N\times 2N}\>,
  \quad
  A_N, B_N  \in \bR^{N\times N}\>.
\eeq
where $N \ge \rNo$ is the number of modes used in the asymptotic regions. The main objective is to construct these finite dimensional matrices $R_N(\beta)$ and $T_N(\beta)$ so that:
\begin{enumerate}
\item calculations are numerically stable and precise,
\item $S_N(\beta)$ satisfies the time reversal symmetry (\ref{eq:S_sim_T}) and the generalised unitarity relation (\ref{eq:S_sim_1} - \ref{eq:S_sim_3}),
\item the sub-block of $R_N(\beta)$ and $T_N(\beta)$ of dimension $N'\in (\rNo,N]$ is calculated with controllable accuracy, where $\rNo = \lfloor k a/\pi\rfloor$ is the number of open modes in the asymptotic region.
\end{enumerate}
The recipe to achieve these assumptions may be separated into two parts. In the first part we cure the numerical instability caused by the maximal element $r(\beta)= \exp(\beta\nu_N)$ in ${\cal F}_N(\beta)$, which are exponentially diverging with increasing $N$. This is achieved by separating the bend into $2^{n_0}$ identical subsections of angle $\beta'= 2^{-n_0}\beta$ so that $r(\beta')\eps_{\rm num}\approx 1$, where $\eps_{\rm num}$ is the numerical precision e.g. $\eps_{\rm num} = 2^{-52}$ in double precision floating point arithmetic. The number $n_0$ is calculated as
\beq
  n_0 = \max\left\{0, \;
 1 + \left\lfloor \log_2 \left(\frac{|\log\eps_{\rm num}|}{\beta \nu_N} \right)\right\rfloor \right\}\>.
  \label{eq:comb_n_steps}
\eeq
The scattering matrix $S_N(\beta')$ of a small subsection of the bend  can be calculated in a very stable way and up to a high precision. By concatenating scattering matrices of subsections together with the recursion
\beq
  S_N(2^{-m+1}\beta ) 
  = S_N(2^{-m}\beta ) \odot S_N(2^{-m}\beta )\>.
  \label{eq:comb_recurs}
\eeq
we obtain the scattering matrix of the whole bend $S_N(\beta)$. The symbol $\odot$ denotes the {\it operation for concatenating scattering matrices} associated to scatterers on the waveguide and is defined in \ref{sec:comb_scatt}.\par
\begin{figure}[!htb]
\centering
\begin{minipage}[c]{12.5cm}
\hbox{
\includegraphics*[bb=50 85 400 302, width=6.6cm]{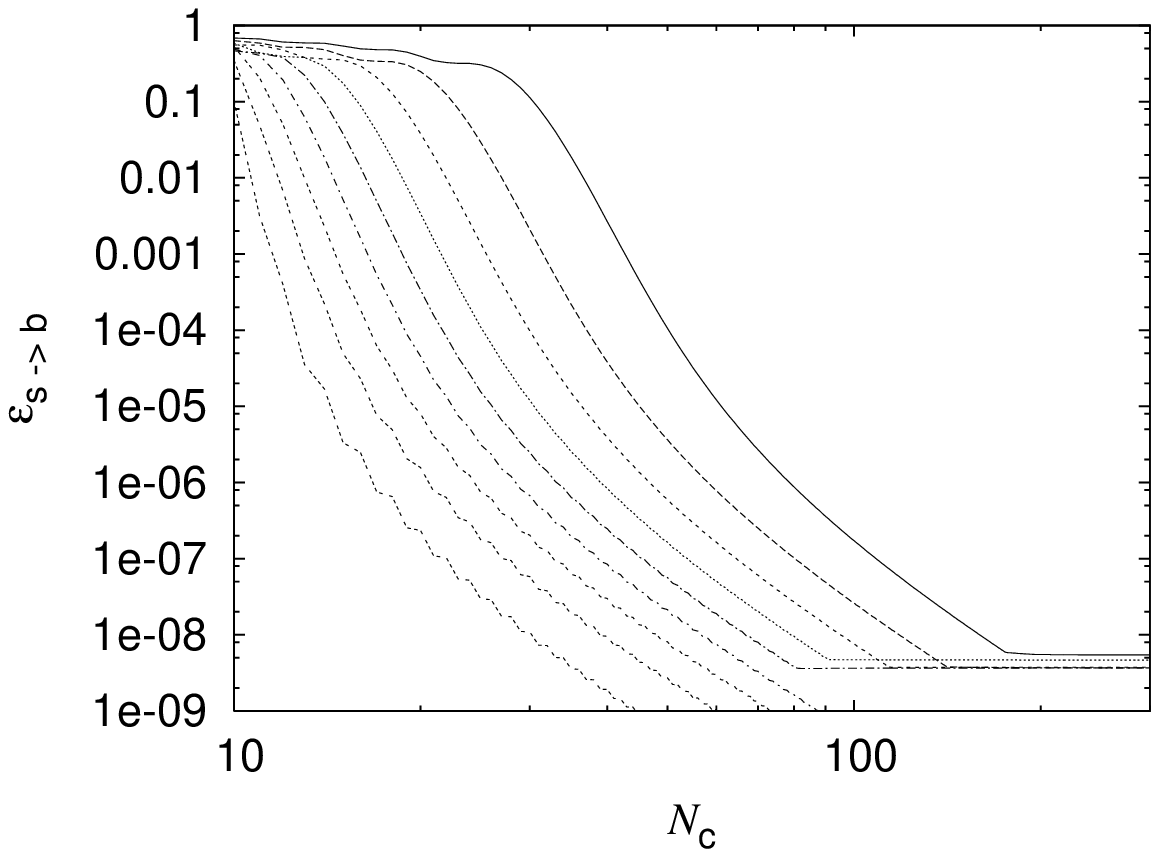}%
\includegraphics*[bb=80 85 400 302, width=6cm]{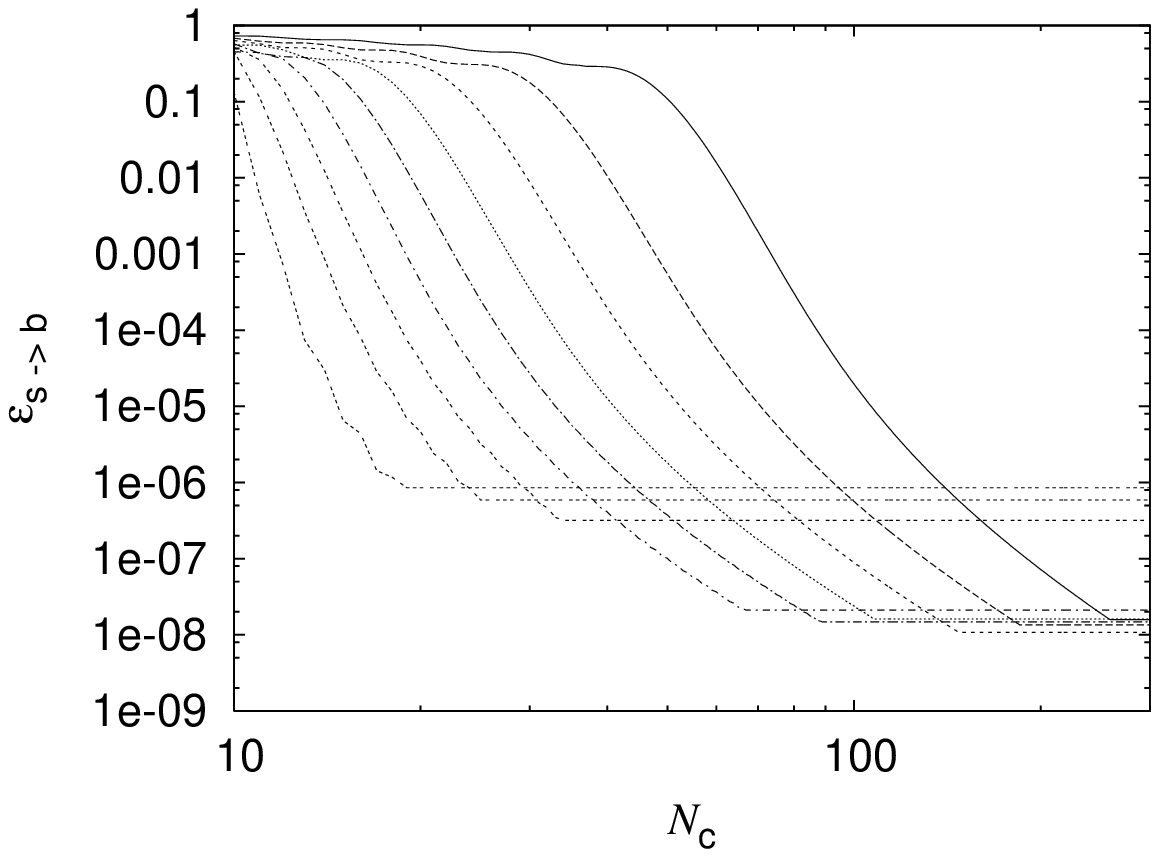}}
\hbox{
\includegraphics*[bb=50 50 400 302, width=6.55cm]{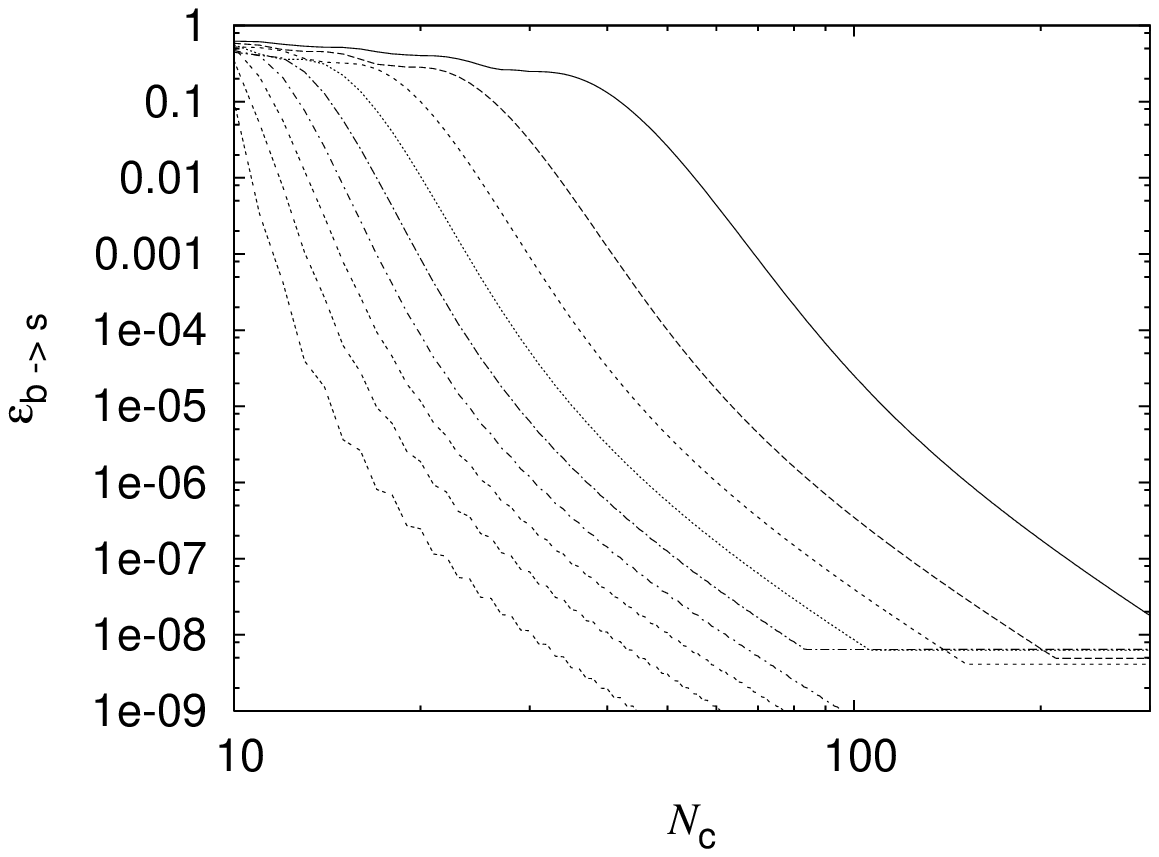}%
\includegraphics*[bb=80 50 400 302, width=6cm]{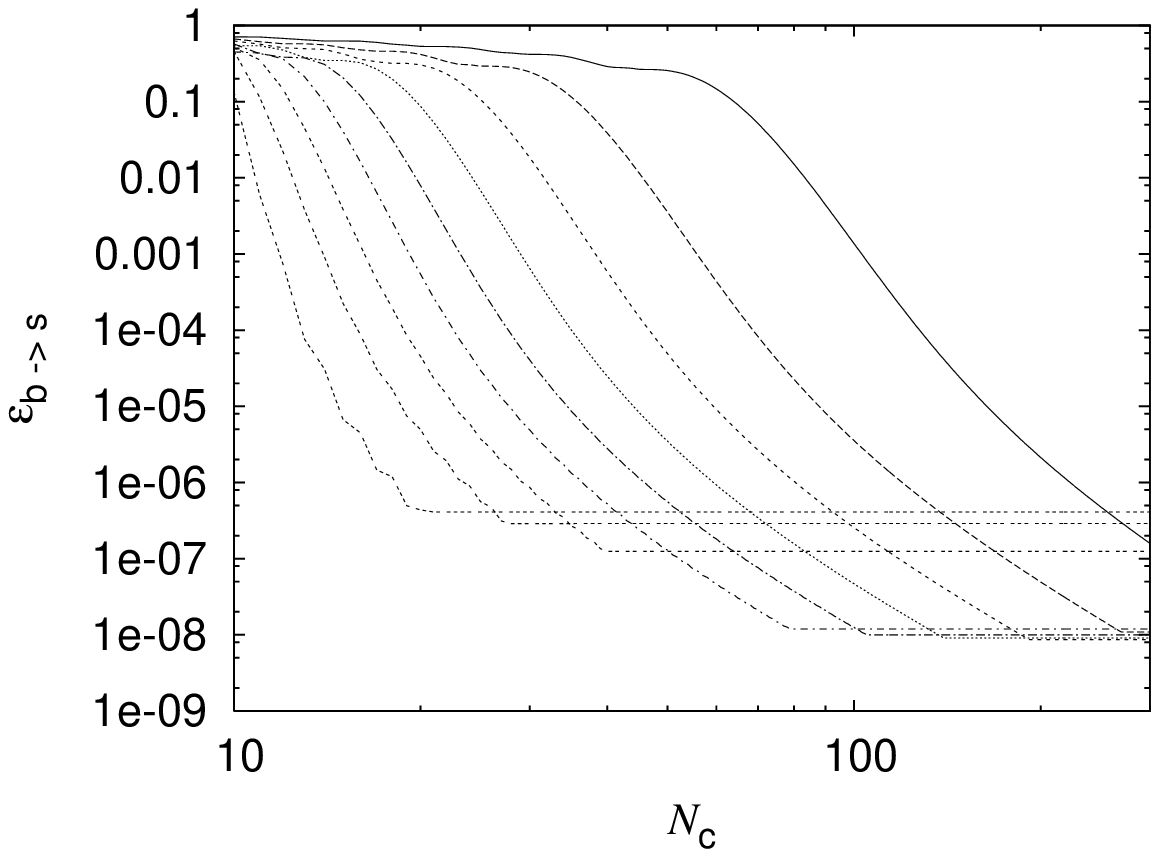}}
\end{minipage}
\begin{minipage}[c]{1.5cm}
\includegraphics*[bb=310 82 397 285, width=1.5cm]{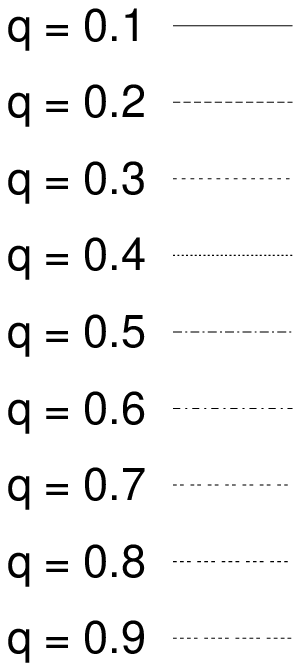}
\end{minipage}
\caption{The transition errors $\eps_{{\rm s}\to {\rm b}}(\rNo+\rNc,N')$  (top) and  $\eps_{{\rm b}\to {\rm s}}(\rNo+\rNc,N')$ (bottom) calculated on the sub-space of dimension $N'=\rNo+10$ vs. the number of closed modes $\rNc$ for $\rNo=10,100$ (left, right).}
\label{pic:id_eps}
\end{figure}
\begin{figure}[!htb]
\centering
\begin{minipage}[c]{12.5cm}
\vbox{\includegraphics*[bb=50 85 400 302, width=6.5cm]{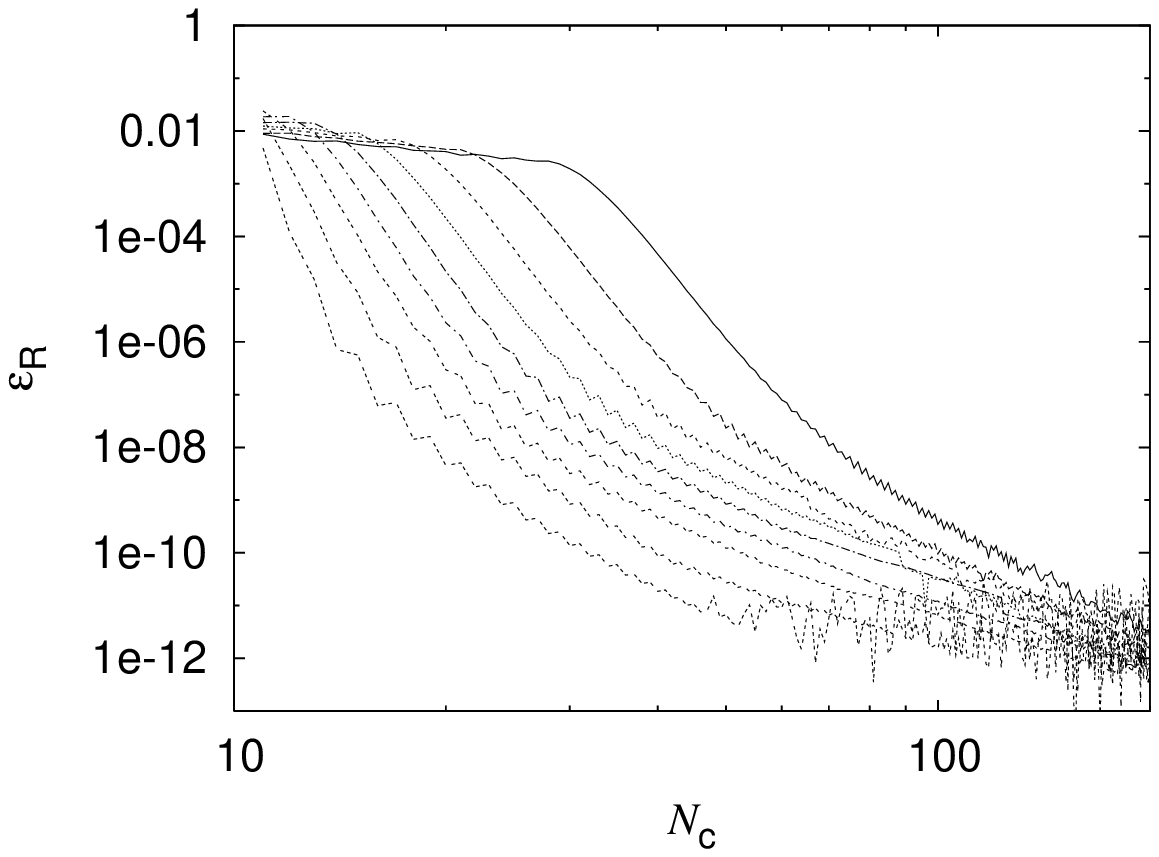}%
\includegraphics*[bb=80 85 400 302, width=6cm]{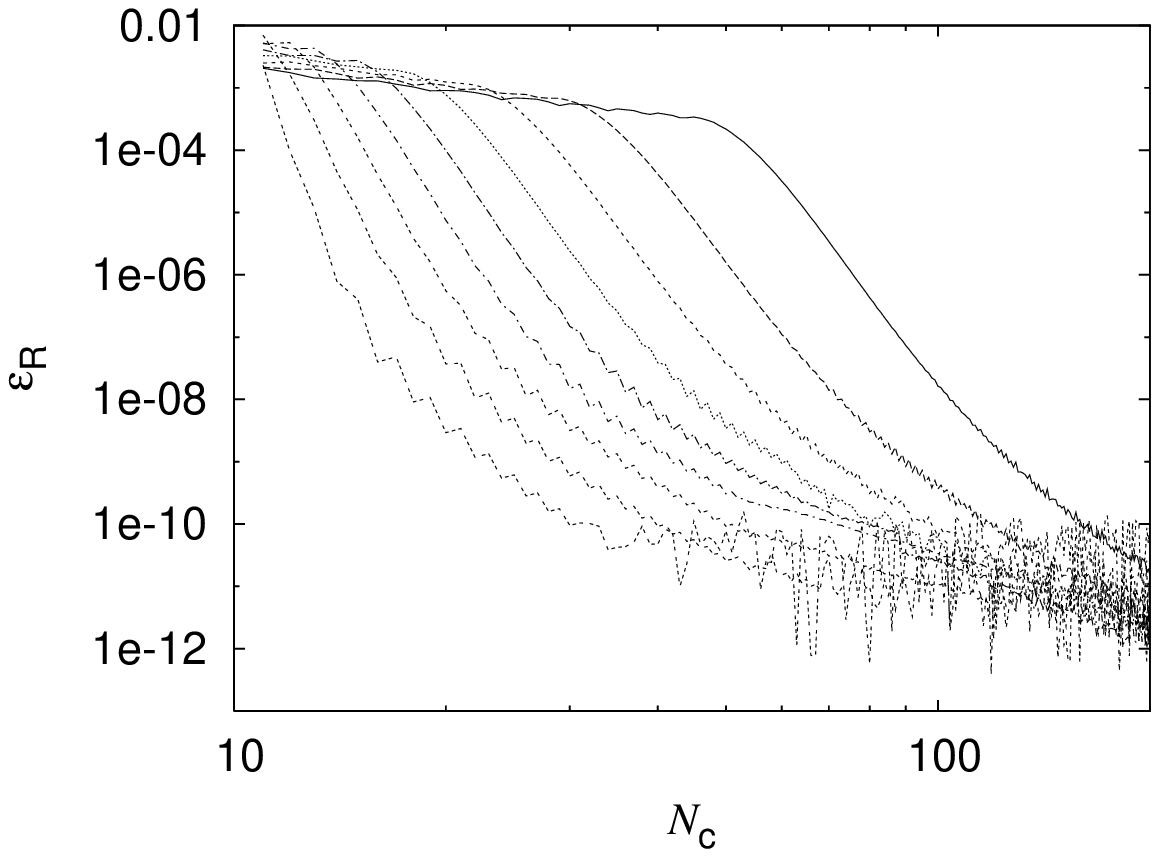}}%
\vbox{\includegraphics*[bb=50 50 400 302, width=6.5cm]{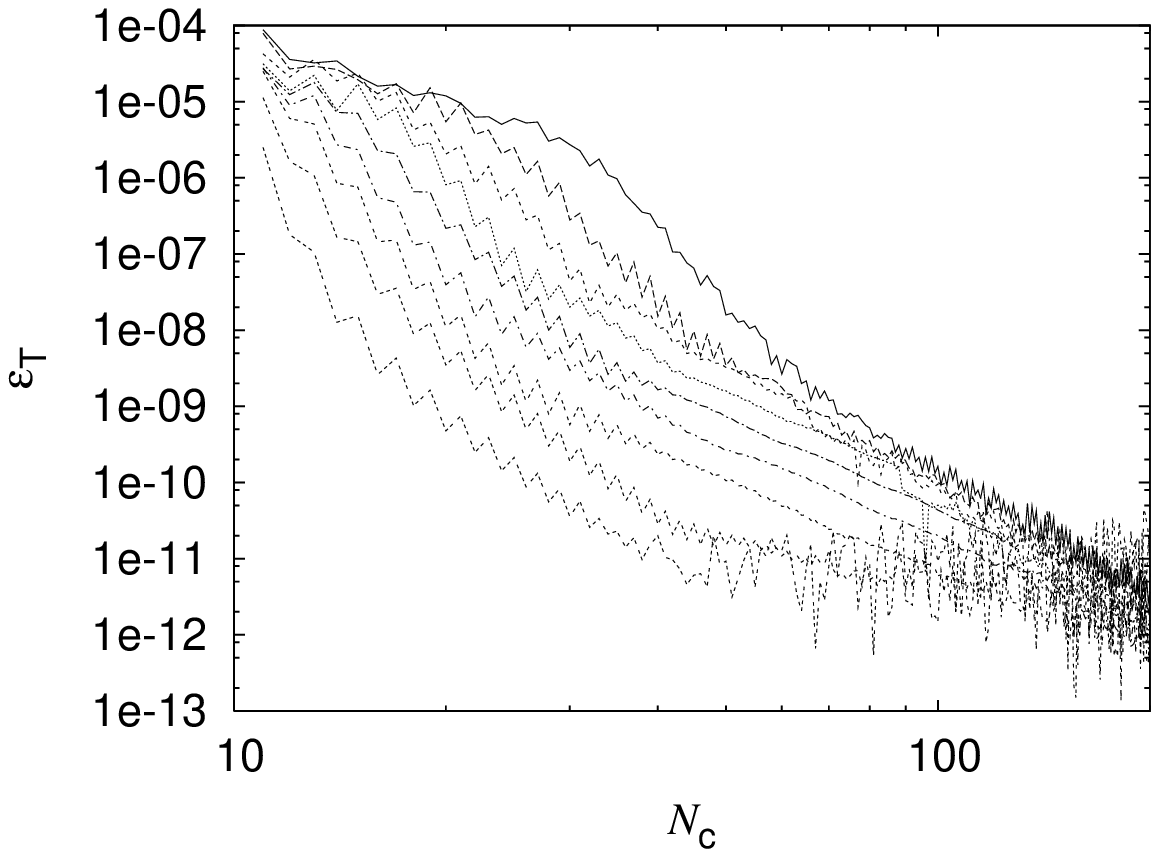}%
\includegraphics*[bb=80 50 400 302, width=6cm]{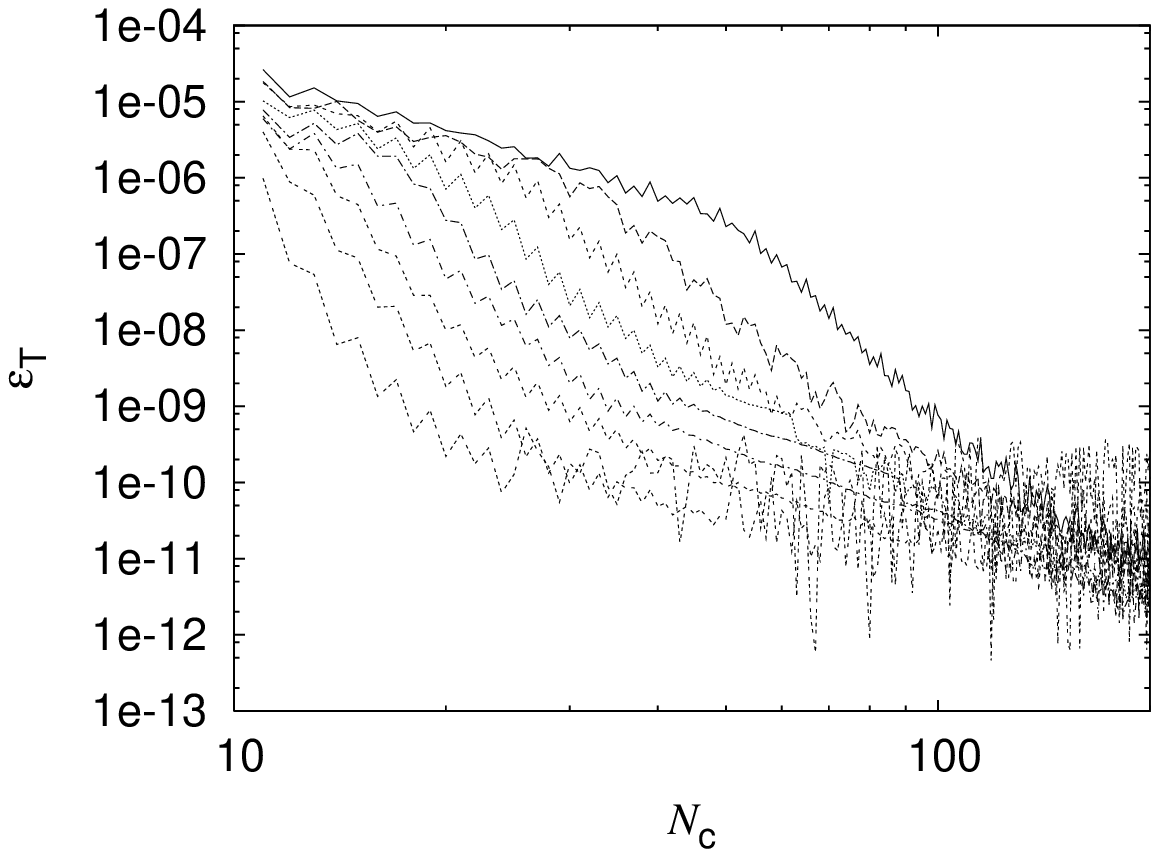}}%
\end{minipage}
\begin{minipage}[c]{1.5cm}
\includegraphics*[bb=317 82 397 285, width=1.5cm]{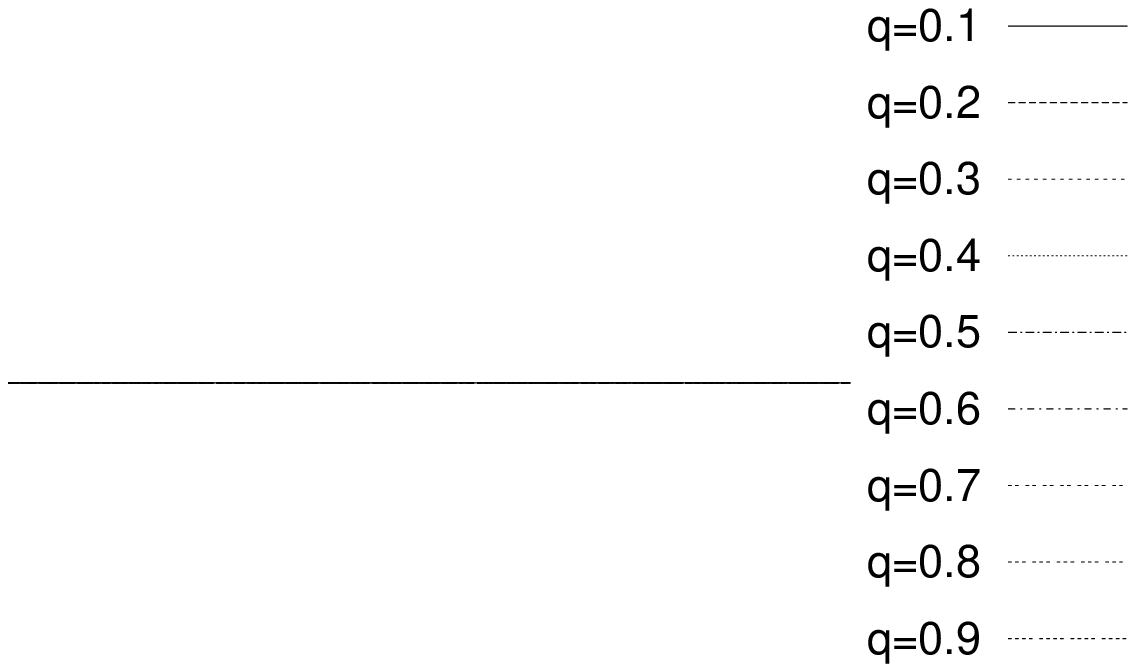}
\end{minipage}
\caption{The measures of convergence $\eps_{\rm R,T}(\rNo+\rNc,\rNo+10)$ (top, bottom) as functions of $\rNc$ for $\rNo = 10,100$ (left, right) at various inner radii $q$ as indicated in the figure. The bend is of angle $\beta=\pi$.}
\label{pic:RT_convergeX}
\end{figure}

In the second part we are discussing the problem that $S_N(\beta')$ diverges with increasing $N$,
which is due to violation of the identity $AB^T = \id$ (\ref{eq:mat_AB_id}) for finite truncated transition matrices $A_N$ and $B_N$. We eliminate the problem by deforming $A_N$ and $B_N$ so that they are non-singular and exactly fulfil the condition $A_N B_N^T = \id$. We make the SVD decomposition of the truncated matrix $A_N = U_N \Sigma_N V^T_N$, modify its singular values $\Sigma_N = \diag\{\sigma_i\}_{i=1}^N$ to $\tilde \Sigma_N = \diag\{\tilde\sigma_i\}_{i=1}^N$ so that they fit in the bounds obtained for infinitely dimensional case (\ref{eq:AB_svd})
\beq
  \tilde\sigma_i = 
  \left \{
  \begin{array}{lll} 
     \sigma_i&:& \sigma_i\in [q^{1\over 2},1] \cr
     \sigma^*&:& {\rm otherwise}
  \end{array}
  \right. \>,
  \label{eq:correct_sigma}   
\eeq
and again generate both matrices
\beq
  \widetilde A_N = U_N \widetilde{\Sigma}_N V^T_N\>,\qquad 
  \widetilde B_N = U_N \widetilde{\Sigma}_N^{-1} V^T_N\>.
  \label{eq:correct_A}  
\eeq
The same procedure can also be done using SVD decomposition of the matrix $B_N$ as a base for generation of both deformed matrices $\widetilde A_N$ and $\widetilde B_N$. The value of $\sigma^*>0$  can be chosen arbitrarily, but the most elegant choice is $\sigma^*=1$. By using matrices $\widetilde A_N$ and $\widetilde B_N$ (\ref{eq:correct_A}) instead of $A_N$ and $B_N$ in $S_N(\beta')$ and consequently in $S_N(\beta)$ these become generalised unitary with a well behaved and physically precise limit $N\to\infty$ at least on the sub-space of dimension $N'$. We first check this by discussing the precision of transition between the modes in the bend and in the straight waveguide on the sub-space of dimension $N'<N$. The {\it error of transition} from the asymptotic region (infinite waveguide) into the bend is quantified by
\beq
\fl\eps_{{\rm s}\to {\rm b}}(N,N')
 = \max_{n, m \in L} 
   |\bra{u_n} \left[\sum_{p=1}^N |U_p)(U_p| - \id \right] \ket{u_m}|  
 = \max_{n, m \in L} 
 |\sum_{p=1}^N  A_{np} B_{mp} - \delta_{nm}|\>,
 \label{eq:id_eps_sb}
\eeq
and for transition in the opposite direction by
\beq
\fl\eps_{{\rm b}\to {\rm s}}(N,N')
 = \max_{p, r \in L} 
   |(U_p| \left[\sum_{n=1}^N \ket{u_n}\bra{u_n} - \id \right] |U_r)|  
 = \max_{p, r \in L}
   |\sum_{n=1}^N A_{nr} B_{np} - \delta_{pr}|\>.
 \label{eq:id_eps_bs}
\eeq
with $L= \{1,2,\ldots N' \}$. The introduced transition errors (\ref{eq:id_eps_sb}) and (\ref{eq:id_eps_bs}) measure the violation of the identity $A B^T =\id$ on the subspace $N'$, when working with the finite number of modes $N$. It is expected and supported by our numerical studies that the errors vanish in the limit $N\to\infty$ at fixed $N'$ and other parameters. In figure \ref{pic:id_eps} we show transition errors as a function of $N$ at a fixed $N'$ for two values of $\rNo$. 
The errors decrease with increasing $\rNc$ down to a certain plateau, which is determined by the precision of mode functions. The most problematic in the precision are the highest few open modes in the bend. A conservative estimate for the plateau is around $10^{-8}$. The transition errors (\ref{eq:id_eps_sb}) and (\ref{eq:id_eps_bs}) increase with decreasing $q$ indicating that we need more modes to achieve equally small error as for higher $q$. 
We did not find any analytic approximation for transition errors. Therefore we numerically estimate the minimal dimension of the functional space $N_{\rm t} = \rNc + \rNo$ needed for transition errors to be smaller than some $\eps$, defined as
\beq
N_{\rm t} (N', \eps)  = 
  \min\{ N : \eps_{{\rm s}\to {\rm b}} (N,N')  < \eps  \textrm{ and }
             \eps_{{\rm b}\to {\rm s}} (N,N')  < \eps \}\>,
  \label{eq:N_geom}	     
\eeq
where $N'= \rNc' + \rNo$ is the dimension of the observed sub-space. In our numerical analysis we set $\eps \approx 10^{-7}$. We check the convergence of $R_N(\beta)$ and $T_N(\beta)$ with increasing $N$ on some fixed sub-space of dimension $N'<N$. The matrices $R_N$ and $T_N$ are calculated using the method of dividing the bend into sub-sections together with deforming the transition matrices. The convergence is measured through the relative difference of matrices at subsequent changes of the dimension $N$ 
\beq
\fl\hspace{1cm}  \eps_{\rm R} (N,N') = 
  \frac{\|R_{N+1}(\beta) - R_N(\beta) \|_{N'}}{\|R_N(\beta)\|_{N'}}\>,
  \quad     
  \eps_{\rm T} (N,N') = 
  \frac{\|T_{N+1}(\beta) - T_N(\beta) \|_{N'}}{\|T_N(\beta)\|_{N'}}\>,
  \label{eq:conv_cauchy_measure}
\eeq
where we introduce a matrix norm $\| A \|_M = \max_{i,j\in [1, M]} |A_{ij}|$ on the sub-space of dimension $M$. The expressions (\ref{eq:conv_cauchy_measure}) give upper bounds for the deviations of matrices from their asymptotic forms 
\beqa
  \|R_{\infty}(\beta) - R_N(\beta) \|_{N'} \le 
  C_1(N') \sum_{M=N+1}^\infty \eps_{\rm R}(M,N')\>, 
  \label{eq:conv_abs_measure_1}\\
  \|T_{\infty}(\beta) - T_N(\beta) \|_{N'} \le 
  C_2(N') \sum_{M=N+1}^\infty \eps_{\rm T}(M,N')\>, 
  \label{eq:conv_abs_measure_2}
\eeqa
with expressions $C_1 (N')=\max_{N\ge N'} \|R_N(\beta)\|_{N'}$ and $C_2 (N')=\max_{N\ge N'} \|T_N(\beta)\|_{N'}$, which are of the order of magnitude 1. We have numerically studied quantities $\eps_{\rm R, T}(N,N')$ as functions of $N$ at fixed $N'$ and the results are shown in figure \ref{pic:RT_convergeX}.
In the case $N = N' = \rNo$ we are talking about the open-open block of the scattering matrix, which is sometimes called the semi-quantal approximation \cite{bogomolny:nonlin:92}. We see that $\eps_{\rm R}$ and $\eps_{\rm T}$ decrease with increasing total number of modes $N=\rNo+\rNc$ in a similar fashion as transition errors down to some plateau around $10^{-12}$. The plateau is almost equal to the machine precision, which is surprisingly better than the transition errors.\par
The presented method for the calculation of the scattering matrix and its accuracy control works well for all $q>0$. Nevertheless, a treatment of high curvature cases $q\to 0$ are difficult as we need to consider a large number of closed modes to reach a sufficient precision of the scattering matrix. 
However, this is feasible to achieve by our method in a stable and controlled way.
\subsection{Quantum transport across the bend}
\begin{figure}[!htb]
\vbox{\small\hspace{4.7cm}$q=0.2$\hspace{3.8cm}$q=0.6$}
\vskip2pt
\vbox{
\centering
\includegraphics*[width=5cm]{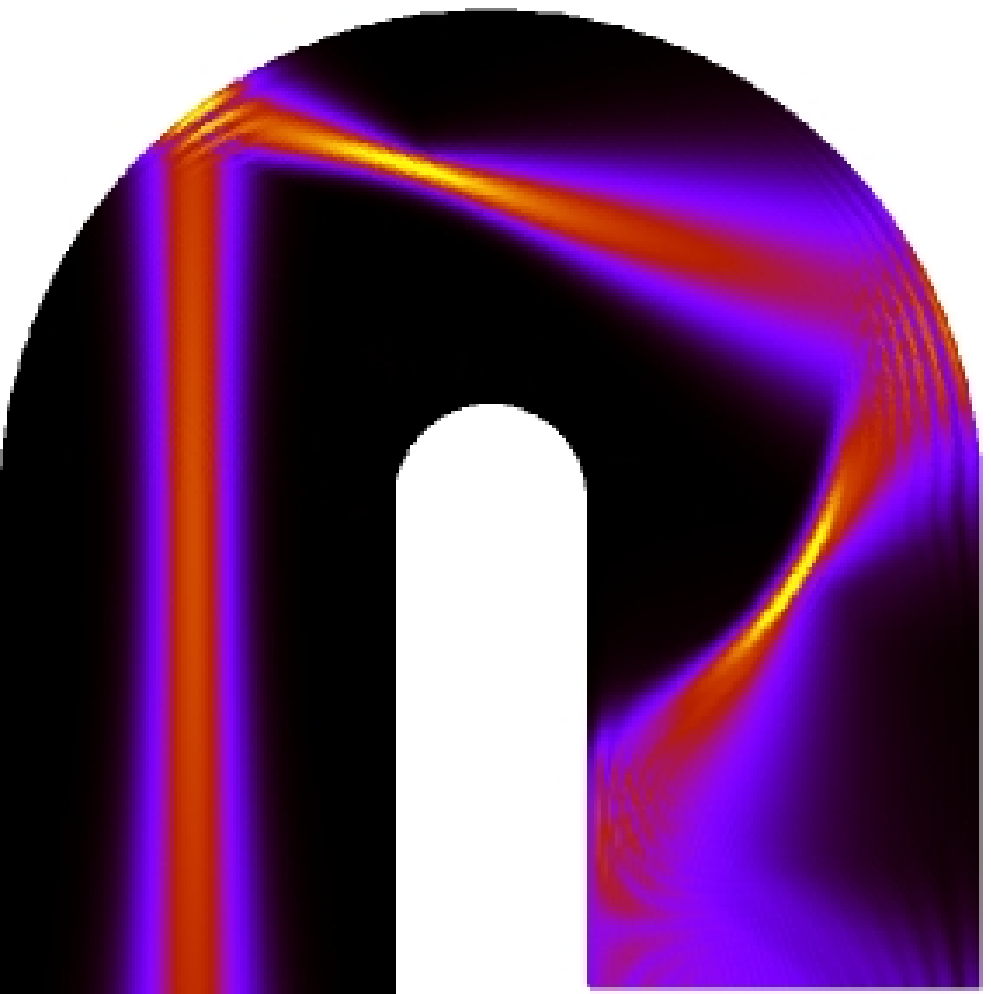}\hskip1pt%
\includegraphics*[width=5cm]{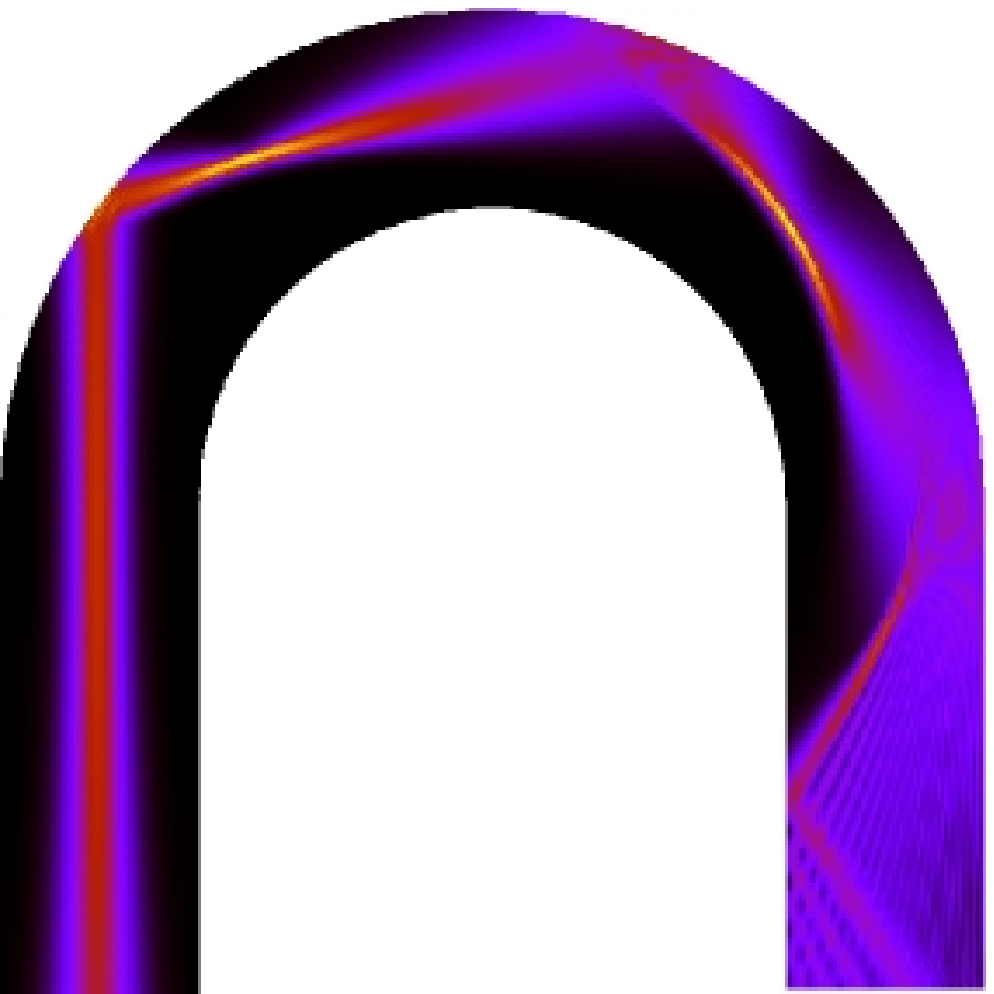}%
}
\centering
\hskip5mm\includegraphics*[bb = 80 60 388 90, width=10cm]{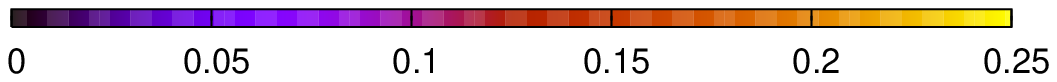}
\caption{Scattering of an incoming ray of Gaussian shape with unit probability flux at wavenumber $k$, which supports 100 open modes, and for two inner radii $q$ = 0.2, 0.6 (left, right).}
\label{eq:qm_bend_figs}
\end{figure}

The scattering matrix of the open billiard describes the stationary quantum transport of a particle over the bend. We discuss here the most important and obvious measures of the transport, which are the reflection probability and the {\it Wigner-Smith delay time}. Since the classical particle cannot scatter back \cite{horvat:jpa:unidir04} we are particularly interested in the reflection probability as a genuine quantum (wave) property. This is a measure of quantum tunnelling between the two classically invariant components of the phase space corresponding to right and left going waves. The scattering of an incoming Gaussian ray over the bend is illustrated in figure \ref{eq:qm_bend_figs}. The ray follows the classical trajectories, but as it is of finite width, its parts are scattered differently when hitting the curved wall. The parts of the ray travel different lengths and interfere among themselves.
\par
\begin{figure}[!htb]
\vbox{\small\hspace{3cm}$R_N$\hspace{4cm}$T_N$\hspace{4.5cm}$T_{\rm class}$}
\vbox{
\centering
\includegraphics*[height=5cm]{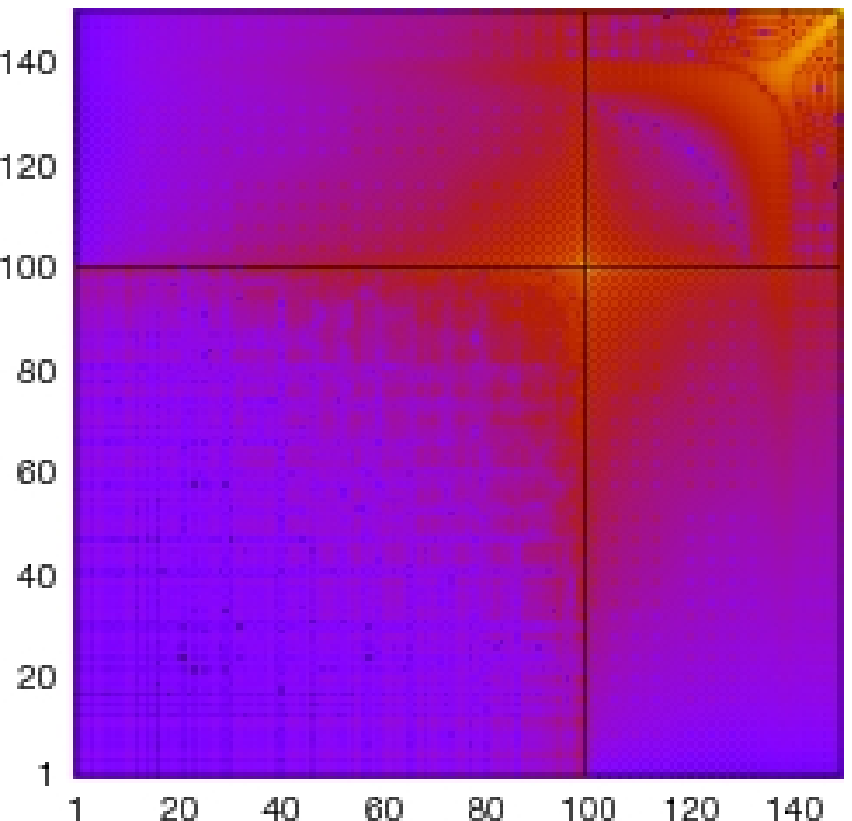}\hskip1pt%
\includegraphics*[height=5cm]{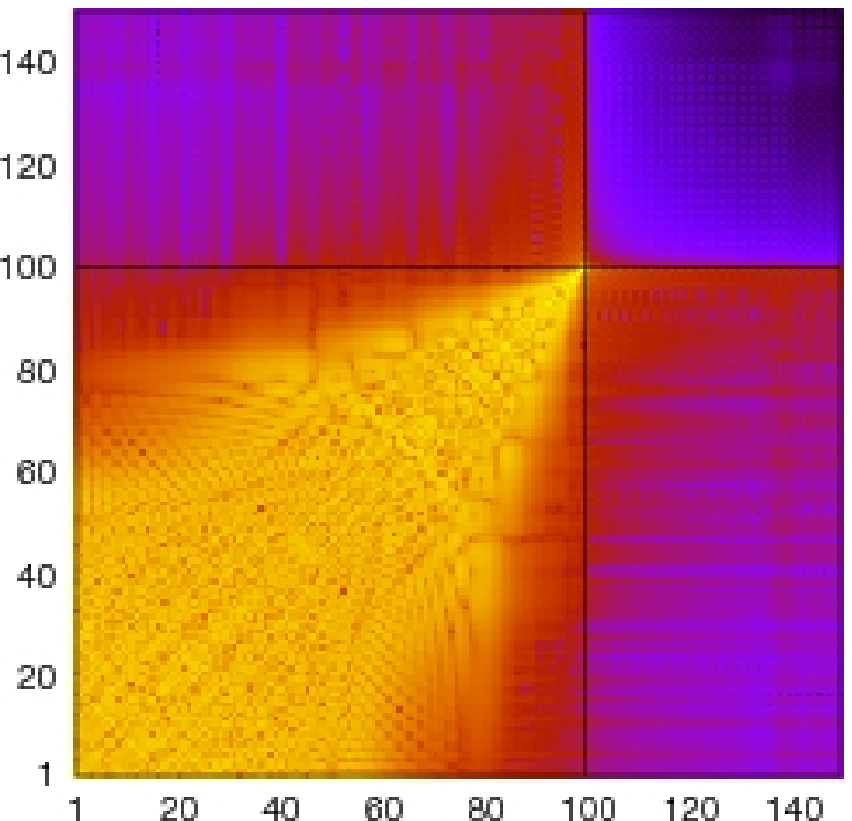}\hskip10pt%
\includegraphics*[height=3.5cm]{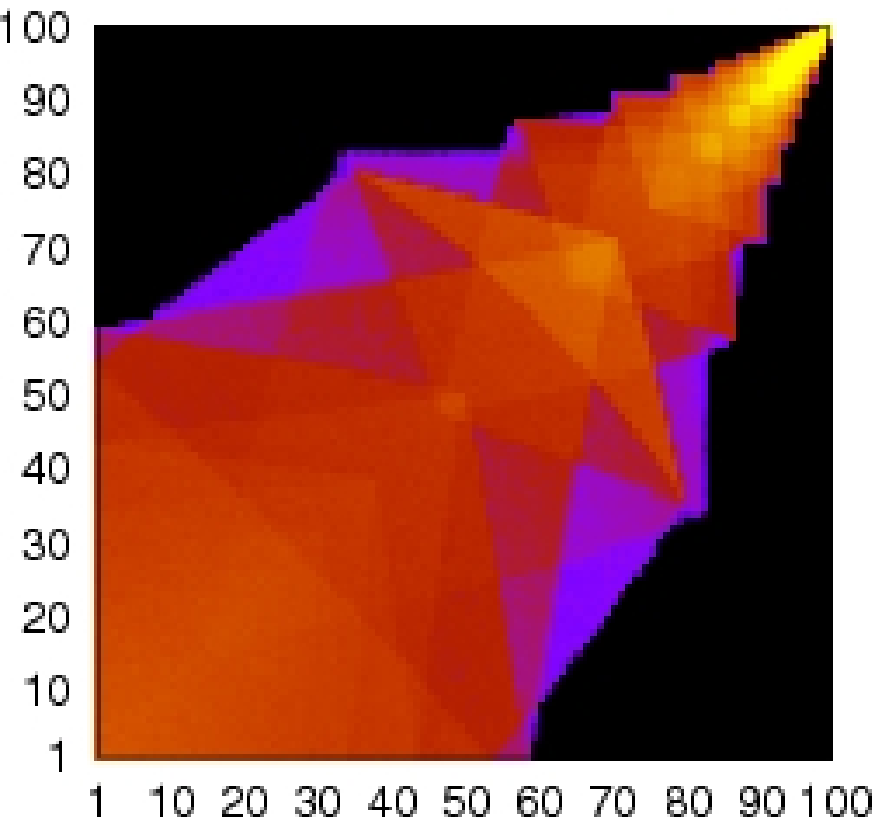}%
}
\vspace{2mm}
\vbox{\hspace{1cm}%
\includegraphics[bb=77 60 383 82, width=9cm]{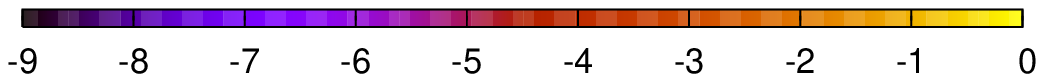}%
\hspace{1.2cm}%
\includegraphics[bb=114 57 272 82, width=4cm]{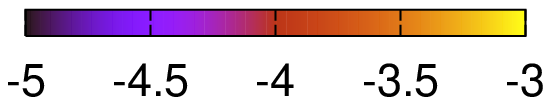}}
\caption{The density plot of the scattering matrices $\log_{10}|(R_N)_{nm}|$ and $\log_{10}|(T_N)_{nm}|$ and the classical analog of the later $\log_{10}|(T_{\rm class})_{nm}|$ calculated at inner radius $q=0.6$, wavenumber $k = 100.5\frac{\pi}{a}$ and with the total number of modes $N=150$.}
\label{pic:S_matrix}
\end{figure}

We are discussing the scattering over the bend at some fixed wavenumber $k$ and inner radius $q$. The scattering properties are contained in the transmission matrix $T$ (\ref{eq:scatt_T}) and the reflection matrix $R$ (\ref{eq:scatt_R}).  The wavefunction over the asymptotic region is described in $N$ modes, where $N \ge \rNo = \lfloor ka/\pi \rfloor$. We consider 
an incoming wave coming to the bend from the left side written in the asymptotic region as
\beq
  \psi_{\rm in}({\bf r}) = \sum_{n=1}^N a_n e_n^+({\bf r})\>.
  \label{eq:incoming}
\eeq
By introducing a vector of complex coefficients $a = \{a_n\}_{n=1}^{\rNo}$ we can write the transmitted
and reflected probability flux,  $j_{\rm T}$ and $j_{\rm R}$ respectively, in an elegant form
\beq
  j_{\rm R} = a^\dag \Pi a\>,\qquad
  j_{\rm T} = a^\dag \Sigma a\>,\qquad 
  j_0 = a^\dag a = j_{\rm R} + j_{\rm T}\>,
 \label{eq:ref_trans_prop_current}
\eeq
where we introduce matrices $\Pi$ and $\Sigma$ calculated from open-open mode blocks of
 $R$ and $T$:
\beq
  \Pi = R_\roo^\dag R_\roo\>,\quad 
  \Sigma = T_\roo^\dag T_\roo\>,\;\textrm{  where }\;
  \Pi + \Sigma = \id\>.
  \label{eq:ref_trans_amp_rel}
\eeq
The average transport properties are given by the first and the second moment of the probability currents $j_{\rm R, T}$ averaged over an ensemble of incoming states $a$. The ensemble represents states (vectors $a$) uniformly distributed over the 2N-dimensional sphere of radius $j_0$ \cite{newton:book:02,prosen:jpa:02}.  The {\it average probability currents} are given by
\beq
  {\cal R} = \frac{\ave{j_{\rm R}}_\alpha}{j_0}
           = \frac{1}{\rNo} \tr\{\Pi\}\>,
  \qquad	   
  {\cal T} = 1 - {\cal R}\>,
\eeq
and the {\it standard deviations of probability currents} (giving fluctuations within an ensemble) 
are written as
\beq
  \sigma^2_{\rm R} =
  \frac{\ave{(j_{\rm R} - \ave{j_{\rm R}}_\alpha)^2}_\alpha}{j_0^2} = 
  \frac{1}{\rNo+1} 
  \left[\frac{\tr\{\Pi^2\}}{\rNo} - {\cal R}^2\right] =   \sigma^2_{\rm T}\>.
\eeq
In the following we thus discuss the average reflection $\cal R$ and the dispersion of reflection 
$\sigma^2_{\rm R}$. An approximation of the transmission matrix $T$ can be determined from the semi-classical calculations, whereas for the reflection matrix $R$ it can not, as the reflection in the bend is a purely quantum phenomenon. The gross structure of matrices $R$ and $T$, similarly as of matrices $A$ and $B$, does not change significantly with increasing wavenumber $k$. In figure \ref{pic:S_matrix} we show the density plot of matrices $R$ and $T$ with $\rNo=100$ open modes.
The high probabilities in the matrix $T$ have a classical correspondence, which is revealed through the calculation of the classical scattering matrix $T_{\rm class}$ \cite{mendez:pre:02}. Both, the classical and the quantum transmission matrices feature similar patterns, but due to the quantum interference, we can not establish a clear correspondence. In the matrix $T$ we have a large area of high values so we can expect that transmission probability of individual modes should be high. It is important to notice the area in the reflection matrix 
of high intensity is concentrated around the last open mode with the index $\rNo$. \par
\begin{figure}[!htb]
\centering%
\vbox{\small\hspace{3cm}\hbox to9cm{before\hspace{2.5cm}middle\hspace{2.5cm}after\hfil}}
\vbox{%
\begin{minipage}[c]{1cm}\small $R_N:$ \end{minipage}%
\begin{minipage}[c]{10mm}
\hbox to10cm{%
\includegraphics*[width=3.9cm,origin=c]{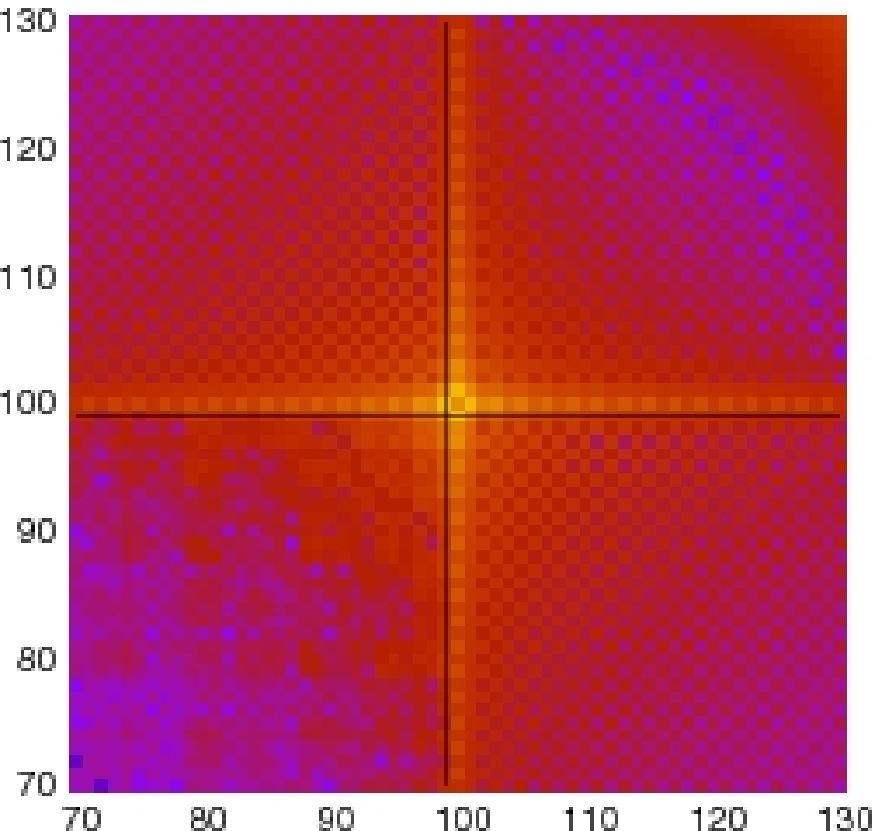}%
\hskip1pt
\includegraphics*[width=3.9cm,origin=c]{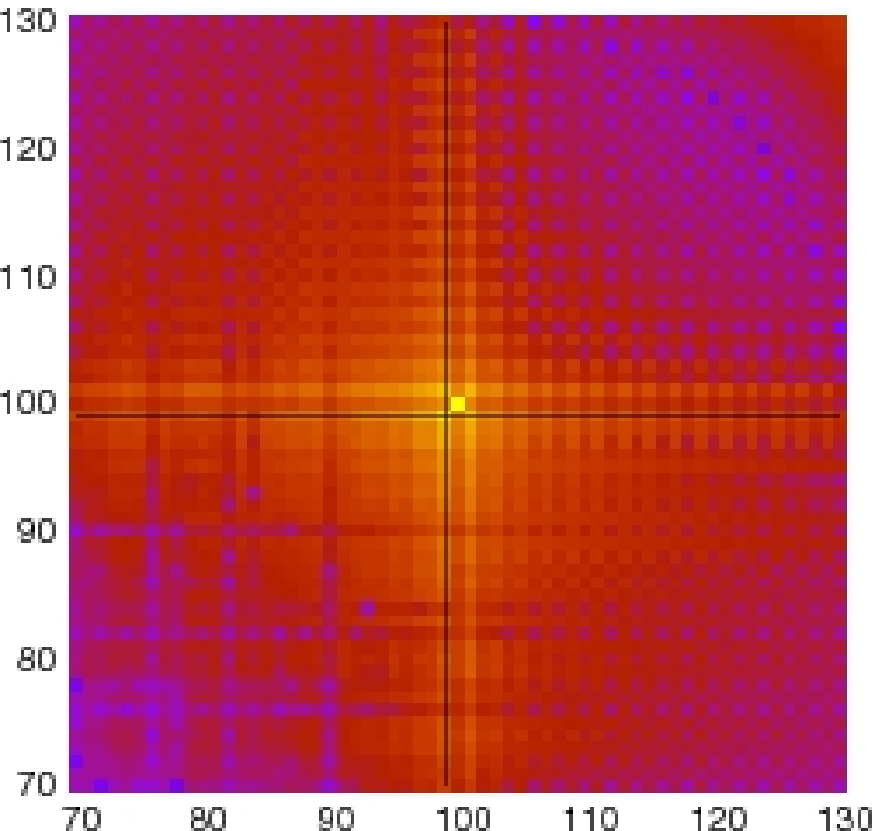}%
\hskip1pt%
\includegraphics*[width=3.9cm,origin=c]{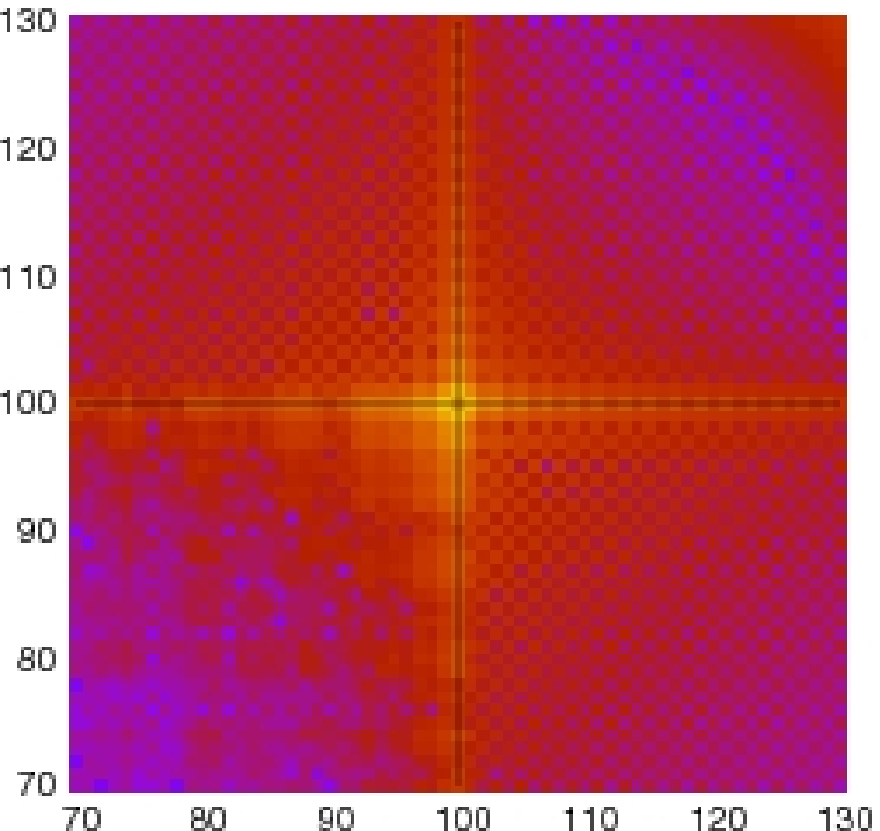}}
\end{minipage}}
\vbox{%
\begin{minipage}[c]{1cm}\small $T_N:$ \end{minipage}%
\begin{minipage}[c]{10mm}
\hbox to10cm{%
\includegraphics*[width=3.9cm,origin=c]{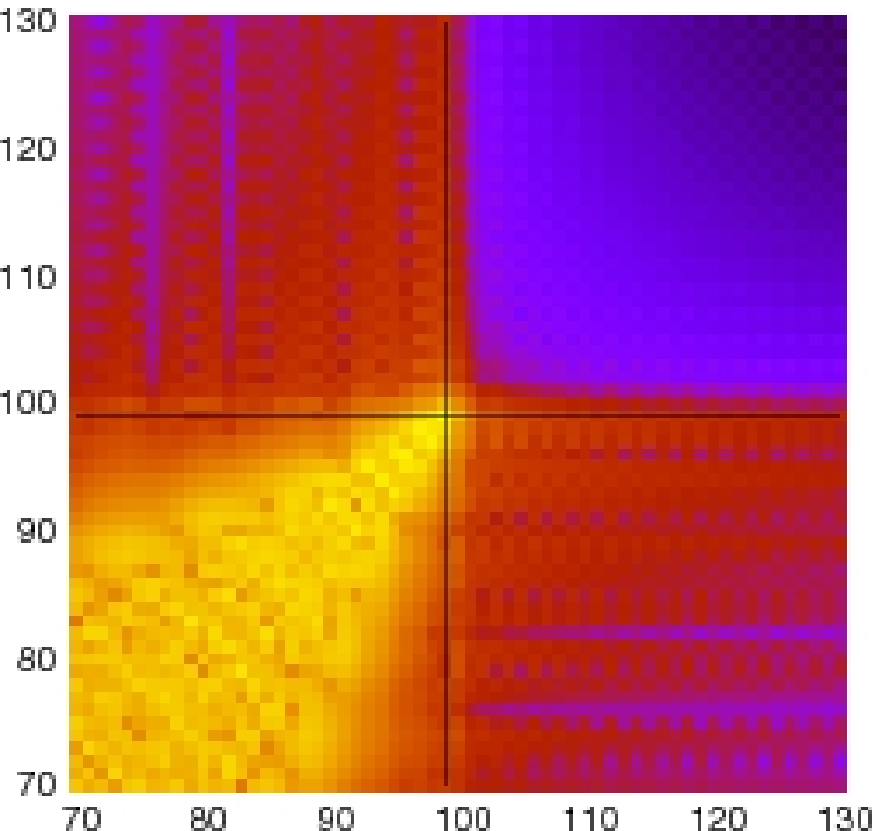}%
\hskip1pt%
\includegraphics*[width=3.9cm,origin=c]{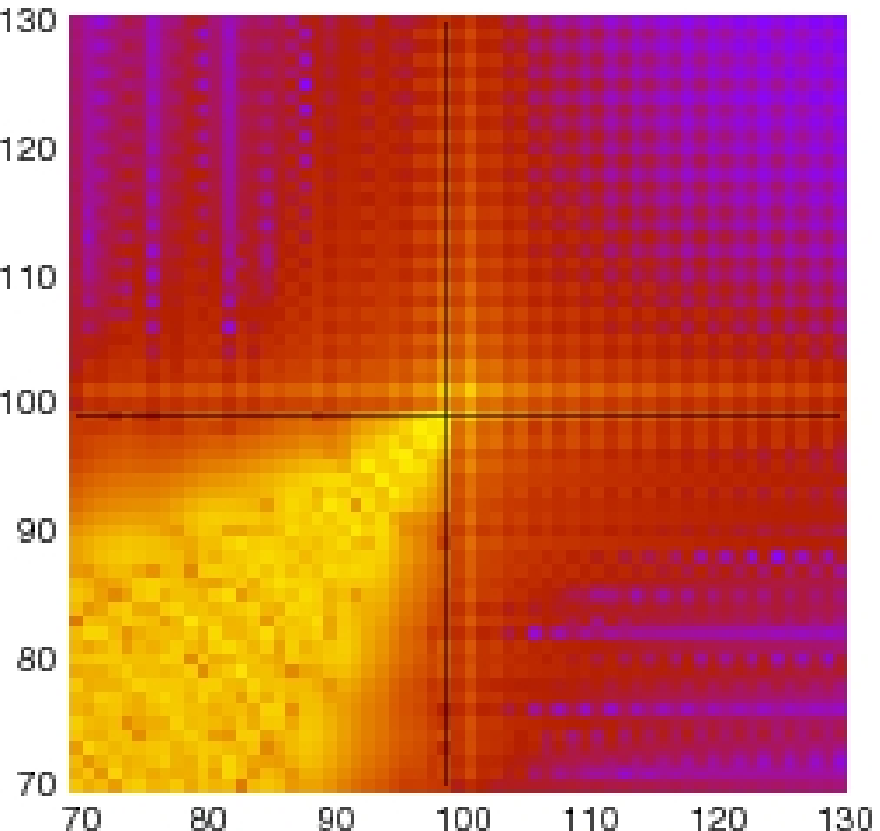}%
\hskip1pt%
\includegraphics*[width=3.9cm,origin=c]{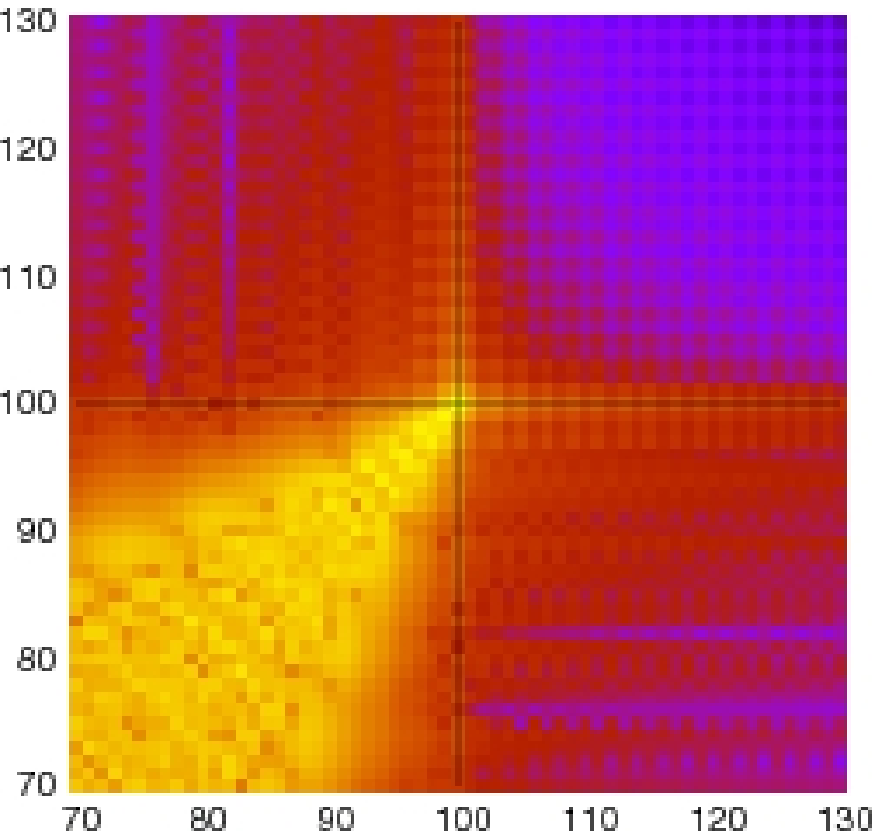}}
\end{minipage}}
\vbox{\includegraphics*[bb=77 56 383 90, width=10cm]{S_pallete.eps}}
\caption{The density plots of the reflection matrix $\log_{10}|(R_N)_{nm}|$ and the transmission matrix $\log_{10}|(T_N)_{nm}|$ around the point $(n,m) = (\rNo,\rNo)$ at wavenumbers $k = (100-10^{-3})\frac{a}{\pi}$ (before), $k\doteq 100 \frac{a}{\pi}$ (in the middle) and $k = (100+10^{-3})\frac{a}{\pi}$ (after the resonance). The solid lines denote indices $n= \rNo$ and $m = \rNo$.}
\label{pic:S_matrix_res}
\end{figure}

At the so called resonant wavenumbers $k_m = \frac{\pi}{a} m$ ($m\in\bN$) a new open mode appears in the asymptotic region and causes a strong increase in the reflection matrix elements at open modes with high indices. This is demonstrated in figure \ref{pic:S_matrix_res}, where we show the scattering matrices $R$ and $T$ around the highest open mode calculated calculated at $k\approx k_{100}$ and at $q=0.6$.  The changes are centred around the index $\rNo$ and significantly influence the average transport. \par
\begin{figure}[!htb]
\centering
\includegraphics[width=7.5cm]{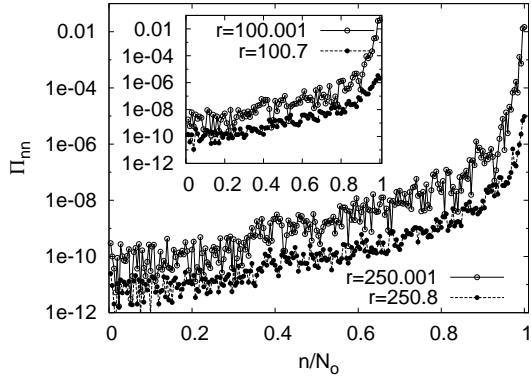}
\caption{The diagonal matrix elements $\Pi_{nn}$ at wave-numbers $k=r\frac{\pi}{a}$ as indicated in the figure, and fixed inner radius $q=0.6$. Note that $\rNo = \lfloor r\rfloor$.}
\label{pic:R_mat_elem}
\end{figure}

\begin{figure}[!htb]
\centering%
\vbox{%
\includegraphics*[width=7.5cm]{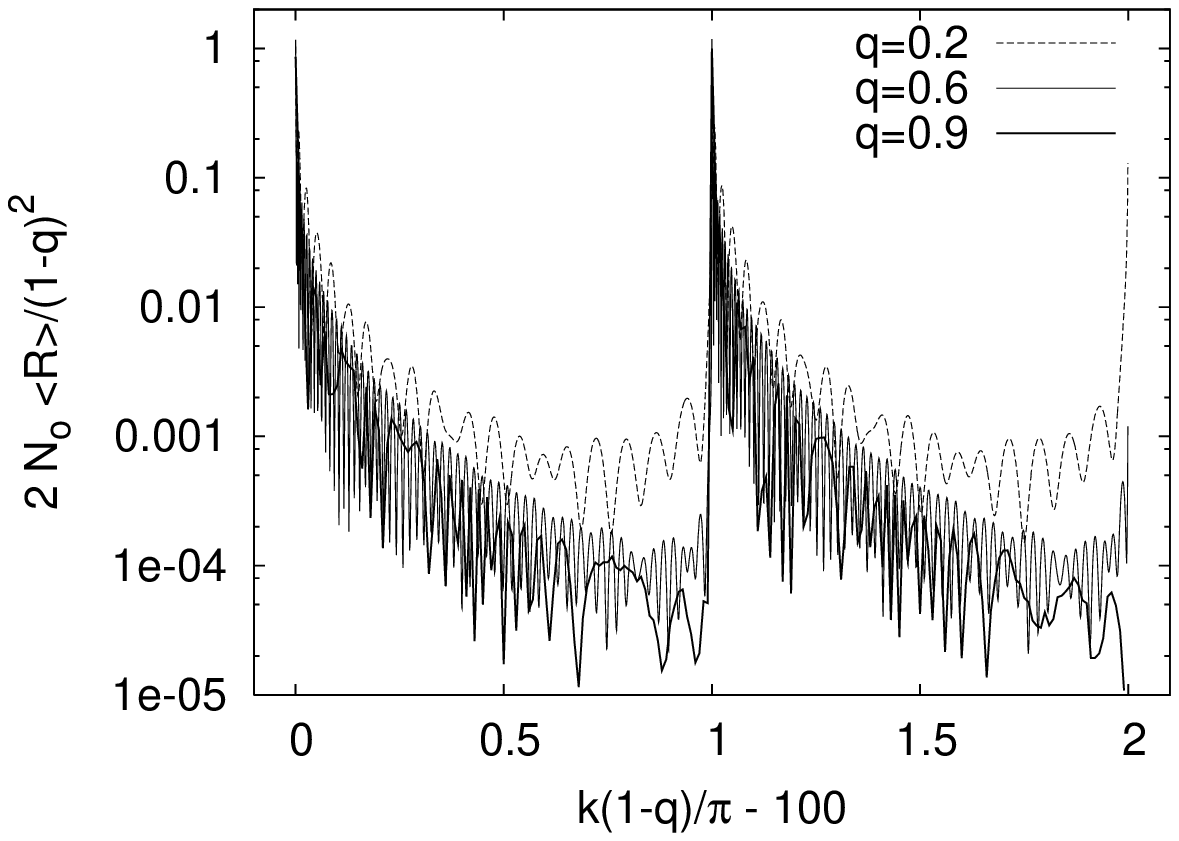}%
\includegraphics*[width=7.5cm]{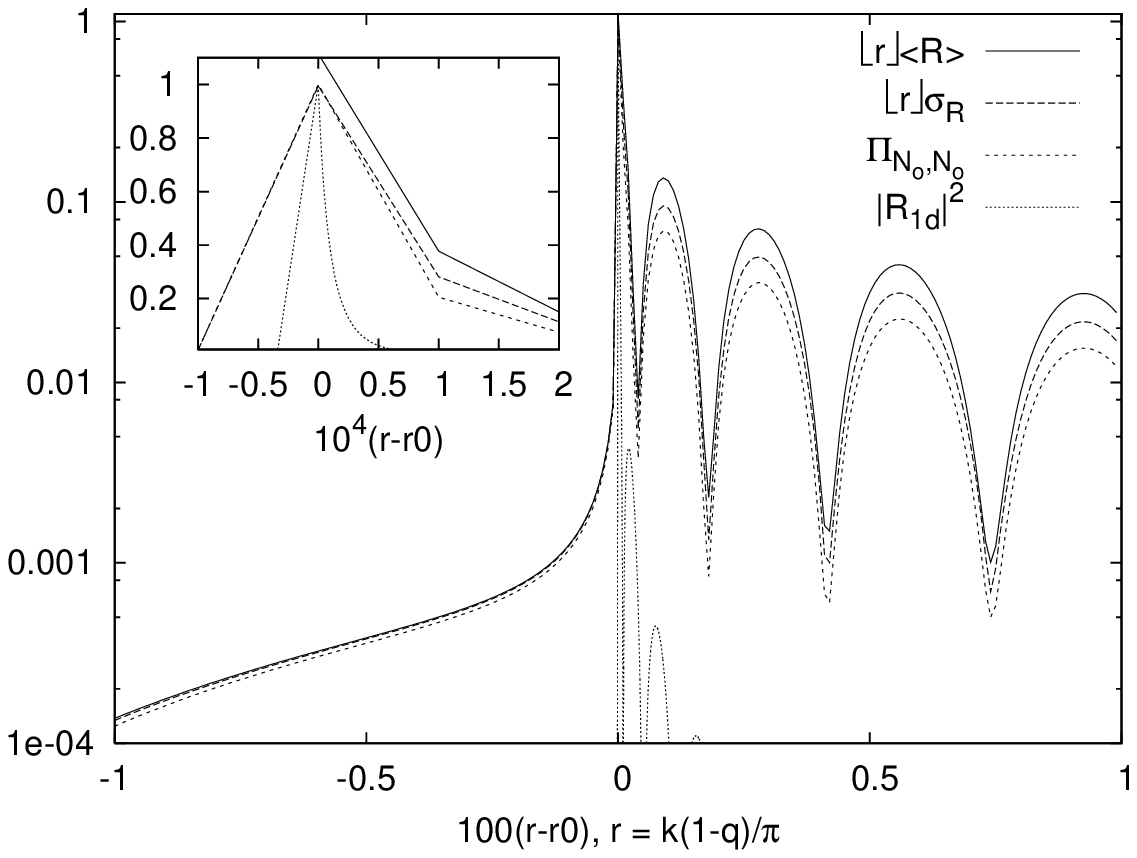}}%
\hbox to16cm{\hfil(a)\hfil(b)\hfil} 
\caption{Reflection a bend of an angle $\beta=\pi$ as a function of the wavenumber $k$ in the region of a reflection resonance $k=101 \frac{\pi}{a}$. In (a) we show the average reflection measure for different values of $q$ (indicated in the figure), while in (b) we show a zoom-in around the resonance and 
compare the average reflection and its standard deviation with the reflection in the last open mode
(inset shows even a more very narrow region of the resonance).}
\label{pic:reflect_res}
\end{figure}

\begin{figure}[!htb]
\centering
\includegraphics*[width=7cm]{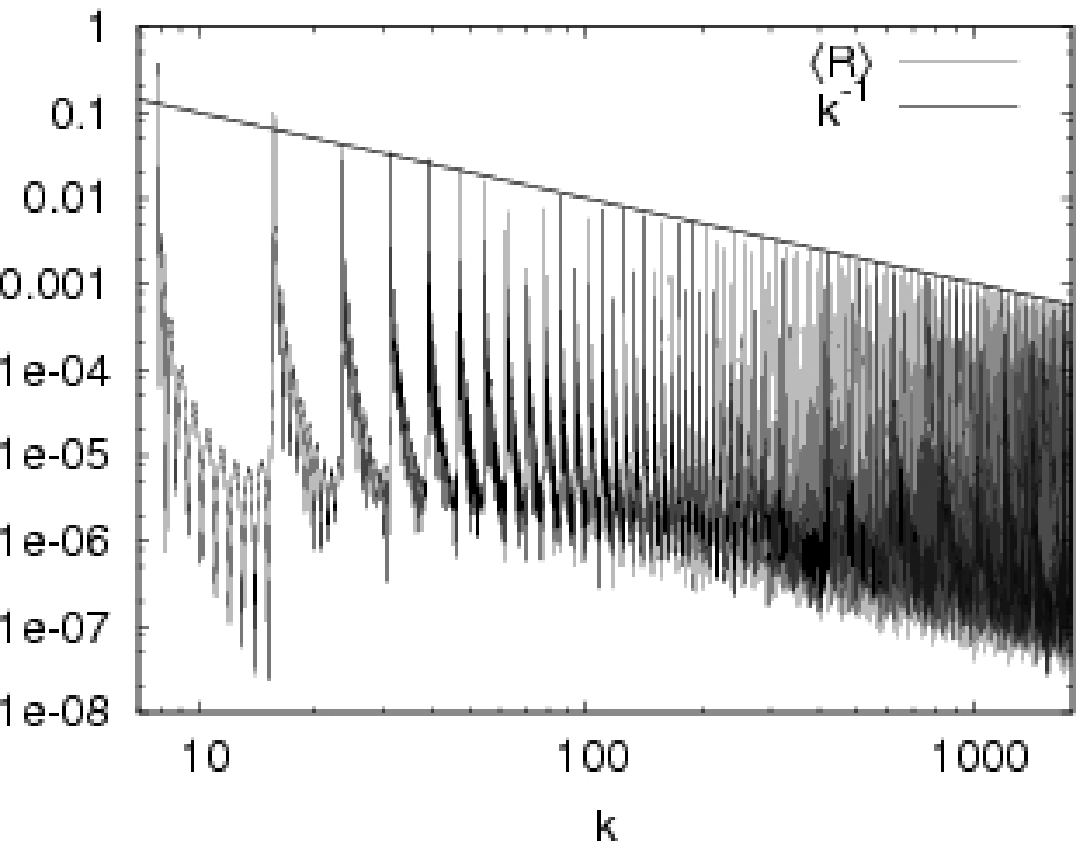}\hskip5pt%
\includegraphics*[width=7cm]{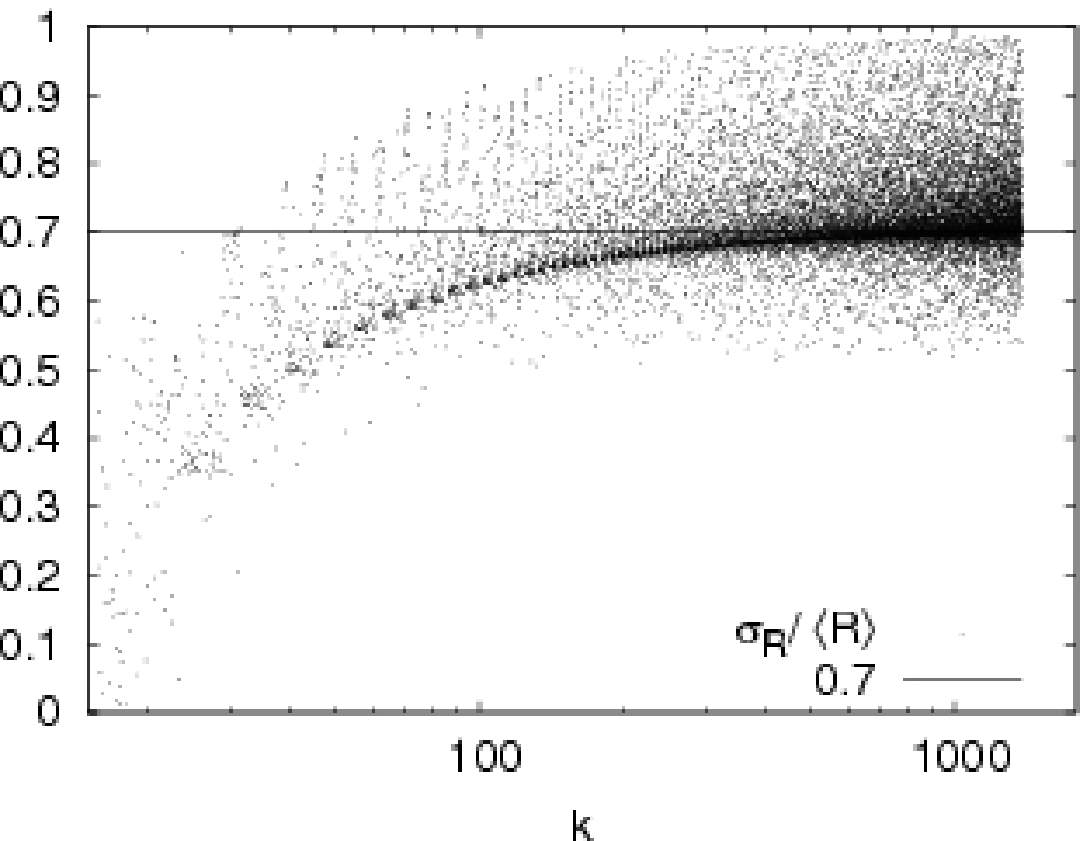}%
\caption{The average reflection $\cal R$ and the relative deviation of reflection $\sigma_{\rm R}$ over a larger interval of wave-numbers $k$ at $q=0.6$. }
\label{pic:R_gross}
\end{figure}

\begin{figure}[!htb]
\centering
\includegraphics[width=7.5cm]{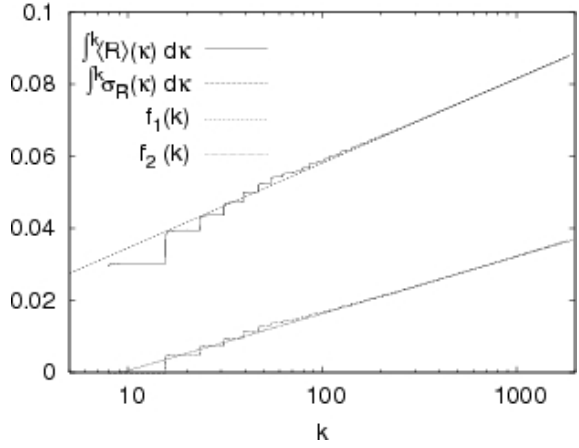}
\caption{The integral of the average reflection $\cal R$ and deviation $\sigma_{\rm R}$ over larger interval of wave-numbers $k$ of a bend with a radius $q=0.6$ and an angle $\beta=\pi$. The inserted lines are $f_1(k) = 0.01103 + 0.01022 \log k$ and $f_2(k) = -0.01553 + 0.006924 \log k$.}
\label{pic:R_int_gross}
\end{figure}

In order to clarify the contributions to the total reflection we plot in figure \ref{pic:R_mat_elem} the reflection probability of individual modes $\Pi_{nn}$ for wavenumbers near and far from the resonance. We see that the highest open mode has the strongest reflection and the reflection probability of ``all'' modes increases at resonant wavenumbers $k=k_{\rNo}$. In particular the reflection of highest open mode is almost perfect $\Pi_{\rNo, \rNo}\approx 1$. In the vicinity of the resonance we could effectively approximate the average reflection as ${\cal R}\approx \Pi_{\rNo,\rNo}/\rNo$ (see figure \ref{pic:reflect_res}).
The resonant wavenumbers $k_{\rNo}$ are important markers for anomalously strong reflection. This is illustrated in figure \ref{pic:reflect_res}, where we plot the average reflection $\cal R$ as a function of the wave-number around $k=k_{100}$. We see the average reflection $\cal R$ has a strong sharp maximum at resonant wavenumbers and decreases in an irregular oscillating manner with increasing wavenumber until crossing the next resonant wavenumber. The frequency of irregular oscillations increases with increasing wavenumber. From numerical results we see that $\cal R$ decreases with increasing $q$. In narrow channels $a\to 0$ at large wave-numbers we showed with a perturbative approach (see \ref{sec:pert_scatt}) that
\beq
{\cal R} \sim \frac{a^2}{\rNo}\>,
\eeq
which is confirmed numerically. The resonant behaviour around the resonant wavenumber can be partially explained by neglecting all open modes expect the one with the highest index $\rNo$. Such system can be treated as an independent 1d scatterer ($d=1$) with the reflection and transmission matrix elements reading
\beq
  R_{\rm 1d} 
  = -\frac{\sin(\beta \nu_{\rNo})} {\sin (\beta \nu_{\rNo} + \ii \mu_{\rNo})}\>,
  \quad
  T_{\rm 1d} 
  = \frac{\sin(\ii \mu_{\rNo})} {\sin (\beta \nu_{\rNo} + \ii \mu_{\rNo})}\>,
  \label{eq:R_T_resonance}
\eeq
with the phase shift 
\beq
  \mu_{\rNo} = 2\atanh\left( K_{\rNo} \right)\>,\quad 
   K_p = \frac{g_p A_{p,p}}{\nu_p B_{p,p}} \in [0,1]\>.
  \label{eq:R_T_phase_shift}
\eeq
At $k= k_{\rNo}$, the mode number $g_{\rNo} \sim \mu_{\rNo}$ and the phase shift $\mu_{\rNo}$ become zero yielding a perfect reflection in a 1d scattering model, $R_{\rm 1d} = -1$ and $T_{\rm 1d}= 0$. This treatment is meaningful, because the matrices $A$ and $B$ are approximately diagonal at $(\rNo,\rNo)$ with an algebraic decay of matrix elements when we move away from the diagonal. If the modes were strictly independent we would have $\Pi_{\rNo, \rNo}=\|R_{\rm 1d}\|^2$, but the algebraic tails in matrices $A$ and $B$ make this solution to hold only as a rough approximation as can be seen in figure \ref{pic:reflect_res}.
In figure \ref{pic:R_gross} we study measures of reflection $\cal R$ and $\sigma_{\rm R}$ over a larger range of wave-numbers $k$. From figure \ref{pic:R_gross}.a we see that $\cal R$ strongly oscillates with peaks at resonant wave-number $k_n$ and its upper bound decreases proportionally to $k^{-1}$, as predicted. The numerical results in figure \ref{pic:R_gross}.b indicate that $\sigma_{\rm R} < {\cal R}$ and ${\cal R} \sim \sigma_{\rm R}$ as $k$ goes to infinity.
To get rid of oscillations and get an overall average behaviour of $\cal R$ and $\sigma_{\rm R}$ we calculate their cumulative integrals with respect to the wave-number. The results are shown in figure \ref{pic:R_int_gross} and yield the following dependence
\beq
 \int_{k_0}^k {\cal R}\, \dd \kappa \propto
 \int_{k_0}^k \sigma_{\rm R}\, \dd \kappa = O(\log(k))\>.
 \label{eq:int_measures_asymp}
\eeq
This indicates together with previous conclusions that the reflection measures, averaged over small wave-number ranges, indeed scale as
\beq
 {\cal R}\sim \sigma_{\rm R} = O(k^{-1})\>,\quad k\to\infty\>.
 \label{eq:measures_asymp}
\eeq
It seems that this relation (\ref{eq:measures_asymp}) is valid for an arbitrary inner radius $q$ and represents a new and very useful information for the study of wave-guides and general billiards that include bends.\par
\begin{figure}
\centering
\vbox{%
\includegraphics*[width=7cm]{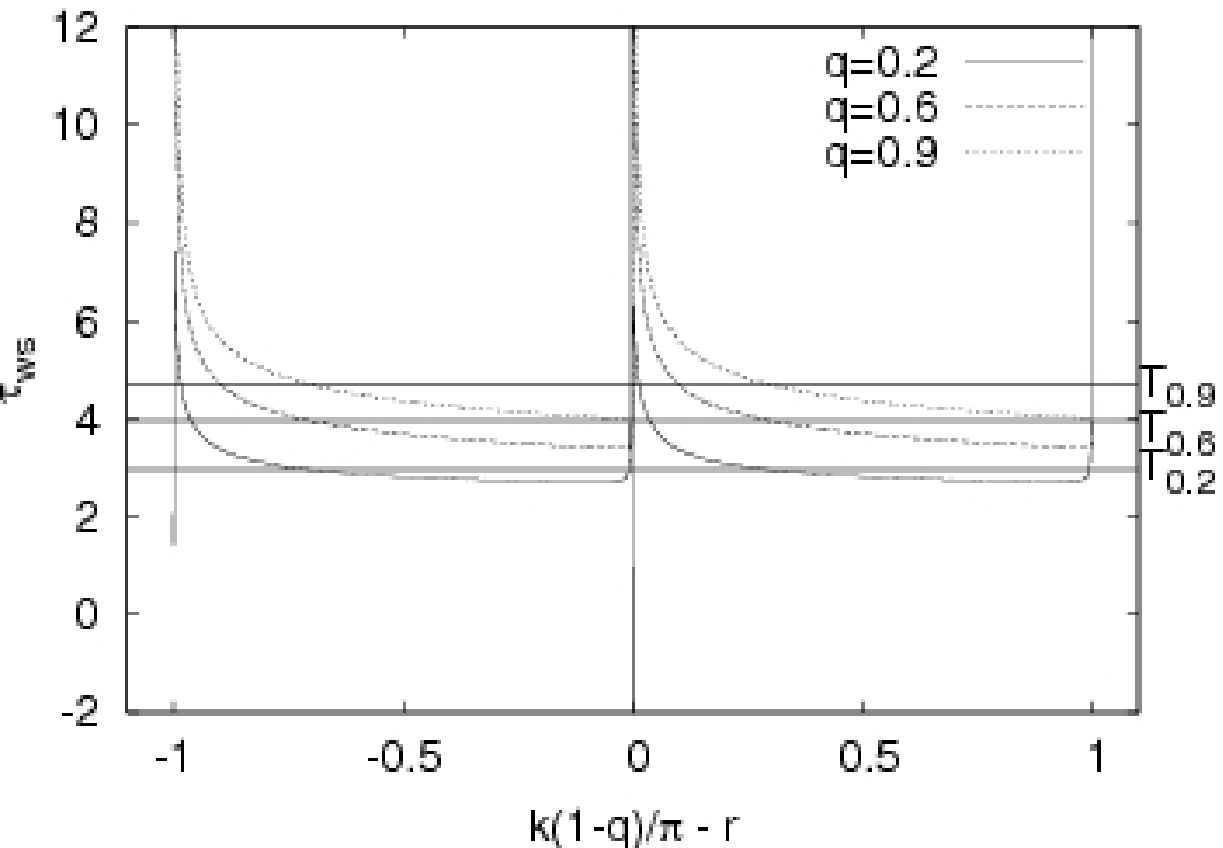}\hskip7pt%
\includegraphics*[width=7cm]{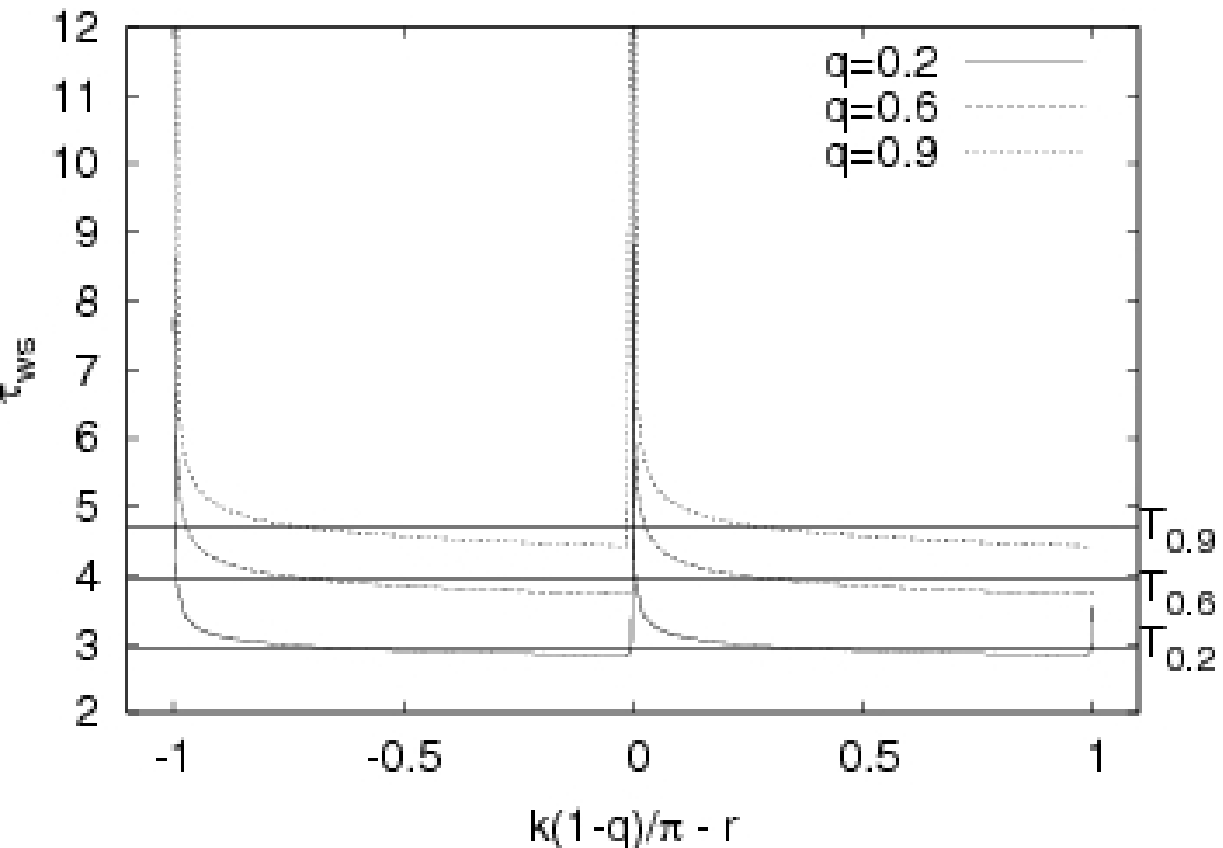}}
\caption{The Wigner-Smith delay time $\tau_{\rm ws}$ for the bend of angle $\beta=\pi$ around the reflection resonance at $r$ = 11, 101 (left,right). The solid horizontal lines with the labels $T_{0.2,0.6,0.9}$ represent the classical delay times at $q=0.2,0.6,0.9$, which are approximately given with the formula $T_q =2.45863 + 2.48696 q$.}
\label{pic:wigner_time}
\end{figure}

Another insight into the scattering properties gives the Wigner-Smith delay time $\tau_{\rm ws}$ \cite{wigner:phys_rev:55, smith:phys_rev:60}, which is the quantum analog of the geometric length travelled by a wave. In the semi-classical limit, where we could apply geometric optics, $\tau_{\rm ws}$ is equal to the average geometric length of classical trajectories over the open billiard. By using the hermitian variant of the {\it lifetime matrix}
\beq
  Q =  S_\roo^\dag \frac{\dd S_\roo}{\dd \ii k}\>,
  \label{eq:lifetime_matrix}
\eeq
the {\it Wigner-Smith delay time} is defined as
\beq
 \tau_{\rm ws} 
 = \frac{1}{2 \rNo} \tr \{Q\} 
 = \frac{1}{\rNo} 
   \tr\left\{R_\roo^\dag \frac{\dd R_\roo}{\dd \ii k}\right\} +
   \frac{1}{\rNo} 
   \tr\left\{T_\roo^\dag \frac{\dd T_\roo}{\dd \ii k}\right\}
 \>,
\eeq
where we have used block symmetries of our matrix $S$ (\ref{eq:bend_scatt_def}).  $\tau_{\rm ws}$ can be thought of as an average  delay time corresponding to particular modes, which are defined as
\beq
 \tau^n_{\rm ws} = \frac{1}{\rNo}
 \sum_{m=1}^{\rNo}
 \Im\left\{\left[R_\roo^\dag\right]_{nm} \left[R_{\rm  oo}'\right]_{mn} +
 \left[T_\roo^\dag\right]_{nm} \left[T_{\rm  oo}'\right]_{mn}\right\}\>,
\label{eq:time_part_mode}
\eeq
with derivative defined as $(\bullet)' = \dd/ \dd k$. Numerical results shown in figure \ref{pic:wigner_time} point to a similar dependence of Wigner-Smith delay time $\tau_{\rm ws}$ on the wave-number $k$ as the average reflection $\cal R$, just the oscillations are smoothed out. The time $\tau_{\rm ws}$ strongly increases near the resonance wave-number $k=k_{\rNo}$ due to intense changes in the scattering matrices $R$ and $T$ in the area around the index of the newly open mode. Qualitatively we can explain the singular behaviour by treating the highest open mode in the resonance regions 
within 1d scattering model in which the delay time is given by 
\beq
  \tau^{\rm 1d}_{\rm ws} 
  = \Im\left\{T_{\rm 1d}^* T_{\rm 1d}' + R^*_{\rm 1d} R_{\rm 1d}'\right\}
  = \frac {\beta \sinh(2 \mu_{\rNo})\nu_{\rNo}' - \sin (2\beta \nu_{\rNo}) \mu_{\rNo}'}
          {\cosh (2\mu_{\rNo}) - \cos (2 \beta \nu_{\rNo})}\>.
\label{eq:time_one_mode}
\eeq
The first term in the numerator of equation (\ref{eq:time_one_mode}) corresponds to the transmission and the second term to the reflection. By slowly increasing the wavenumber across the region of the resonance we can notice three different regimes: before, in the vicinity and after the resonant wave-number. Slightly before the resonance $k < k_{\rNo}$, a new real mode appears in the bend (see formula (\ref{eq:Nb_asym})) making the propagation across the bend very slow. From the formula (\ref{eq:time_one_mode}) we learn that this results in a large transmission time and consequently in a large time delay $\tau^{\rm 1d}_{\rm ws}$. In the instance of crossing the reflection resonance a new mode appears in the straight wave-guide, which causes a square-root singularity $\tau^{\rm 1d}_{\rm ws}\sim (k-k_{\rNo})^{-{1\over 2}}$ for $k> k_{\rNo}$ and its sign is determined by $2\beta \nu_{\rNo}$. This reflection term has a short-scale influence to the behaviour of the time delay and can enhance or reduce its size. Obviously, this is a very non-classical situation. By going further away from the resonance wave-numbers the reflection contribution to the time delay is levelled by an increasing transmission term due to a very slow propagation of the mode in the asymptotic region, which again increases the transition time. So we can experience one or two peaks of the time delay in the vicinity of the reflection resonance. Away from the reflection resonance the time delay drops even below the classical time. The latter we assume is due to reflection phenomena which reduces the classically expected phase shift. The presented 1d scattering model has only an instructive purpose and does not represent any useful quantitative approximation, similarly as was the case in the discussion of the reflection.
\section{Conclusions}

We present mathematical, numerical and physical background of the non-relativistic 2D scattering of a quantum particle on a circular bend connected to infinite straight waveguides. We discuss mathematical properties and derive numerical recipes for accurate and reliable calculation of the mode functions and the corresponding mode numbers in a bend. We take a special care of closed (evanescent) modes in the bend. The obtained modal structure and its properties are incorporated in a robust and stable numerical scheme for computing the scattering matrix with a controllable precision. Our numerical apparatus is applied to the study of transport properties. We focus mainly on the reflection, which is a purely quantum (wave) phenomenon since the back-reflection of classical rays is not possible. Our study is particularly focused on the possibility to investigate the (semi-classical) regime of very large wave-numbers. Some of the obtained physical properties of the scattering problem can be explained analytically. In addition, we present results on the Wigner-Smith delay in the bend. The obtained transport properties can be useful in discussing and predicting properties of open billiards (or wave guides) composed of arbitrary combination of bends and straight segments.

\section*{Acknowledgements}
MH would like to thank Prof. Dr. Nico Temme for references on literature considering cross-products of Bessel functions at imaginary orders. Useful discussions with M. \v Znidari\v c as well as the financial support by Slovenian Research Agency, grant J1-7347 and programme P1-0044, are gratefully acknowledged. 
%


\appendix
\section{The symmetry of mode numbers} \label{sec:mode_symm}
\begin{figure}[!htb]
\centering
\includegraphics[width=7cm]{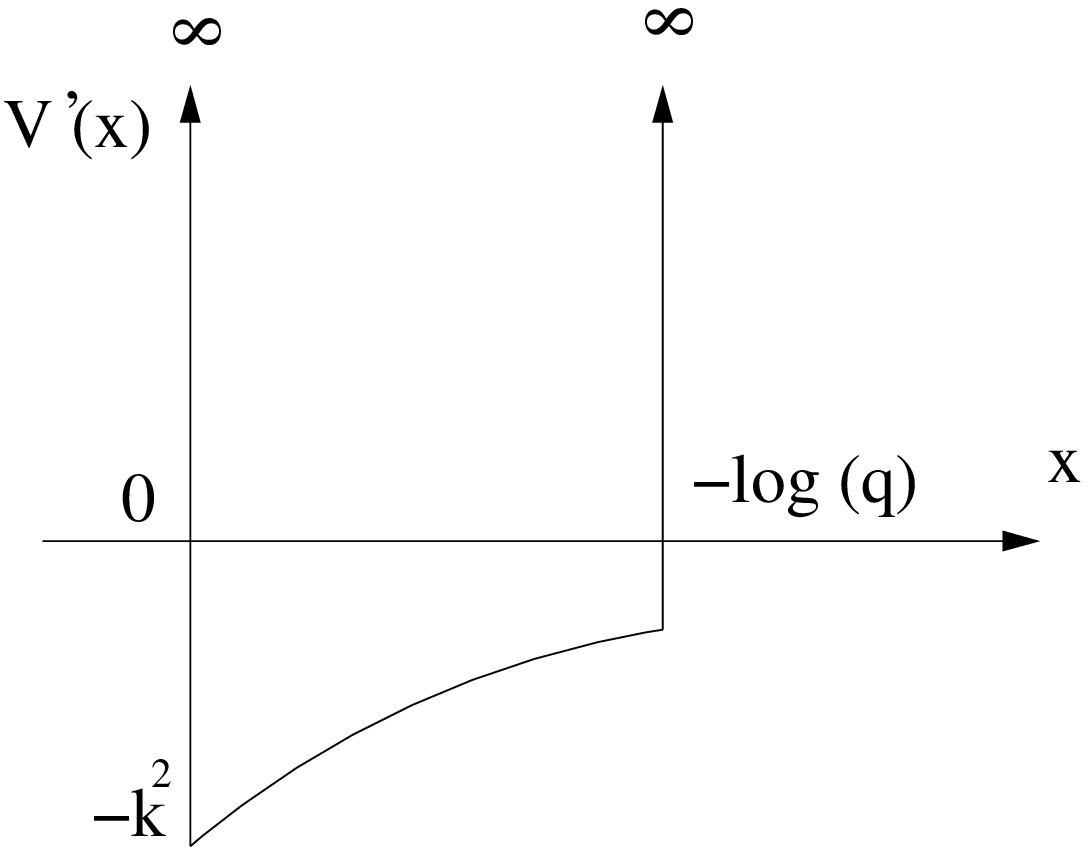}
\caption{The analog of the quantum potential in the eigenvalue equation for the mode functions in the bend.}
\label{pic:bess_pot}
\end{figure}

We prove the symmetry (\ref{eq:bcp_symm1}), by changing the variable $r = e^{-x}$ and transforming the Bessel equation (\ref{eq:eigen_bend}) and the corresponding boundary condition into the equation
\beq
  \frac{\dd^2 Z}{\dd x^2} + \left(k^2 e^{-2x} -\nu^2\right)Z = 0\>,
  \qquad 
  Z|_{x=0,\log q} = 0\>,
\eeq
which can be interpreted as a one-dimensional quantum mechanical eigenvalue problem, with the Hamiltonian $\hat H$ and potential $V'$:
\beq
\fl
  \hat H Z = -\nu^2 Z\>,\quad
  \hat H = -\frac{\dd^2}{\dd x^2} + V'(x)\>,\quad
  V'(x) = \left \{ 
  \begin{array}{lll} 
    -k^2 e^{-2x} &:& x\in[0,- \log q] \cr 
    \infty &:& \textrm{elsewhere}
  \end{array}
  \right.\>.
  \label{eq:schrod}
\eeq
Because the Hamiltonian $\hat H$ is a Hermitian operator, the eigenvalues are real $-\nu^2\in\bR$ yielding $\nu\in\bR\cup\ii\bR$. From the form of the potential $V'(x)$, depicted in figure \ref{pic:bess_pot}, we see that there are only finite number of real mode numbers $\nu^2>0$ and infinite number of imaginary mode numbers $\nu^2<0$. The independent solution of equation (\ref{eq:schrod}) are orthogonal with respect to the measure $\dd x = r^{-1}\,\dd r$ and so the weight function between mode functions $U_p(r)$ in the bend is $w(r) = r^{-1}$. 
\section{The number of modes in the straight and the bent waveguide} \label{sec:mode_diff}
We discuss the number of open modes $N_{\rm b}(k,q)=\card \Re\{ {\cal M}_{k,q,+}\}$ in the bend and its deviation from the number of modes in the straight waveguide $N_{\rm s} = \lfloor ka/\pi\rfloor$. The mode numbers in ${\cal M}_{k,q,+}$ continuously slide with increasing $k$ and fixed $q$ from the imaginary to the real axis by crossing the point $\nu=0$. This dynamics is depicted in figure \ref{pic:bess_value}. This means that $N_{\rm b}(k,q)$ is equal to the number of zeros $x$ of $Z_{0,x}(q)$ up to the value $k$ 
\beq
  N_{\rm b}(k,q) 
  = \card\Re \{{\cal M}_{k,q,+} \}
  = \card \{ x \le k: Z_{0,x}(q) = 0 \}\>.  
\eeq
By using substitutions $U = r^{-{1\over 2}} \varphi$ and $r = q + a x$ the mode problem in the bend (\ref{eq:bend_cross_eq0}) is in the case $\nu=0$ transformed into a 1d stationary Schr\"odinger equation $(\gamma=q/(1-q))$
\beq
   -\frac{\dd^2 \varphi}{\dd x^2} + 
   V(x) \varphi = e \varphi\>,
   \quad 
   V(x) = -[2 (x + \gamma)]^{-2}\>,\quad  x\in[0,1]\>,
\label{eq:schrod_1d}
\eeq
with the eigen-energy dentoed by $e=(ka)^2$. The discrete set of eigen-energies is ordered as $e_{n+1} > e_n$, $n\in\bN$. By setting $V=0$ in expression (\ref{eq:schrod_1d}) we obtain the mode problem for appropriately rescaled straight waveguide. In the eigen-energies in this case se $(\pi n)^2$, $n\in\bN$. Then taking into account the (empirical) fact $e_n \le (\pi n)^2 < e_{n+1}$ we can conclude
\beq
  0 \le N_{\rm b}(k,q) - N_{\rm s}(k,q) \le 1\>, \qquad 
  \forall q \in (0,1)\>.
\eeq
This means that at certain $k$ and $q$ we can have in the bend one open mode more, but not less than in the straight waveguide. In the semi-classical limit $k\to\infty$ the eigen-energies $e_n$ can be obtained using the Debye approximation valid for $qk\gg 1$. In this way we get a relation between the eigenvalues $e$ and its counting number $N_{\rm b}$
\beqa
   2 \pi N_{\rm b} &=&
       \left(4 k^2 + 1\right)^{1\over 2} - 
       \left(4(q k)^2 + 1\right)^{1\over 2}\nonumber \\
       &-&\atan \left((4k^2 + 1)^{-{1\over 2}} \right) 
        +  \atan \left((4(q k)^2 + 1)^{-{1\over 2}}\right)\>,
  \label{eq:WKB_real_0}
\eeqa
which yields with asymptotic expansion in $k$ the expression
\beq
 N_{\rm b}(k,q) = \frac{k a}{\pi} + 
              \frac{a}{8\pi kq} + 
	      \frac{a(1+q)}{64 \pi (qk)^2} + 
	      O\left(a(qk)^{-3}\right)\>.
\label{eq:Nb_asym}
\eeq
We see that $N_{\rm b}$ and $N_{\rm s}$ are close to each other for high wave-number and and not too small inner radius $q$. 

%
\section{The method of concatenating scattering matrices} \label{sec:comb_scatt}

Here we outline a method to concatenate the scattering matrices \cite{mayer:phys_rev:99} associated to scatterers on sectioned wave guides. Let us assume to have two scatterers labelled by A and B and with scattering matrices $S_{\rm A}$ and $S_{\rm B}$, respectively.
\beq
 S_{\rm A,B} = 
 \mymat{r_{\rm A,B}^\rL}{t_{\rm A,B}^\rR}
       {t_{\rm A,B}^\rL}{r_{\rm A,B}^\rR}
  \in\bC^{2N\times2N}\>.
\eeq
By combining both scatterers A and B in the order AB we build a ``larger'' scatterer with the scattering matrix $S$. The matrix $S$ is calculated from matrices $S_{\rm A, B}$ by a nonlinear operation $\odot: \bC^{2N\times2N}\times \bC^{2N\times2N}\to \bC^{2N\times2N}$ defined as
\beq
S = S_{\rm A} \odot S_{\rm B} = 
\mymat{r^\rL}{t^\rR}{t^\rL}{r^\rR} \in\bC^{2N\times2N}\>,
\label{eq:S_merge}
\eeq
which explicitly reads
\beqa
 &r^\rL = r_{\rm A}^\rL + 
           t_{\rm A}^\rR r_{\rm B}^\rL L^{-1} t_{\rm A}^\rL\>,\quad 
 &t^\rL = t_{\rm B}^\rL L^{-1}t_{\rm A}^\rL\>, \label{eq:S_merge_1}\\
 &r^\rR = r_{\rm A}^\rR + 
           t_{\rm A}^\rL r_{\rm B}^\rR {L'}^{-1} t_{\rm A}^\rR\>,\quad 
 &t^\rR = t_{\rm B}^\rR {L'}^{-1}t_{\rm A}^\rR\>,\label{eq:S_merge_2}
\eeqa
where we define $L = 1 - r_{\rm A}^\rR r_{\rm B}^\rL$ and $L' = 1 - r_{\rm B}^\rL r_{\rm A}^\rR$. Note that a bend on a straight waveguide can be treated as a scatterer. By combining bends of angles $\gamma'$ and $\delta'$ with scattering matrices $S(\gamma')$ and $S(\delta')$, respectively, we get a bend of angle $\gamma'+\delta'$ with the scattering matrix $S(\gamma'+\delta')$. The latter matrix can be obtained from matrices $S(\gamma')$ and $S(\delta')$ by the formula
\beq
  S(\gamma'+\delta') 
  = S(\gamma') \odot S(\delta') 
  = S(\delta') \odot S(\gamma')\>.
\label{eq:S_bend_merge}
\eeq
\section{Perturbative calculation of the scattering matrix for narrow bent wave-guide}
\label{sec:pert_scatt}
We present a semi-classical approximation, for $k\gg 1$, of a scattering matrix corresponding to a single bend on a straight wave-guide of width $a$, as one shown in figure \ref{pic:schema}. Here we are discussing only narrow channels $a \ll 1$, where the influence of closed modes on the scattering diminishes. Therefore closed modes are neglected in our calculations. We are working at wavenumbers $k$, where in all regions of the open-billiard the number of open modes is equal. This enables us to write the reflection and the transmission matrix in the following simpler form 
\beq
  R = -[C_+ - \ii S_+]^{-1} (C_- + \ii S_-)\>,
  \qquad 
  T = 2  [C_+ - \ii S_+]^{-1}\>, 
 \label{eq:semiclaas_scatt}
\eeq
where we use the diagonal matrices ${\cal C} = \Re \{{\cal F}\}$,  ${\cal S}= \Im \{{\cal F}\}$ and $G =  \diag \{g_n\}_{n=1}^{\rNo}$ to express the introduced matrices
\beqa
 C_{\pm} &=&  G^{1\over 2} A {\cal C} B^T G^{-{1\over 2}}
            \pm G^{-{1\over 2}} B {\cal C} A^T G^{1\over 2}\>,
 \label{eq:semiclaas_C} \\
 S_{\pm} &=&  G^{1\over 2} A {\cal S} A^T G^{1 \over 2}
            \pm G^{-{1\over 2}} B{\cal S} B^T G^{-{1\over 2}}\>.	    
 \label{eq:semiclaas_S} 
\eeqa
We proceed by rescaling the variables to dimensionless form by the following substitutions
\beq
\fl\hspace{1cm}
  y = a \xi\>,\quad 
  r = q + a\xi\>,\quad 
  \kappa = a k\>,\quad
  h_n = a g_n =\sqrt{\kappa^2 - (\pi n)^2}\>,\quad 
  \nu_p = \alpha v_p\>,
  \label{eq:semiclass_substitutions}
\eeq
with a new transverse coordinate $\xi\in[0,1]$, and geometric properties being described by the parameter $\alpha = a/q \ll 1$. The transition matrices are then expressed as
\beqa
  A_{np} = q^{1\over 2} Q_{np}\>,\quad 
  Q_{np} =
  \frac{\int_0^1 \dd \xi\; b_n(\xi) \phi_p(\xi) 
	                    (1+\alpha \xi)^{-{1\over2}}}
       {\sqrt{\int_0^1 \phi_p(\xi)^2 (1+\alpha \xi)^{-2}}}\>,
  \label{eq:trans_mat_asym_Q}\\
  B_{np} = q^{-{1\over 2}} P_{np}\>,\quad 
  P_{np} =
  \frac{\int_0^1 \dd \xi\; b_n(\xi) \phi_p(\xi) 
                          (1+\alpha \xi)^{-{3\over2}}}
       {\sqrt{\int_0^1 \phi_p(\xi)^2 (1+\alpha \xi)^{-2}}}\>.
  \label{eq:trans_mat_asym_P}
\eeqa
with $b_n(\xi)=\sqrt{2} \sin(\pi n \xi)$. The eigen-pairs $(v_p,\phi_p(\xi))$ are defined by the following differential equation and the boundary condition: 
\beq
  \frac{\dd^2 \phi_p}{\dd \xi^2}  + 
  \left(\kappa^2 - 
        \frac{v_p^2 - \frac{\alpha^2}{4}}{(1+\alpha \xi)^2}
  \right)\phi_p = 0 \>, \qquad 
  \phi_p(0)= \phi_p(1) = 0\>.
  \label{eq:mode_equation_dimless}
\eeq
We can easily recognize that the solutions of equation (\ref{eq:mode_equation_dimless}) converge in the limit $\alpha\to 0$ to $v_p = h_p$ and $\phi_p(\xi) = b_p(\xi)$. We assume that the solutions can be expanded in a power series of variable $\alpha$. The eigen-pairs can then be obtained using the standard perturbation theory with the perturbation parameter $\alpha$. The rescaled mode numbers are written as
\beq
  v_p^2 =  h_p^2  
    \left [1 + 
           \alpha + 
	   \alpha^2 \left(\frac{1}{3} - \frac{1}{2 (\pi p)^2 }\right)\right] + 
    \alpha^2 \left(\frac{1}{4}  + O(h_p^4) \right) + O(\alpha^3)\>,
  \label{eq:mode_num_asym}
\eeq
and the rescaled mode functions read as
\beq
 \phi_{p}(\xi) = \sum_n V_{np}\, b_n (\xi)\>,\quad
  V_{np} = \delta_{np} + 
  \alpha \frac{8 n p\, h_p^2}{\pi^4 (n^2 - p^2)^3} \delta_{n+p}^{\rm odd} + O(\alpha^2)\>,
  \label{eq:mode_fun_asym}
\eeq
where we use the symbol $\delta_n^{\rm odd} = (1: n \textrm{ is odd};\;0 : \textrm{otherwise})$. By plugging the mode functions $\phi_p (\xi)$ (\ref{eq:mode_num_asym}) into transition matrices $Q$ (\ref{eq:trans_mat_asym_Q}) and $P$ (\ref{eq:trans_mat_asym_P}) we obtain
\beqa
 Q_{np} = \delta_{np} + 
          \alpha \left(F_{np} + \frac{1}{4} \delta_{np}\right) + 
	  O(\alpha^2)\>,
 \label{eq:trans_mat_asym_pert_Q} \\
 P_{np} = \delta_{np} 
          - \alpha \left(F_{pn} + \frac{1}{4} \delta_{np}\right) + 
	  O(\alpha^2)\>,
 \label{eq:trans_mat_asym_pert_P}\\ 
 F_{np} = \frac{8 n p}{\pi^2 (n^2 - p^2)^2} 
 \left( \frac{h_p^2}{\pi^2(n^2 - p^2)}  - \frac{1}{4} \right) \delta_{n+p}^{\rm odd}\>.
 \label{eq:trans_mat_asym_pert_F}
\eeqa 
Note that the rescaled transition matrices $Q$ and $P$ satisfy the known identity $Q P^T = P Q^T = \id$. We insert the expressions for $Q$ (\ref{eq:trans_mat_asym_pert_Q}) and $P$ (\ref{eq:trans_mat_asym_pert_P}) back into  $S_\pm$ (\ref{eq:semiclaas_S}) and $C_\pm$ (\ref{eq:semiclaas_C}) and write the reflection and the transmission matrix as
\beqa
\fl R = -\frac{\alpha}{2} {\cal F} 
    \left[ H^{\frac{1}{2}} \left([F,{\cal C}]+\ii (F {\cal S})^{\rm s} \right) H^{-\frac{1}{2}}  + 
           H^{-\frac{1}{2}}\left([F^T,{\cal C}]+\ii (F^T {\cal S})^{\rm s}\right) H^{\frac{1}{2}} \right] + O(\alpha^2)\>,
\label{eq:R_asym_pert}\\
\fl T = {\cal F} - \frac{\alpha}{2} 
       \left(H^{\frac{1}{2}}[{\cal F},F]H^{-\frac{1}{2}} + 
	     H^{-\frac{1}{2}}[F,{\cal F}]H^{\frac{1}{2}}\right)
+ O(\alpha^2)\>.
\label{eq:T_asym_pert}
\eeqa 
where we have introduced the symbol $(A)^{\rm s} = A + A^T$ and the diagonal matrix $H=\diag\{h_n\}_{n=1}^{\rNo}$. The approximations of the reflection matrix $R$ (\ref{eq:R_asym_pert}) and the transmission matrix $T$ (\ref{eq:T_asym_pert}) are valid far away from the resonant condition $ak = \pi n$, because we assumed that $|g_m| > \alpha $ for all $m \le \rNo$. We conclude that the strength of reflection scales as ${\cal R}\sim \alpha^2$ and that narrow channels can be treated as perturbed straight wave-guides.
\end{document}